\documentclass[letterpaper]{article}
\usepackage{xr}
\usepackage[table]{xcolor}

\usepackage{tikz}
\usepackage{pgf}
\usepackage{amsmath,mathtools, booktabs, makecell, tabularx}
\usepackage{float}
\floatstyle{plaintop}
\restylefloat{table}
\usepackage{bbm}
\usepackage{enumerate}
\usepackage[authoryear, round]{natbib}
\usetikzlibrary{arrows,positioning,automata,calc,fit,shapes.geometric,arrows.meta}
\usepackage{geometry}
\usepackage{abstract}
\usepackage{multirow}   
\usepackage{color, colortbl}
\usepackage{caption}
\usepackage{subcaption}
\usepackage{longtable}
\usepackage[T1]{fontenc}
\usepackage{chngcntr}
\usepackage{apptools}
\usepackage{cancel}
\usepackage{rotating}

\usepackage{mathtools}
\mathtoolsset{showonlyrefs,showmanualtags}

\definecolor{darkred}{RGB}{100,0,0}
\definecolor{darkgreen}{RGB}{0,100,0}
\definecolor{darkblue}{RGB}{0,0,150}
\usepackage{xr}

\usepackage{xr-hyper}
\usepackage[pdftex,colorlinks]{hyperref}
\hypersetup{colorlinks=true, linkcolor=darkred, citecolor=darkgreen, urlcolor=darkblue}

\usepackage{subfiles}
\usepackage{scalerel,stackengine}
\usepackage{lipsum}

\usepackage[english]{babel}
\usepackage{amsthm}
\usepackage{bbm}
\usepackage{blindtext}
\usepackage{dsfont}
\usepackage{algorithm}
\usepackage{algorithmic}
\usepackage[utf8x]{inputenc}

\usepackage{ulem}

\makeatletter
\def\BState{\State\hskip-\ALG@thistlm}
\makeatother
\usepackage{amssymb}
\usepackage[inline]{enumitem}
\usepackage{url} 

\DeclareFontFamily{OT1}{pzc}{}
\DeclareFontShape{OT1}{pzc}{m}{it}{<-> s * [1.10] pzcmi7t}{}
\DeclareMathAlphabet{\mathpzc}{OT1}{pzc}{m}{it}
\usepackage{geometry}
\usepackage[pdftex,colorlinks]{hyperref}

  \hypersetup{
colorlinks =true,
citecolor    = blue,
citebordercolor = violet,
filebordercolor=blue,
linkbordercolor=blue
}

\makeatletter
\newcommand*{\addFileDependency}[1]{
  \typeout{(#1)}
  \@addtofilelist{#1}
  \IfFileExists{#1}{}{\typeout{No file #1.}}
}
\makeatother

\theoremstyle{plain}
\newtheorem{theorem}{Theorem}
\newtheorem{lemma}{Lemma}

\newtheorem{assumption}{Assumption}

\newcommand{\ee}{\end{eqnarray}}
\newcommand{\ba}[1]{\begin{align}#1\end{align}}
\newcommand{\ban}[1]{\begin{align*}#1\end{align*}}
\newcommand{\eqnn}{\begin{eqnarray*}}
\newcommand{\een}{\end{eqnarray*}}
\newcommand{\ea}{\end{align}}
\newcommand{\be}{\begin{eqnarray}}

\definecolor{orange}{RGB}{255,127,0}

\newcommand{\red}[1]{\textcolor{red}{#1}}

\newcommand{\logit}{\text{logit}}
\newcommand{\wh}{\hat}
\newcommand{\whb}{\hat{\beta}}
\newcommand{\wt}{\widetilde}
\newcommand{\nn}{\nonumber}
\newcommand{\var}{\mbox{Var}}
\newcommand{\betadr}{\hat{\beta}_{dr}}

\DeclareMathAlphabet{\mathpzc}{OT1}{pzc}{m}{it}

\newtheorem{innercustomthm}{Theorem}
\newenvironment{customthm}[1]
  {\renewcommand\theinnercustomthm{#1}\innercustomthm}
  {\endinnercustomthm}

\definecolor{LightCyan}{rgb}{0.88,1,1}

\def\spacingset#1{\renewcommand{\baselinestretch}%
{#1}\small\normalsize} \spacingset{1}

\renewcommand{\baselinestretch}{1.66}

\title{Doubly Robust Estimation under Possibly Misspecified Marginal Structural Cox Model}
\author{Jiyu Luo\thanks{Herbert Wertheim School of Public Health and Human Longevity Science, University of California, San Diego, La Jolla, CA 92093-0112, USA. E-mail: \href{mailto:jil130@ucsd.edu}{\textsf{jil130@ucsd.edu}}.},
~~~Denise Rava \thanks{Department of Mathematics, University of California, San Diego, USA, La Jolla, CA, 92093. E-mail: \href{mailto:drava@ucsd.edu}{\textsf{drava@ucsd.edu}}.},
~~~Jelena Bradic \thanks{Department of Mathematics and Halicioglu Data Science Institute, University of California,
San Diego, USA, La Jolla, CA, 92093. E-mail: \href{mailto:jbradic@ucsd.edu}{\textsf{jbradic@ucsd.edu}}.}
~~~and~~Ronghui Xu \thanks{Herbert Wertheim School of Public Health and Human Longevity Science, 
Department of Mathematics and
Halicioglu Data Science Institute, University of California, San Diego, La Jolla, CA 92093-0112, USA. E-mail: \href{mailto:rxu@health.ucsd.edu}{\textsf{rxu@health.ucsd.edu}}.}}

\date{}

\begin{document}
\maketitle
\begin{abstract}
In this paper we address the challenges posed by non-proportional hazards and informative censoring, offering a path toward more meaningful causal inference conclusions.
We start from the marginal structural Cox model, which has been widely used for analyzing observational studies with survival outcomes, and typically relies on the inverse probability weighting method. The latter hinges upon a propensity score model for the treatment assignment, and a censoring model which incorporates both the treatment and the covariates.
 In such settings model misspecification can occur quite effortlessly, 
and the Cox regression model's non-collapsibility has historically posed challenges when striving to guard against model misspecification through augmentation. 
We introduce an augmented inverse probability weighted estimator which, enriched with doubly robust properties, paves the way for integrating machine learning and a plethora of nonparametric methods, effectively overcoming the challenges of non-collapsibility. The estimator extends naturally to estimating a time-average treatment effect when the proportional hazards assumption fails. We closely examine its theoretical  and practical performance, showing that it satisfies both the assumption-lean and the well-specification criteria discussed in the recent literature  \citep{buja:etal:2019a, buja:etal:2019b, buja:etal:2019c}. Finally, its application to a dataset 
reveals insights into the impact of mid-life alcohol consumption on mortality in later life.   
\end{abstract}
\noindent
{\bf Keywords}:  
AIPW, Causal hazard ratio, Causal inference, Machine learning, Time-averaged treatment effect.

\section{Introduction}

\subsection{Background}

The marginal structural Cox model \citep{hernan2001marginal, hernan:robins:book} has been widely used in observational studies with survival outcomes to estimate the causal hazard ratio; see, for example, \citet{cole2003effect,feldman2004administration,sterne2005long,hernan2006comparison} and \citet{buchanan2014worth}, 
among many others. While the interpretation of the hazard function for causal inference has recently been under debate \cite[and references therein]{prentice2022intention, ying2023defense}, the Cox model formulation continues to be broadly utilized and can be easily adapted to derive more commonly agreed-upon interpretable quantities, such as survival probabilities.
 
The common approach to estimating  parameters under the marginal structural Cox model, i.e.~the causal estimands, has been 
inverse probability of treatment weighting (IPTW) and inverse probability of censoring weighting \citep[IPCW]{hernan2001marginal}; for the rest of the paper we will refer to it as the Cox-IPW estimator. Both weighting schemes require estimation of quantities related to the conditional distribution of treatment assignment and the conditional distribution of censoring given covariates. Parametric or semiparametric models for these conditional distributions are often subject to misspecification, leading to inconsistent estimators of the estimands. More flexible approaches such as machine learning or nonparametric methods, on the other hand, give rise to estimators that converge to the true estimands at slower than $\sqrt{n}$ rates \citep{BCH2013}.

To overcome the above drawbacks of inverse probability  weighting (IPW) approaches,  
 augmented inverse probability weighting (AIPW) methods have been developed  
  \citep[]{RRZ1995, scharfstein1999adjusting, robins2000marginal,robins2000marginalsnm,robins2000profile,RobinsRotnitzky01, van2003unified, bang2005doubly, tsiatis2006semiparametric}. These methods often exhibit so-called doubly robust  properties,  to be elaborated on in more details later.   
In particular, 
\cite{robins1998marginal} derived a generic class of semiparametric estimators for the parameters of marginal structural  models with a focus on efficiency, and without being robust against possible misspecification of the propensity score. 
A main challenge in developing doubly robust  estimators under the marginal structural Cox model is 
the non-collapsibility of the Cox regression model \citep{mart:vans}, i.e. the Cox model formulation including the proportional hazards assumption
 typically  no longer holds when a covariate is integrated out from the model, a fact  also well-known since the 1980s \citep{lancaster, gail:etal:84, ford:etal, xu:96}.
This gives rise to the difficulty of specifying a conditional survival outcome model that is needed in a doubly robust approach, and at the same time  compatible with the marginal structural Cox model which defines the causal estimand \citep{tchetgen2012parametrization}. 

In the use of the Cox proportional hazards model a major concern is the 
violation of the proportional hazards assumption. This also applies to the marginal structural Cox model. 
In such cases, the partial likelihood estimator 
 has been known to be swayed by the nuisance censoring distribution, even in the absence of confounding bias \citep{xu:96, XO2000}. While the doubly robust property helps to guard against possible misspecification of the so-called working models, little has been investigated in the causal inference literature when the model used to define the causal estimand is misspecified.

\subsection{Overview of the paper}

In this paper we derive an AIPW 
 estimator under the  marginal structural Cox model. New to our approach is the {joint augmentation} of the estimating functions for both the log hazard ratio and the nuisance baseline hazard function under the Cox model.
Unlike previous attempts using the partial likelihood under the Cox model, 
this joint augmentation gives rise to  estimating functions with independent and identically distributed (i.i.d.) constructs and enables contemporary developments from semiparametric theory to be applied. 
The augmentation leads to working models for the treatment assignment given the covariates, i.e.~the propensity score, the failure time and the censoring time given the treatment and the covariates. 
In order to circumvent  the non-collapsibility  problem described above, 
and specify a conditional failure time model that is compatible with the original marginal structural Cox model, 
data adaptive machine learning  or nonparametric methods are needed. 
We show that with cross-fitting  the resulting estimator has 
rate doubly robust property which allows $\sqrt{n}$ inference in the presence of slower than $\sqrt{n}$ convergence rate of the working models \citep{rotn:etal:21, hou:etal:2021}. 

Also new in this paper is the consideration of possibly misspecified marginal structural Cox model. 
 In place of the proportional hazards assumption on the distributions of the two potential failure time outcomes, we consider a general time-varying log hazard ratio.  We show that the AIPW estimator developed in this paper converges to a well-defined and a {\it well-specified} time-averaged treatment effect under the potential outcomes framework. We establish rate double robustness under this general time-varying log hazard ratio, which contains the marginal structural Cox model as a special case. 

In the following 
after reviewing related work in the literature, in Section 2  we 
define the notation, the model and the assumptions, 
and augment the Cox-IPW estimator of both structural parameters, namely the log hazard ratio and the infinite dimensional baseline hazard function. We study the estimand of the AIPW approach under misspecified marginal structural Cox model in Section 3 and show that it has the interpretation of a time-averaged causal effect. 
The asymptotic properties of the AIPW 
estimator  are established in Section 4. 
Through simulations of Section 5 we show that our estimator outperforms the existing IPW-Cox estimator both in terms of finite sample bias and variance, 
and we apply our estimator to data from a cohort of Japanese men in Hawaii followed since the 1960s in order to study the effect of mid-life alcohol exposure on late life mortality. We conclude with a discussion in the last section. The proofs of all the theoretical results are given in  
 the Supplementary Material.

\subsection{Related work}

For survival outcomes, AIPW approaches have been studied outside the Cox model. 
\cite{rotnitzky2005inverse} introduced an augmented IPCW  
method tailored for censored survival data. Works by \cite{zhang2012contrasting}, \cite{bai2017optimal}, and \cite{sjolander2017doubly} produced  doubly robust  estimators for  a contrast between the expected transformed potential failure times.  \cite{yang2020semiparametric} designed a doubly robust  estimator for structural accelerated failure time models. Both \cite{petersen2014targeted} and \cite{zheng2016doubly} derived targeted maximum likelihood estimators (TMLE) with doubly robust  properties by discretizing time and framing the failure time as a binary outcome, and \cite{rytgaard2022targeted} extended them to continuous time with possible competing risks and focuses on cumulative quantities like the survival probabilities.  
Within the additive hazards model, \cite{dukes2019doubly} and \cite{hou:etal:2021} presented doubly robust  estimators for hazard differences across low and high dimensions, and \cite{rava2023doubly} extended these to competing risks.

Outside the causal inference context 
another significant application of IPCW emerges when there is  violation of the proportional hazards assumption. 
 Several studies, including \cite{xu:96, XO2000, BKG2012, HH2012, NG2017, NG2021}, have worked on correcting the bias caused by a nuisance censoring mechanism using IPCW. It is worth noting, however, that not all these works explicitly use the term `IPCW'. Some opt for (conditional) survival distribution increments as weights, but mathematically, they align with the inverse probability of censoring weights. Among these works \cite{xu:96} and \cite{XO2000} assumed the censoring distribution to be  independent of the regressors in the model. \cite{BKG2012} allowed the censoring distribution to be different between the treatment groups but otherwise independent of the covariates.
  \cite{NG2017} allowed the censoring distribution to depend on the covariates, and introduced a survival tree method to estimate the conditional censoring distribution given the covariates. 

Informative censoring has recently received much attention in applications such as clinical oncology 
\citep{campigotto2014impact, templeton2020informative, olivier2021informative}.
Meanwhile efforts have been made in the statistical community in order to select covariates to account for censoring \citep{van2021principled}, to find transformations that render the relevant model and parameters identifiable \citep{dere:vank:2021}, and to apply copula type approaches \citep{czado:vank:2023}. 


Finally, there has been discussion in the literature about model robustness and assumption-lean estimation \citep[with discussion]{buja:etal:2019a, buja:etal:2019b}, in the sense that models are approximations \citep{box1979robustness} to perhaps much more complex reality. 
 The emphasis here is not on the possibly wrong working models for any nuisance parameters, but on the model that defines the estimand itself. While the original contributions \citep{buja:etal:2019a, buja:etal:2019b} center around parametric models, the Cox model serves as an obvious example of semiparametric models in the discussion \citep{whit:etal:2019, buja:etal:2019c}. It is clear that the problem has not been solved; our work here 
contributes to finding an assumption-lean solution to a well-specified estimand in that context. 
We will provide further discussion on this aspect after we describe our approach below. 

\section{Doubly Robust Score}
 
\subsection{Marginal Structural Cox Model}

Let \(A\) be a binary treatment. Define \(T(0)\) and \(T(1)\) as the potential failure times for a subject under $a=0$ and $1$, respectively. Let $\lambda_{T(a)}(t)$ denote the hazard function of the potential failure time $T(a)$, $a \in \{0,1\}$.
The marginal structural Cox model  \citep{hernan2001marginal}  for the potential outcomes  posits that
\be\label{msm}
\lambda_{T(a)}(t)=\lambda_0(t)\exp(\beta a),
\ee
where \(\lambda_0(t)\) represents an unknown baseline hazard function, and $\beta $ serves as the causal log hazard ratio, delineating the contrast between the potential failure time outcome distributions when \(a=1\) and \(a=0\).
The potential failure time \(T(a)\) might be right-censored by  \(C(a)\). Define \(\Delta(a)=I\{T(a) \leq C(a)\}\)  where $I(\cdot)$ is an indicator function, 
and \(X(a)=\min\{T(a),C(a)\}\). 
We use $T,C,X,\Delta$ to indicate the  observed counterparts
 once the treatment is received. 
Denote  $Z $   a vector of $p$-dimensional  covariates. 

 We adopt the standard causal inference assumptions  \citep{hernan:robins:book}, and `$\perp$' below indicates statistical independence.

\begin{assumption}[SUTVA] \label{assump1}
The potential outcomes of one subject are not affected by the treatment assignment of the other subjects, and there are no hidden versions of the treatments.      
\end{assumption}

\begin{assumption}[Consistency] \label{assump2}
    $T = AT(1) + (1-A)T(0)$, and $C = AC(1) + (1-A)C(0)$.
\end{assumption}

\begin{assumption}[Exchangeability]  \label{assump3}
    $(T(a), C(a)) \perp A \mid Z$, for $a = 0,1$.
\end{assumption}

\begin{assumption}[Strict Positivity]  \label{assump4}
There exists $0<\epsilon<1$ such that 
$\epsilon < P(A=1 | Z=z) < 1- \epsilon$,  
$P(C>\tau|A=a, Z=z) > \epsilon$, 
$P(T>\tau|A=a, Z=z) > \epsilon$ 
for all values of $a$ and $z$, where $\tau$ is a maximum follow-up time. 
\end{assumption}

Although model \eqref{msm} does not incorporate the covariates \(Z\), we adopt an informative censoring assumption that allows the censoring time to be dependent on $Z$.

\begin{assumption}[Informative Censoring] \label{assump5}
    $T(a) \perp C(a)\mid Z$, for $a = 0,1$.
\end{assumption}

In the following we start with the full data,  which include both potential outcomes for $a = 0,1$ and do not involve  censoring. We create our full data estimating functions based on the full data martingale as per model \eqref{msm} and its increments.  We then apply IPW which leads to the identification of the causal estimands using the observed data only. Finally, we develop novel joint augmentation which gives the desired doubly robust property. 


\subsection{Full-data score}

When we consider both the counterfactual outcome and censoring as missing data,  the full data, 
using the notion in \cite{tsiatis2006semiparametric},
is $(T(0), T(1), Z)$. From this, we can define for $a=0, 1$,  the full data counting process $N_T^a(t) = I(T(a) \le t)$, and the full data at-risk process $Y_T^a(t) = I(T(a) \ge t)$. It can be shown that 
\be
M_T^a(t; \beta, \Lambda) = N_T^a(t) - \int_0^t Y_T^a(u) e^{\beta a} d\Lambda(u) \nn
\ee
is a full data martingale with respect to the filtration $\mathcal{F}_t^a=\left\{N_T^a(u), Y_T^a(u^+):  0 \leq u \leq t\right\}$ under model \eqref{msm},   where $\Lambda(t) = \int_0^t \lambda_0(u) du $ \citep{FH1991}.
We start by constructing a full data score function, i.e.~an estimating function we would use if we were able  to  observe a single copy of  the full data.   
Using the  martingale property, we define the
 full data scores for $\Lambda(t)$ and $\beta$ as follows:
\begin{eqnarray}
\label{eq:full1}
D_1^f(t;\beta, \Lambda) &=& \sum_{a=0}^1 dM_T^a(t;\beta, \Lambda),\\
D_2^f(\beta, \Lambda) &=& \sum_{a=0}^1 \int_0^{\tau} a \cdot dM_T^a(t;\beta, \Lambda).
\label{eq:full2}
\end{eqnarray}
Note that $D_1^f(\beta, \Lambda, t)$ is a martingale difference function that is often used in survival analysis; see for example, \cite{LY2004}. 
For each $t$, the true values of $\beta$ and $\Lambda(t)$ satisfy
$
E\{D_1^f(\beta, \Lambda, t)\} = 0 ~~\text{and}~~ E\{D_2^f(\beta, \Lambda)\} = 0. 
$
In addition,  
it can be readily verified that for a random sample of size $n$, these would give the well-known Breslow's estimate of $\Lambda(t)$, as well as the partial likelihood score for $\beta$. 


\subsection{IPW score}

The above full data are {never} observed. Instead we have 
 the observed counting process $N(t) = I(X \le t, \Delta = 1)$, and the observed at-risk process  $Y(t) = I(X \geq t)$.
Define
\be
    M(t;\beta, \Lambda)     = N(t) - \int_0^t Y(u)e^{\beta A}d\Lambda(u). \label{observed.M}
\ee
Note that $ M(t;\beta, \Lambda)$ in general is  not a martingale under model \eqref{msm}, creating theoretical challenges in designing an effective estimation scheme. 
To bridge the divide between the full data  and the observed data, inverse probability weighting is commonly employed. This  involves weighting an observation by its inverse probability of being sampled from the target population \citep{horv:thom:52}, resulting in a pseudo-random sample representing the desired population. Specifically, in the presence of non-randomized treatment in observational studies, as well as informative censoring dependent on the covariates, \cite{hernan2001marginal} applied IPTW and IPCW to obtain consistent estimates of the parameters, provided that the relevant models are correctly specified .

Let $\pi(Z) = pr(A=1 | Z)$ and $\tilde \pi(A,Z)=\pi(Z)^A\{1-\pi(Z)\}^{1-A}$. 
 In addition, let $S(t;A,Z) = P(T > t|A,Z)$ and $S_c(t;A,Z) = P(C >t|A,Z)$ denote the conditional survival function of $T$ and  $C$, respectively. 
 We now have the IPW scores:
\be
D_1^w(t;\beta, \Lambda, \pi, S_c) &=& \frac{dM(t;\beta, \Lambda)}{\tilde \pi(A,Z) S_c(t;A,Z)}, 
\label{d1w} \\
D_2^w(\beta, \Lambda, \pi, S_c) &=& \int_0^{\tau} \frac{A \cdot dM(t;\beta, \Lambda)}{\tilde \pi(A,Z)S_c(t;A,Z)}. \label{d2w}
\ee
It can also be readily verified that for a random sample of observed data, $\sum_{i=1}^n D_{1i}^w(t;\beta, \Lambda, \pi, S_c)=0$ gives a weighted Breslow's estimate of $\Lambda(t)$ and,  after profiling out $\Lambda(\cdot)$, $\sum_{i=1}^n D_{2i}^w(\beta, \Lambda, \pi, S_c)$ gives the weighted  partial likelihood score for $\beta$.
The Supplementary Material  contains a formal proof of identifiability of these causal estimands via the IPW scores \eqref{d1w} and \eqref{d2w}. 

\subsection{Augmented IPW score}\label{der}

The IPW score is unbiased when the weights, or equivalently, \(\pi(Z)\) and \(S_c(t;A,Z)\) are known \citep{hernan2001marginal}. In practice these quantities are typically unknown. Propensity score models and conditional censoring models are often employed to estimate the respective conditional distributions. When these models are misspecified, the resulting estimate of the causal hazard ratio becomes inconsistent.
 
To protect against possible misspecification of the models, semiparametric  theory has been developed   to augment the IPW score \citep{tsiatis2006semiparametric}. The resulting 
 AIPW score possesses the  doubly robust properties that will be described in details later. 
 In particular, \cite{van2003unified} augmented the IPTW score function for a binary treatment, and \cite{rotnitzky2005inverse} augmented the IPCW score function for a survival parameter of interest. Here we apply these approaches together to the full data martingale increments, in order to simultaneously account for confounding and informative censoring. 
We will show that the resulting AIPW score is doubly robust. 

Denote the counting process for censoring events $N_c(t) = I(X \le t, \Delta =0)$, and $\Lambda_c(t;A,Z) = - \int_0^t S_c(u;A,Z)^{-1} d S_c(u;A,Z) $ the cumulative hazard function of $C$ given $A,Z$. 
Define \( M_c(t;A, S_c) = N_c(t) - \int_0^t Y(u) d\Lambda_c(u;A,Z) \); 
then it is a martingale with respect to its natural history filtration.
Following \cite{zhang2012double} and \cite{LX2022}, 
define also the censor-free counting process $N_T(t) = I(T \le t)$, and the censor-free  at-risk process $Y_T(t) = I(T \ge t)$. Let $M_T(t; \beta, \Lambda) = N_T(t) - \int_0^t Y_T(u)e^{\beta A}d\Lambda(u) $. 
We note that $M_T$ here is not a martingale, in the presence of confounding.

 We proceed to simultaneously augment both equations \eqref{d1w} and \eqref{d2w}, leading to AIPW estimate of $\beta$ and  $\Lambda$ jointly. 
 {Joint augmentation mitigates potential bias that may emerge when augmenting distinct model components separately. By considering the interdependencies among augmented elements, it enhances the precision of parameter estimates, thereby bolstering the accuracy and reliability of inferential outcomes.} 
 Our simultaneous augmentation leads to the following new estimating equations: 
 $$D_1 (t; \beta, \Lambda, \eta) =0 \mbox{ and } D_2 (\beta, \Lambda, \eta) =0, $$
where $\eta$ denotes the vector of three nuisance functions $\pi, S$ and $S_c$ (whenever possible), 
\begin{eqnarray}
D_1 (t; \beta, \Lambda, \eta)  
&= &
\frac{dM(t;\beta, \Lambda)}
{\tilde \pi(A,Z)S_c(t;A,Z)} 
- \frac{E\{dM_T(t;\beta, \Lambda)|A,Z\}}
{\tilde \pi(A,Z) 
} + \sum_{a=0}^1  
{E\{dM_T(t;\beta,\Lambda)| A=a,Z\}}
\nn \\ 
& + &\sum_{a=0}^1  w^a(A,Z) \int_0^t \frac{dM_c(u;a, S_c)}{S_c(u;a,Z)} 
E\{dM_T(t;\beta,\Lambda)|T \ge u, A=a, Z\}, 
\nn 
\end{eqnarray}
with $w^a(A,Z)=  A^a (1-A)^{1-a}  \{ \tilde \pi(A,Z)\}^{-1}$, and 
\begin{eqnarray}
D_2 (\beta, \Lambda, \eta) 
&= &\int_0^\tau \bigg[ \frac{A\cdot dM(t;\beta, \Lambda) }{\pi(Z)S_c(t;A,Z)} 
- \frac{A\cdot E\{dM_T(t;\beta, \Lambda)|A,Z\}}{\pi(Z)
} + 
{E\{dM_T(t;\beta,\Lambda)| A=1,Z\}} 
\nn \\ 
& +& \frac{A}{\pi(Z)} \int_0^t \frac{dM_c(u;1, S_c)}{S_c(u;1,Z)}
E\{dM_T(t;\beta,\Lambda)|T \ge u, A=1, Z\} \bigg]. 
\nn
\end{eqnarray}
Simplifying  the conditional expectations above, leads to the newly proposed AIPW scores: 
\be\label{eq:aipw1}
   D_{1}(t; \beta, \Lambda, \eta) &=& d\mathcal{N}^{(0)}(t;\eta) - \Gamma^{(0)}(t;\beta,\eta) d\Lambda(t), \\
    D_{2}(\beta, \Lambda, \eta) &=& \int_0^\tau d\mathcal{N}^{(1)}(t;\eta) - \Gamma^{(1)}(t;\beta,\eta) d\Lambda(t),  \label{eq:aipw2}
\ee
where for $l=0, 1$,  we have
\be
    d\mathcal{N}^{(l)}(t;\eta) &=& \frac{  A^l dN(t) }{\tilde \pi(A,Z)S_c(t;A,Z)} + \frac{A^l dS(t;A,Z)}{\tilde \pi(A,Z)} \nonumber\\
    &-& 
   \sum_{a=0}^1 a^l \left\{1 + w^a(A,Z)J(t;a,S,S_c) \right\} dS(t;a,Z), 
    \\
    \Gamma^{(l)}(t;\beta,\eta) &=& \frac{  A^l Y(t)e^{\beta A} }{\tilde \pi(A,Z)S_c(t;A,Z)} - \frac{A^l S(t;A,Z)e^{\beta A}}{\tilde \pi(A,Z)} \nonumber\\
    &+& 
     \sum_{a=0}^1 a^l e^{\beta a} \left\{1 + w^a(A,Z) J(t;a,S,S_c)\right\}S(t;a,Z), 
\ee
and 
$
J(t;a,S,S_c) = \int_0^t {dM_c(u;a,S_c)} / \{S(u;a,Z)S_c(u;a,Z)\}
$.
Note that the first term in $d\mathcal{N}^{(l)}(t; \cdot)$ is the IPW version of  $ A^l dN(t) $, and the rest  are   augmentation for both  IPTW and IPCW used in the first term; these expressions are parallel to those derived in \cite{tsiatis2006semiparametric} and \cite{bai2017optimal} for the simple mean of a (transformed) failure time. 
Similarly,  $ \Gamma^{(l)}(t; \cdot) $ can be seen as the augmented weighted $  A^l Y(t)e^{\beta A}$.
We also note that the observed $ A^l dN(t) $ and $  A^l Y(t)e^{\beta A}$ for $l=0, 1$ correspond to the building blocks of the original full data  score \eqref{eq:full1} - \eqref{eq:full2}: 
$d N_T^a(t)$, $a\cdot d N_T^a(t)$, $ Y_T^a(u) e^{\beta a}$ and $a\cdot  Y_T^a(u) e^{\beta a}$, respectively.  
In this way the AIPW score \eqref{eq:aipw1} - \eqref{eq:aipw2} parallels the full data score  via weighting and then augmentation. 
 

The theorem below introduces a crucial doubly robust property inherent to the newly introduced population score, elucidating the significance of the new joint augmentation. This theorem marks an initial stride towards the potential establishment of a rate or model  double robust properties, and to the best of our knowledge, it stands as a novel contribution. It underscores that the population equations are centered when either one of the two key conditions holds (but not necessarily both): the first condition pertains to the accurate model specification for the conditional distribution of the failure time, while the second condition concerns the correct model specification of both missingness mechanisms, the conditional censoring  as well as the treatment assignment probability.  
In the following the superscript `$^o$' denotes the true value of a parameter.

\begin{theorem}\label{AIPW:thm:dr}
Under Assumptions \ref{assump1}-\ref{assump5}, if either $S = S^o$, or both $S_c = S_c^o$ and $\pi = \pi^o$, then for all $t$,
$E\{D_1(t;\beta^o,\Lambda^o, \pi,S, S_c)\} = 
E\{D_2(\beta^o,\Lambda^o,\pi,S, S_c)\} = 0. 
$
\end{theorem}


\section{Misspecified Cox model}

The proportional hazards assumption under model \eqref{msm} might be violated. 
Without this assumption a saturated model for the  causal log hazard ratio that may change over time is
 \be\label{eq:betat}
\lambda_{T(a)}(t)=\lambda_0(t)\exp\{\beta(t) a\}.
\ee
Obviously $\beta(t) = \log\{ \lambda_{T(1)}(t) / \lambda_{T(0)}(t) \}$. 
Under the saturated model \eqref{eq:betat}, instead of estimating $\beta(t)$ at every $t$, in practice it is often of interest to estimate an average log hazard ratio:
\be\label{eq:ave}
\beta^* = \frac{ \int_0^\tau \omega(t) \beta(t) dt }{ \int_0^\tau \omega(t) dt }, 
\ee
where $ \omega(t) >0 $ is a weight function. 
If $\beta(t) =\beta_0$, then $\beta^* =\beta_0$. 

Under model \eqref{eq:betat} we may still consider for the full data
\be
M_T^a(t; \beta, \Lambda) = N_T^a(t) - \int_0^t Y_T^a(u) e^{\beta a} d\Lambda(u), \nn
\ee
where $\Lambda(\cdot)$ is right-continuous, non-decreasing with $\Lambda(0) = 0$ and $\Lambda(\tau) < \infty$. 
Note that $ M_T^a(t; \beta, \Lambda)$
is no longer  a full data martingale for any constant $\beta$ (and $\Lambda$) 
if the proportional hazards assumption is violated, i.e.~model \eqref{msm} does not hold. 

In order to understand what the full data score \eqref{eq:full1} and \eqref{eq:full2} 
 estimate under the saturated model \eqref{eq:betat}, we have the following lemma. 
\begin{lemma}\label{TATE:lem:full data}
Under \eqref{eq:betat} the full data equations $E\{D_1^f(t;\beta, \Lambda) \} = 0$ for each $t$ and $E\{D_2^f(\beta, \Lambda)\} = 0$ have unique solution $\beta^*$ and $\Lambda^*(t)$ satisfying  
\be
h(\beta^*) &:=& \int_0^\tau \left\{  {\cal E}(\beta(t), t) - {\cal E}(\beta^*, t)
\right\} \cdot \sum_{a=0}^1dF_a(t) = 0, \label{eq:soln1} \\
    \Lambda^*(t) &=& \int_0^t \frac{ \sum_{a=0}^1 dF_a(t)}{ \sum_{a=0}^1 S_a(t)e^{\beta^* a}},   \label{eq:soln2} 
\ee
where  $S_a(t) = 1- F_a(t) = P(T(a) >T)$, and 
$ {\cal E}(\beta, t) = 
e^{\beta } S_1(t) / {\sum_{a=0}^1 e^{\beta a} S_a(t)}$. 
\end{lemma}

\begin{figure}[htbp]
\begin{center}
\begin{tabular}{cccc}
\includegraphics[width=0.21\linewidth]{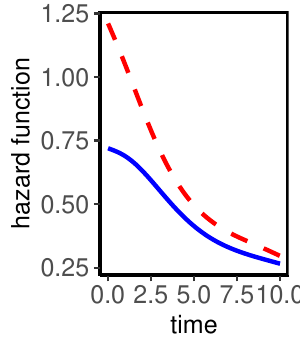} 
&\includegraphics[width=0.21\linewidth]{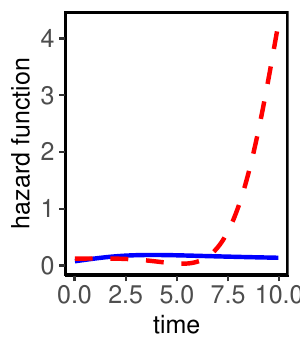} 
&\includegraphics[width=0.21\linewidth]{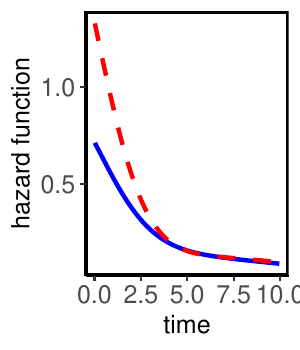} 
&\includegraphics[width=0.21\linewidth]{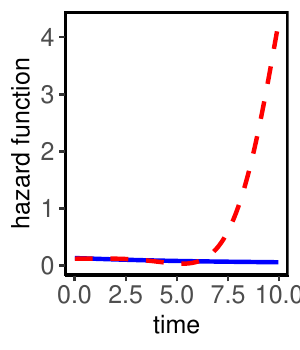}  \\
\includegraphics[width=0.21\linewidth]{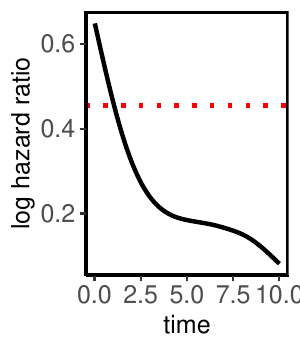} 
&\includegraphics[width=0.21\linewidth]{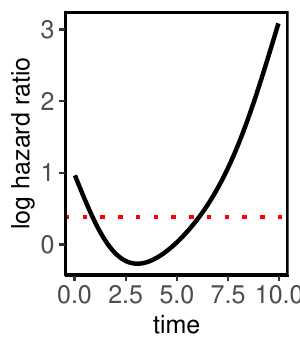} 
&\includegraphics[width=0.21\linewidth]{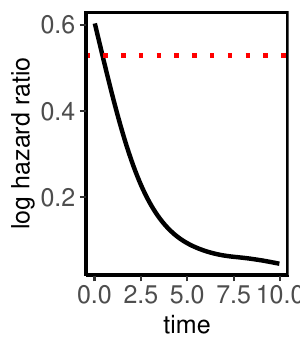} 
&\includegraphics[width=0.21\linewidth]{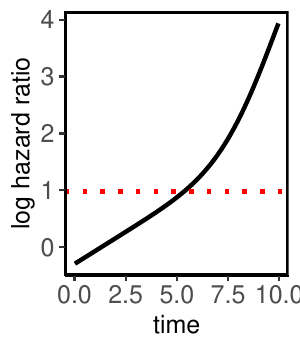} \\
~~~~~~~(a) & ~~~~~~~(b) & ~~~~~~~(c) & ~~~~~~~(d) 
\end{tabular}
\caption{Top row represents the hazard functions for  $a=0$ (blue) and  $a=1$ (red). In the bottom row  we have the corresponding log hazard ratio $\beta(t)$ with {red} dotted line indicating $\beta^*$.  We use the following scenarios: (a) $\log\{T(a)\} \sim \mbox{N}(-0.5a,1)$ for $a = 0,1$; 
(b) $\log\{T(0)\} \sim N(1.5,1)$, $T(1) \sim \mbox{Unif}(0,10)$; 
(c) $\log\{T(a)\} = -a + \epsilon$ for $a = 0,1$ with $\epsilon \sim $ logistic; 
(d) $\log\{T(0)\} = 2 + \epsilon$ with $\epsilon\sim $ logistic, and $T(1) \sim \mbox{Unif}(0,10)$.
}\label{fig:illus}
\end{center}
\end{figure}

It can be immediately verified from \eqref{eq:soln1} and \eqref{eq:soln2} that if $\beta(t) =\beta_0$, then $\beta^* =\beta_0$ and $\Lambda^*(\cdot) = \Lambda(\cdot)$ under the marginal structural Cox model \eqref{msm}. 
In Figure \ref{fig:illus} we provide some examples of  $T(a)$ distributions and the corresponding 
$\beta^*$ values. As illustrated in the figure as well as evident from Lemma \ref{TATE:lem:full data}, the estimand $\beta^*$ is {\it well-specified} \citep{buja:etal:2019b, buja:etal:2019c} as a functional of the potential outcome distributions only; 
it does not depend on the treatment assignment mechanism, or the covariate distribution, or the censoring mechanism. 

To help further understand its interpretation, 
from \eqref{eq:soln1}  using the mean value theorem or first-order Taylor expansion we can see that \eqref{eq:ave} holds with 
$\omega(t) = v (\tilde\beta(t), t)\cdot \sum_{a=0}^1 f_a(t)$, where 
 $$v (\beta, t) = \frac{\partial}  { \partial \beta} {\cal E}(\beta, t) 
 = \frac{ e^{\beta } S_0(t)S_1(t) }{ \{ S_0(t) + e^{\beta } S_1(t) \}^2 }, $$  
 $f_a(t) = dF_a(t)/dt $ and $\tilde\beta(t)$ lies between $\beta(t)$ and $\beta^*$. 
The above is in fact equivalent to \cite{XO2000} 
under 1:1 randomized treatments, where  \eqref{eq:soln1} becomes
\be\label{eq:beta*}
\int_{0}^{\tau}\left\{E_{\beta(t)}(A|T=t) - E_{\beta^*}(A|T=t) \right\} dF(t) =0, 
\ee
and $F(t)$ is the CDF of $T$. In this case we also have that 
$\omega(t) = v (\tilde\beta(t), t)\cdot  f(t)$, 
 $v (\beta(t), t) = \var(A | T=t)$ which changes very mildly over $t$ for a finite $\tau$, and $f(t) = dF(t)/dt $. 

Finally, since model \eqref{eq:betat} is saturated, we have as special cases 
when $T(a)$ follows the semiparametric transformation model
\be\label{eq:transf}
    g(T(a)) = \gamma a + \epsilon,
\ee
for $a = 0,1$, where $g(\cdot)$ is an unspecified strictly increasing function, and  
$\epsilon$ is a member of the $G^\rho$ family 
 \citep{harrington1982class}. 
Similar to \cite{xu2001semiparametric} it can be shown that
if $\tau$ is large so that $P(T(a)<\tau) = 1$ for $a=0,1$, then 
$\beta^* = -{\gamma}/(\rho + 1)$ \citep{L2023}.
In this way estimation of the average log hazard ratio $\beta^*$ leads immediately to the estimation of the causal effect $\gamma$ under the semiparametric transformation model \eqref{eq:transf}, as $\rho\ge 0$ is assumed known under the model.

Once we understand what the full data score \eqref{eq:full1} and \eqref{eq:full2}  estimate, it is then perhaps not surprising that with observed  data, the AIPW score \eqref{eq:aipw1} and \eqref{eq:aipw2} identify  the causal estimand $\beta^*$ and $\Lambda^*$ under model \eqref{eq:betat}. This is shown in the following theorem, together with 
the doubly robust property just as previously stated under model \eqref{msm} but now without the proportional hazards assumption. 

\begin{customthm}{1$'$} \label{thm:dr:avebeta}
  Under Assumptions \ref{assump1}-\ref{assump5}, if either $S = S^o$, or both $S_c = S_c^o$ and $\pi = \pi^o$, then for all $t$,
$E\{D_1(t;\beta^*,\Lambda^*, \pi,S, S_c)\} = 
E\{D_2(\beta^*,\Lambda^*,\pi,S, S_c)\} = 0. 
$
\end{customthm} 

The above result stands in stark contrast to the previous literature under the Cox regression model, where the partial likelihood serves as the standard approach under the proportional hazards assumption, as long as the censoring distribution is independent of the failure time, given the covariates in the regression model. The marginal structural model, on the other hand, does not include any covariates, so IPCW is applied from the start, and AIPW leads to improved robustness and possibly efficiency. In this way, as will be shown in the next section our joint augmentation provides an umbrella approach that gives valid inference for $\beta^o$ under the marginal structural Cox model \eqref{msm}, but also at the same time, valid inference for the causal estimand $\beta^*$ (and $\Lambda^*$) under the saturated model \eqref{eq:betat}. This is particularly relevant for practical applications where, as we will illustrate later, if we believe that model \eqref{msm} holds, we are estimating the log hazard ratio $\beta^o$; in the case that model \eqref{msm} fails, we are estimating the time-averaged log hazard ratio $\beta^*$. The result therefore remains interpretable under possibly misspecified marginal structural Cox model, hence satisfying both the assumption-lean and the well-specification criteria discussed in \cite{buja:etal:2019a, buja:etal:2019b, buja:etal:2019c}. 
 

\section{Asymptotic Double Robust Properties}

\subsection{Estimation}

In the following we assume the general model \eqref{eq:betat}, with model \eqref{msm} as a special case where $\beta^*=\beta^o$. 
Given i.i.d.~observations $(X_i,\delta_i, A_i,Z_i)$, $i=1, ..., n$,  
we may estimate $(\beta^*, \Lambda^*)$ by solving
\begin{equation} \label{TATE:EE1}
\frac{1}{n} \sum_{i=1}^n D_{1i}(t; \beta, \Lambda, \eta)= 0  \ \mbox{and } \ \frac{1}{n}  \sum_{i=1}^n D_{2i}(\beta, \Lambda, \eta)=0.  
\end{equation}
For $l = 0,1,$ define 
$
    \mathcal{S}^{(l)}(t;\beta, \eta) = n^{-1} \sum_{i =1}^n  \Gamma_i^{(l)}(t;\beta,\eta).$
 Solving for $\Lambda$ in 
  \eqref{TATE:EE1} we  obtain a newly proposed AIPW estimate 
\be\label{TATE:lambda aipcw}
\wt{\Lambda}(t;\beta, \eta) = \frac{1}{n} \sum_{i=1}^n \int_0^t \frac{ d\mathcal{N}^{(0)}_{i}(u;\eta)}{\mathcal{S}^{(0)}(u;\beta, \eta)}.
\ee
With it, we can go back to \eqref{TATE:EE1} and  obtain a new AIPW estimate of $\beta$ that solves the estimating equation $U(\beta; \eta)=0$,  with
\be\label{TATE:beta aipcw}
    U(\beta; \eta) = \frac{1}{n} \sum_{i=1}^n \int_0^{\tau} d\mathcal{N}^{(1)}_{i}(t; \eta)  - \bar{A}(t;\beta, \eta) d\mathcal{N}^{(0)}_{i}(t; \eta),
\ee
and  $\bar{A}(t;\beta, \eta) = \mathcal{S}^{(1)}(t;\beta, \eta)/\mathcal{S}^{(0)}(t;\beta, \eta)$. 
We note that $U(\beta; \eta)$ has a parallel expression to the partial likelihood score function under the Cox model,  observing that $\bar{A}(t;\cdot)$ is  the sum of augmented  $A_i e^{\beta A_i} Y_i(t)$ divided by 
 the sum of augmented  $e^{\beta A_i} Y_i(t)$,  and 
$d\mathcal{N}_i^{(l)}(t;\eta)$ is the augmented $A_i^l dN_i(t)$ for $l=0, 1$ as mentioned before. 
As demonstrated below, these augmentations yield the doubly robust property for inference in large samples. 
The above derivation and result is to the best of our knowledge  new to the literature. 

\begin{algorithm}
\caption{Jointly Augmented Cox DR Estimator}
\begin{algorithmic}
\label{AIPW:alg:1}
\REQUIRE  A sample of $n$ observations, $(X_i,\delta_i, A_i,Z_i)_{i=1}^{n}.$
\STATE Split the full sample into $k$ folds indexed by $\mathcal{I}_1, \mathcal{I}_2, \ldots, \mathcal{I}_k$.
\FOR{each fold indexed by $m$}
\STATE Let $\mathcal{I}_{-m} \coloneqq \{1, \ldots, n \}\setminus \mathcal{I}_m$
\STATE Estimate nuisances   $ \hat \eta ^{(-m)} = \Bigl(\wh{\pi}^{(-m)}$, $\wh{S}^{(-m)},  \wh{S}_c^{(-m)} \Bigl)$ using the out-of-fold sample  $\mathcal{I}_{-m}$.
\STATE Obtain $U_m(\beta, \hat \eta ^{(-m)})$, the $m$-th fold estimating equation for $\beta$, 

$\qquad \qquad $ by profiling  out $\Lambda$ from the in-fold equation $\sum_{i \in \mathcal{I}_m} D_{1i}(t;\beta, \Lambda,\hat \eta ^{(-m)}) = 0$.

\STATE Construct $\wt{\Lambda}_m(t;\whb, \hat \eta ^{(-m)} )$ as in \eqref{TATE:lambda aipcw}.
\ENDFOR 
\RETURN $\whb$ as  the solution to $U_{cf}(\beta) =0$ with
\[
   U_{cf}(\beta) =  \frac{1}{k}\sum_{m=1}^k U_m(\beta, \hat \eta ^{(-m)} )   \quad \mbox{and} \quad
     \wh{\Lambda}(t) = \frac{1}{k} \sum_{m=1}^k \wt{\Lambda}_m(t;\whb, \hat \eta ^{(-m)} ). 
\]
\end{algorithmic}
\end{algorithm}

In the above the nuisance functions $\pi(z)$, $S(t;a,z)$ and $S_c(t;a,z)$ are  unknown 
 in practice and need to be estimated. 
We utilize the cross-fitting procedure that is commonly considered in the literature 
\citep{hasminskii1978nonparametric, bickel1982adaptive, robins2008higher, CCDDHNR2018} 
and, as will be shown later, provides root-$n$ inference for $\beta^*$. 
The estimation procedure is summarized in Algorithm~\ref{AIPW:alg:1}, where the details of the relevant quantities are further given in the Supplementary Material, although the notation used here should also be self-explanatory. 
Following the fit, under the marginal structural Cox model \eqref{msm} if needed we may also 
 estimate the survival probabilities, or one minus the risk, $P(T(a) >t ) $ by $ \exp\{- \wh{\Lambda}(t) \exp(\whb a) \}$.

\subsection{Asymptotic properties and inference}


We first focus on rate double robustness and describe some additional assumptions.
Let $O^\dag$ denote a random sample  $\{(X^\dag_i, \Delta^\dag_i, A^\dag_i, Z^\dag_i), i=1, \ldots, n\}$ used for estimating $\wh{\pi}$, $\wh{S}$ and $\wh{S}_c$. Let $(X, \Delta, A, Z)$ be a data point independent of $O^\dag$ and drawn from the same distribution as $O^\dag$. Define below for $f = S, S_c$, where $E^\dag$ denotes expectation  with respect to $(X^\dag, \Delta^\dag, A^\dag, Z^\dag)$, and $E$ denotes expectation  with respect to $(X, \Delta, A, Z)$ conditional on $O^\dag$,
\begin{eqnarray*}
    \left\|\wh{\pi} - \pi^*\right\|^2_\dag &=& E^\dag \left [ E \left[   \wh{\pi}(Z) - \pi^*(Z)  \right] ^2 \right],  \\
    \left\|\wh{f} - f^*\right\|^2_\dag &=& E^\dag \left [E \Bigl[   \sup_{t \in [0,\tau], a \in \{0,1\}}  | \wh{f}(t;a,Z) - f^*(t;a,Z)  |  ^2 \Bigl] \right].
\end{eqnarray*}

\begin{assumption}[Uniform Convergence]\label{assump6}
There exist deterministic limits $\pi^*(z)$, $S^*(t;a,z)$ and $S_c^*(t;a,z)$ such that $\|\wh{\pi} - \pi^*\|_\dag = o(1)$, $\|\wh{S} - S^*\|_\dag = o(1)$ and $\|\wh{S}_c - S_c^*\|_\dag = o(1)$.
\end{assumption}

We should point out that the above assumption  does not require any of  the three models  to be correctly specified. Instead, it simply asserts that these models possess population counterparts towards which they converge. 
This condition is generally fulfilled for the Donsker model class, regardless of whether they exhibit biases.
{The following theorem provides doubly robust consistency of $\whb$, in cases where either the outcome model (i.e.~the failure time model) or the missingness model (i.e.~the censoring time model and the treatment assignment model) is correctly specified, but not necessarily both, and our proposed estimator remains valid.} 

\begin{theorem}[Consistency]\label{thm:consistency}
Under Assumptions \ref{assump1}-\ref{assump6} and additional regularity  Assumptions~\ref{assumpA1}-\ref{assumpA4} in the Supplementary Material, if either $S^* = S^o$, or $(\pi^*, S_c^*) = (\pi^o, S_c^o)$, then $\whb \overset{p}{\to} \beta^*$. 
\end{theorem}

We remind the reader that the limit $\beta^*$ to which $\whb$ converges is the true log hazard ratio when the marginal structural Cox model \eqref{msm} holds, but otherwise it is the time-averaged log hazard ratio described in the Section 3.

Now let $O$ denote a random sample  $\{(X_j, \Delta_j, A_j, Z_j), j = 1,\ldots,n\}$ that is independent of $O^\dag$ above, and drawn from the same distribution as $O^\dag$. Recall  that $\eta^o =(\pi^o, S^o, S_c^o)^\top$, and define $ \tilde J_{i}  (t;a)=   J_i(t;a, \hat S, \hat S_c) -   J_i(t;a, \hat S, S_c^o) $ and $ \bar J_{i}  (t;a)=   J_i(t;a,  S^o, \hat S_c) -   J_i(t;a,  S^o, S_c^o) $, with $
J_i(t;a,S,S_c) = \int_0^t {dM_{ci}(u;a,S_c)} / \{S_i(u;a)S_{ci}(u;a)\}
$  defined in Section 2.  Here we use various shorthand notations of the likes of  $\wh{S}_i(t;a) = \wh{S}(t;a, Z_i) $ and $S^o_i(t;a) =S^o(t;a, Z_i)$.
Then, with $d \mathcal{K}_i (t,u;a)= d\wh{S}_i(t;a) d\tilde J_{i}  (u;a) - dS^o_i(t;a) d \bar J_{i}  (u;a) $ we 
 define the {\it cross integral products} 
 $ \mathcal{D}_1^\dag:= \mathcal{D}_1^\dag(\wh{S}, \wh{S}_c; \eta^o) $ and 
 $  \mathcal{D}_2^\dag:=\mathcal{D}_2^\dag(\wh{S}, \wh{S}_c; \eta^o) $  as
 \begin{eqnarray*}
       \mathcal{D}_1^\dag  
  &  =& E^\dag \Bigg[ E \Big[ \max_{a \in \{0,1\}}  \frac{1}{n}\sum_{i=1}^n \Big| \int_0^\tau \int_0^t  \{a - \bar{A}(t;\beta^*, \eta^o)\}  d \mathcal{K}_i (t,u;a) \Big|   \Big] \Bigg], \\
  \mathcal{D}_2^\dag 
  &=&  E^\dag \Bigg[ E \bigg[  \max_{a,l \in \{0,1\}}   \frac{1}{n}\sum_{i=1}^n \bigg| \int_0^\tau \delta(t, \beta^*)   J_i(t;a, S^o,\wh{S}_c)^l \{d\wh{S}_i(t;a) - dS^o_i(t;a) \} \bigg|   \bigg] \bigg], 
\end{eqnarray*}
where  $\delta(t, \beta^*) =\bar{A}(t;\beta^*, \pi^o, S^o, \wh{S}_c) - \bar{A}(t;\beta^*, \eta^o)$
and 
$E$   denotes  expectation  with respect to  $O$ conditional on  $O^\dag$. 

\begin{assumption}[Rate Condition]\label{assump7}\ 
\ba{
\bigl\|\wh{S} - S^o\bigl\|_\dag \left( \left\|\wh{\pi} - \pi^o\right\|_\dag +  \bigl\|\wh{S}_c - S^o_c\bigl\|_\dag \right) + \mathcal{D}_1^\dag(\wh{S}, \wh{S}_c; \eta^o) + \mathcal{D}_2^\dag(\wh{S}, \wh{S}_c; \eta^o)= o(n^{-1/2}). \nn
}
\end{assumption}

Note that in addition to the common  products of error rates as in \cite{CCDDHNR2018,rotn:etal:21, hou:etal:2021},  
we also have the cross integral product terms. These integral terms are needed  
 because we have two nuisance functions that involve time $t$; in contrast, the mixed bias property of \cite{rotn:etal:21} suffices when 
at most one of the nuisance functions involves $t$, as in their cases. 
Similar integral terms in  rate conditions can be found in \cite{wang2022doubly} and \cite{vansteelandt2022assumption}.

\begin{theorem}[Rate Double Robustness]\label{thm:AN}
Under Assumptions \ref{assump1}-\ref{assump7} 
and additional regularity Assumptions \ref{assumpA1}-\ref{assumpA4} in the Supplementary Material, if $(\pi^*, S^*, S_c^*) = (\pi^o, S^o, S_c^o)$, we have
$$
    \wh{\sigma}^{-1} \sqrt{n}(\whb - \beta^*) \overset{d}{\to} N(0,1),
$$
where the expression  for  $\wh{\sigma}^2 := \wh{\sigma}^2(\whb)$ is also provided in the Supplementary Material.
\end{theorem}

Theorem \ref{thm:AN} establishes the rate double robustness property, and as mentioned before  
this solves the compatibility issue due to the non-collapsibility of the Cox regression models via the use of machine learning or other nonparametric approaches to estimate the nuisance functions. 
Traditionally, 
Neyman orthogonal scores \citep{neyman1959optimal} have been considered in the semiparametric literature  which,  
when combined with cross-fitting, gives $\sqrt{n}$ consistent estimators, as long as all nuisance parameters are estimated at faster than $n^{-1/4}$ rate \citep{newey1994asymptotic, rotn:etal:21}. 
However, this $n^{-1/4}$ rate requirement 
still rules out a number of data adaptive machine learning methods \citep{bilodeau2022blair, ogburn2022elizabeth, tang2022yanbo}. 
The rate double robustness result  here improves  upon the $n^{-1/4}$ rate requirement: $\whb$ is consistent and asymptotically normal even if one of $\wh{S}$ or $(\wh{\pi}, \wh{S}_c)$ converges arbitrarily slow, as long as their product error rate is faster than $\sqrt{n}$. 
In practice, 
very few of the machine learning  methods used for time-to-event data have known convergence rates, and in the Simulations section below we will investigate empirically the performance of the methods.

Lastly, if we are satisfied with the saturated model \eqref{eq:betat} and do not insist on model \eqref{msm} being valid, we can explore parametric or semiparametric models for the nuisance functions $\pi$, $S$, and $S_c$, as  the compatibility issue is no longer a concern.
 In this case, no cross-fitting is needed, and  $\wh{\pi}$, $\wh{S}$ and $\wh{S}_c$ are regular and asymptotically linear estimators of  $\pi$, $S$ and $S_c$, respectively. Let $\pi^*$, $S^*$ and $S_c^*$ be their limits; that is, $\|\wh{\pi} - \pi^*\| =  \|\wh{S} - S^*\| =  \|\wh{S}_c - S_c^*\| = o(1)$, where 
$\|\wh{\pi} - \pi^*\|^2 = E\{|\wh{\pi}(Z) - \pi^*(Z)|^2\}$, $\|\wh{S} - S^*\|^2 = E\{ \sup_{t \in [0,\tau], a \in \{0,1 \}}|\wh{S}(t;a,Z) - S^*(t;a,Z)|^2\}$, and $\|\wh{S}_c - S_c^*\|^2 = E\{ \sup_{t \in [0,\tau], a \in \{0,1 \}}|\wh{S}_c(t;a,Z) - S^*_c(t;a,Z)|^2\}$.
Similar to \cite{hou:etal:2021} and \cite{wang2022doubly} 
it can be shown that the solution $\betadr$ to $U(\beta; \hat\eta)=0$ in \eqref{TATE:beta aipcw}
enjoys the classical model doubly robust property: it is consistent and asymptotically normal 
if either $S^* = S^o$ or $(\pi^*, S_c^*) = (\pi^o, S_c^o)$.

\section{Numerical Work}
\subsection{Simulations}

 In this section, we assess the performance of the proposed estimator under the marginal structural Cox model \eqref{msm}. Extended simulations under the general model \eqref{eq:betat} are provided in the Supplementary Material.  The codes developed in this work has been implemented in the R \citep{RCore2021} package \texttt{CoxAIPW} which is  available  on CRAN. The package  allows three choices: for 1) causal inference with non-randomized observational data but no informative censoring (i.e.~AIPTW only), 2) informative censoring without causal inference (AIPCW only), and 3)  both causal inference with  observational data and informative censoring (AIPW as described in this paper).

\begin{figure} [ht]
\centering
\resizebox{0.4\textwidth}{!}{
\begin{tikzpicture} 
    \node[draw] (v0) at (-1.6cm,0cm) {$T(a)$};
    \node[draw] (v1) at (1.6cm, 0cm) {$C(a)$};
    \node[draw] (v2) at (-0cm,0cm) {$A$};
    \node[draw] (v3) at (1cm,-1cm) {$Z$};
    \node[draw] (v4) at (-1cm,-1cm) {$U$};
    \node[draw] (v6) at (-1.3cm,1cm) {$T$};
    \node[draw] (v7) at (1.3cm,1cm) {$C$};
    \draw [->] (v3) edge (v1);
    \draw [->] (v4) edge (v0);
    \draw [->] (v4) edge (v3);
    \draw [->] (v3) edge (v2);
    \draw [->] (v2) edge (v6);
    \draw [->] (v2) edge (v7);
    \draw [->] (v0) edge (v6);
    \draw [->] (v1) edge (v7);
\end{tikzpicture}}
\caption{Diagram for the data generating mechanism} \label{AIPW:dag}
\end{figure}
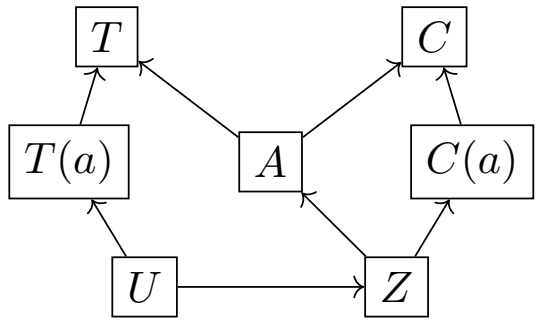

To induce confounding within a marginal structural model, we adopt the approach outlined by \cite{havercroft2012simulating}. This involves simulating latent variables linked to both the covariates and the event outcome. Let $U=(U_1, U_2, U_3)^T$ where $ U_1 \sim \mbox{Unif}(-1,1)$, $U_2 \sim \mbox{Unif}(-1,1)$ and $U_3 \sim \mbox{Unif}(-1,1)$. Let $Z=(Z_1, Z_2, Z_3)^T$ where $Z_1 =0.5U_1 + U_3$, $Z_2 = U_1 + 1.5U_1^2 - 0.5$ and $Z_3 = U_1 + U_2$. 
 Let $T(a) = -\log(0.5U_1+0.5)\exp(a)$ for $a = 0,1$. This gives $T(a)$ which follows  model \eqref{msm} with $\beta^o = -1$ and $\lambda_0^o(t) = 1$. We then simulate $C(a)$ and $A$ according to the four scenarios described in Table~\ref{AIPW:scenario.table}, each giving about $50\%$ per treatment arm and about $40\%$  censored due to loss to follow-up. 
 Finally let $T = AT(1) + (1-A)T(0)$ and $C = AC(1) + (1-A)C(0)$. 
For all simulations we set $\tau=1$; note that this is the maximum time for estimating both the $T$ and $C$ distributions.
  The details of this data generation process are elaborated in Figure~\ref{AIPW:dag}. Each scenario involves 1,000 simulated datasets of \(n=1000\), yielding a margin of error about \(\pm 1.35\%\) for the 95\% confidence intervals. 
  
\begin{table} [ht]
\centering
\small
\caption{Four scenarios for generating $C(a)$ and $A$.}
\label{AIPW:scenario.table}
\setlength{\tabcolsep}{2pt}
\begin{tabular}{@{\extracolsep{5pt}} cccccccc} 
Scenario & Details \\ 
\\
    1  &    $\epsilon \sim \mbox{Unif}(0,1)$     \\
   $C(a)$: Cox &$C(a) = -\log(\epsilon)\exp(0.5 + 0.5a - Z_2 + 0.5Z_3) $\\
$A$: Logistic & $\logit \{\pi(Z) \} = 0.5Z_1 - 0.5Z_2 - 0.5Z_3$  \\
\\
    2  &     $\epsilon \sim \mbox{Unif}(0,1)$        \\
   $C(a)$: Cox & $C(a) = -\log(\epsilon)\exp(0.5 + 0.5a - Z_2 + 0.5Z_3) $\\
$A$: Soft Partition & $\logit \{\pi(Z) \} = -3 \cdot \mathbf{1}\{Z_2 < -0.5\} + 3 \cdot \mathbf{1}\{-0.5 \leq Z_2 < 0.5\} -3 \cdot \mathbf{1}\{Z_2 \geq 0.5\}$  \\
\\
    3  &    $\epsilon \sim \mbox{Unif}(0,1)$ 
     \\
   $C(a)$:  Uniform-Cox & $ C(0) = 1.05\epsilon $, 
 $C(1) = -\log(\epsilon)\exp(3.3 + 3.5Z_3)$ \\
$A$: Logistic & $\logit \{\pi(Z) \} = 0.5Z_1 - 0.5Z_2 - 0.5Z_3$ \\
\\
    4  &     $\epsilon \sim \mbox{Unif}(0,1)$ 
     \\
   $C(a)$: Uniform-Cox  & $ C(0) = 1.05\epsilon $,
  $C(1) = -\log(\epsilon)\exp(3.3 + 3.5Z_3)$ \\
$A$: Soft Partition &  $\logit \{\pi(Z) \} = -3\cdot \mathbf{1}\{Z_2 < -0.5\} + 3\cdot\mathbf{1}\{-0.5 \leq Z_2 < 0.5\} -3\cdot \mathbf{1}\{Z_2 \geq 0.5\}$  \\
\end{tabular}
\end{table}

We contrast the AIPW estimator, using different working models, with the IPW estimator, the partial likelihood estimator under the  naive Cox model, and a full data estimator. 
We estimate the conditional distributions of \(T\) and \(C\) given \(A\) and \(Z\) using Cox regression and the random survival forest  \citep[RSF]{ishwaran2008random} from the R  package \texttt{randomForestSRC}. For \(A\) given \(Z\), we used  logistic regression and the gradient boosted models (GBM) 
 from the R package  \texttt{twang} \citep{ridgeway2022toolkit}. 
RSF settings included the `\texttt{bs.gradient}' split rule, while default settings were maintained for other hyperparameters in both \texttt{twang} and RSF. Five-fold cross-fitting was implemented. For stability, values of \(\wh{S}(t;a,z)\), \(\wh{S}_c(t;a,z)\) were restricted to be at least \(0.05\), and \(\wh{\pi}(z)\) between \(0.1\) and \(0.9\). For each dataset, we calculated model-based and bootstrap standard errors, the latter with 100 replicates, and constructed \(95\%\) confidence intervals using the normal approximation.
 For the full data estimator, the robust sandwich estimator was applied to account for potential intra-subject outcome correlations.

\begin{table}
\def~{\hphantom{0}}
\small
\begin{center} 
\caption{Scenarios 1 - 2 of simulation; true $\beta^o = -1$. \red{Red} indicates that the model is wrong.}  \label{AIPW:simulation.results1} 
\renewcommand{\arraystretch}{0.82} 
\begin{tabular}{cclllll} 
\multicolumn{1}{c}{Scenario} & \multicolumn{1}{c}{Estimator} & \multicolumn{1}{c}{$T$/$C$-$A$ Models}  & \multicolumn{1}{c}{Bias} & \multicolumn{1}{c}{SD} & \multicolumn{1}{c}{SE} & \multicolumn{1}{c}{Coverage}\\  
&&&&&Model/Boot& Model/Boot\\
\\ 
{1}  & \multirow{8}{*}
{AIPW}    &  \red{Cox}/Cox-logit & \multicolumn{1}{c}{~0.002~} & \multicolumn{1}{c}{~0.059~} & \multicolumn{1}{c}{0.060/0.059} & \multicolumn{1}{c}{0.95/0.94} \\  
 &   &  \red{Cox}/Cox-GBM & \multicolumn{1}{c}{~0.003~} & \multicolumn{1}{c}{~0.058~} & \multicolumn{1}{c}{0.061/0.060} & \multicolumn{1}{c}{0.96/0.94} \\  
 &   &  \red{Cox}/RSF-logit & \multicolumn{1}{c}{~0.004~} & \multicolumn{1}{c}{~0.057~} & \multicolumn{1}{c}{0.060/0.059} & \multicolumn{1}{c}{0.96/0.96} \\  
 &   &  \red{Cox}/RSF-GBM & \multicolumn{1}{c}{~0.003~} & \multicolumn{1}{c}{~0.056~} & \multicolumn{1}{c}{0.062/0.060} & \multicolumn{1}{c}{0.97/0.96} \\  
 &   &  RSF/Cox-logit & \multicolumn{1}{c}{~0.002~} & \multicolumn{1}{c}{~0.053~} & \multicolumn{1}{c}{0.053/0.053} & \multicolumn{1}{c}{0.95/0.94} \\  
 &   &  RSF/Cox-GBM & \multicolumn{1}{c}{~0.008~} & \multicolumn{1}{c}{~0.054~} & \multicolumn{1}{c}{0.054/0.055} & \multicolumn{1}{c}{0.95/0.95} \\  
 &   &  RSF/RSF-logit & \multicolumn{1}{c}{~0.005~} & \multicolumn{1}{c}{~0.053~} & \multicolumn{1}{c}{0.053/0.054} & \multicolumn{1}{c}{0.95/0.94} \\  
 &   &  RSF/RSF-GBM & \multicolumn{1}{c}{~0.011~} & \multicolumn{1}{c}{~0.054~} & \multicolumn{1}{c}{0.055/0.056} & \multicolumn{1}{c}{0.95/0.95} \\ 
 \\ 
 &  \multirow{4}{*}{IPW} &   ~~~~Cox-logit & \multicolumn{1}{c}{~0.000~} & \multicolumn{1}{c}{~0.067~} & \multicolumn{1}{c}{~~~-~/0.066} & \multicolumn{1}{c}{~~-~/0.94} \\  
 &   &  ~~~~Cox-GBM & \multicolumn{1}{c}{~0.031~} & \multicolumn{1}{c}{~0.069~} & \multicolumn{1}{c}{~~~-~/0.069} & \multicolumn{1}{c}{~~-~/0.92} \\  
 &   &  ~~~~RSF-logit & \multicolumn{1}{c}{~0.026~} & \multicolumn{1}{c}{~0.060~} & \multicolumn{1}{c}{~~~-~/0.064} & \multicolumn{1}{c}{~~-~/0.95} \\  
 &   &  ~~~~RSF-GBM & \multicolumn{1}{c}{~0.005~} & \multicolumn{1}{c}{~0.062~} & \multicolumn{1}{c}{~~~-~/0.068} & \multicolumn{1}{c}{~~-~/0.97} \\
 \\ 
 &  \red{Naive Cox} &  & \multicolumn{1}{c}{~0.496~} & \multicolumn{1}{c}{~0.100~} & \multicolumn{1}{c}{0.100/0.101} & \multicolumn{1}{c}{0.00/0.00} \\  
 &  Full Data &  & \multicolumn{1}{c}{~0.001~} & \multicolumn{1}{c}{~0.029~} & \multicolumn{1}{c}{0.028/~~-~~} & \multicolumn{1}{c}{0.95/~~-~} \\  
\\ 
{2}  & \multirow{8}{*}
{AIPW}    &  \red{Cox}/Cox-\red{logit} & \multicolumn{1}{c}{~0.260~} & \multicolumn{1}{c}{~0.064~} & \multicolumn{1}{c}{0.068/0.066} & \multicolumn{1}{c}{0.02/0.02} \\  
 &   &  \red{Cox}/Cox-GBM & \multicolumn{1}{c}{~0.018~} & \multicolumn{1}{c}{~0.086~} & \multicolumn{1}{c}{0.093/0.088} & \multicolumn{1}{c}{0.96/0.95} \\  
 &   &  \red{Cox}/RSF-\red{logit} & \multicolumn{1}{c}{~0.268~} & \multicolumn{1}{c}{~0.063~} & \multicolumn{1}{c}{0.069/0.066} & \multicolumn{1}{c}{0.02/0.02} \\  
 &   &  \red{Cox}/RSF-GBM & \multicolumn{1}{c}{~0.033~} & \multicolumn{1}{c}{~0.080~} & \multicolumn{1}{c}{0.090/0.084} & \multicolumn{1}{c}{0.95/0.94} \\  
 &   &  RSF/Cox-\red{logit} & \multicolumn{1}{c}{~0.050~} & \multicolumn{1}{c}{~0.071~} & \multicolumn{1}{c}{0.056/0.061} & \multicolumn{1}{c}{0.80/0.83} \\  
 &   &  RSF/Cox-GBM & \multicolumn{1}{c}{~0.003~} & \multicolumn{1}{c}{~0.073~} & \multicolumn{1}{c}{0.073/0.075} & \multicolumn{1}{c}{0.94/0.94} \\  
 &   &  RSF/RSF-\red{logit} & \multicolumn{1}{c}{~0.052~} & \multicolumn{1}{c}{~0.071~} & \multicolumn{1}{c}{0.056/0.061} & \multicolumn{1}{c}{0.80/0.83} \\  
 &   &  RSF/RSF-GBM & \multicolumn{1}{c}{~0.009~} & \multicolumn{1}{c}{~0.073~} & \multicolumn{1}{c}{0.073/0.075} & \multicolumn{1}{c}{0.95/0.94} \\ 
 \\ 
 &  \multirow{4}{*}{IPW} &   ~~~~Cox-\red{logit} & \multicolumn{1}{c}{~0.164~} & \multicolumn{1}{c}{~0.093~} & \multicolumn{1}{c}{~~~-~/0.092} & \multicolumn{1}{c}{~~-~/0.58} \\  
 &   &  ~~~~Cox-GBM & \multicolumn{1}{c}{~0.103~} & \multicolumn{1}{c}{~0.118~} & \multicolumn{1}{c}{~~~-~/0.106} & \multicolumn{1}{c}{~~-~/0.83} \\  
 &   &  ~~~~RSF-\red{logit} & \multicolumn{1}{c}{~0.177~} & \multicolumn{1}{c}{~0.084~} & \multicolumn{1}{c}{~~~-~/0.088} & \multicolumn{1}{c}{~~-~/0.48} \\  
 &   &  ~~~~RSF-GBM & \multicolumn{1}{c}{~0.134~} & \multicolumn{1}{c}{~0.107~} & \multicolumn{1}{c}{~~~-~/0.101} & \multicolumn{1}{c}{~~-~/0.74} \\ 
 \\ 
 &  \red{Naive Cox} &  & \multicolumn{1}{c}{~0.579~} & \multicolumn{1}{c}{~0.119~} & \multicolumn{1}{c}{0.125/0.125} & \multicolumn{1}{c}{0.00/0.00} \\  
 &  Full Data &  & \multicolumn{1}{c}{~0.001~} & \multicolumn{1}{c}{~0.029~} & \multicolumn{1}{c}{0.028/~~-~~} & \multicolumn{1}{c}{0.95/~~-~} 
\end{tabular} \end{center}  \end{table}

Tables~\ref{AIPW:simulation.results1} and \ref{AIPW:simulation.results2} show the bias, standard deviation (SD), standard error (SE) and coverage probabilities of the  estimators under Scenarios 1 -- 4, respectively. Additional plots to visualize the results are provided in the Supplementary Material. We see that 
under Scenario 1 all eight AIPW estimators have small biases and good coverage probabilities, even when the conditional Cox model for $T$ is wrong (marked in red). 
This is no longer the case under Scenarios 2 and 3 when both the conditional Cox model for $T$ and one of the models for $C$ or $A$ is wrong. Note that under Scenario 2, when RSF is used for the conditional \(T\) model and logistic regression is incorrectly applied for the propensity score model, the biases are relatively substantial, with coverage only at around 80\% in both cases. We think that this has to do with the fact that  rate double robustness when machine learning is involved requires all nuisance estimates to be consistent, at least theoretically. Under Scenario 4 
 the RSF/RSF-GBM estimator  using all machine learning  methods has the smallest bias and is the only one with correct coverage. 
The IPW estimators perform well under Scenario 1 with correctly specified working models for \(C\) and \(A\). However, their performance is generally poor in Scenarios 2 through 4, including the RSF-GBM estimator, which exhibits slower convergence when machine learning methods are used for weight estimation.

\begin{table} 
\def~{\hphantom{0}}
\small
\begin{center} 
\caption{Scenarios 3 - 4 of simulation; true $\beta^o = -1$. \red{Red} indicates that the model is wrong.} \label{AIPW:simulation.results2} 
\renewcommand{\arraystretch}{0.82} 
\begin{tabular}{cclllll} 
\multicolumn{1}{c}{Scenario} & \multicolumn{1}{c}{Estimator} & \multicolumn{1}{c}{$T$/$C$-$A$ Models}  & \multicolumn{1}{c}{Bias} & \multicolumn{1}{c}{SD} & \multicolumn{1}{c}{SE} & \multicolumn{1}{c}{Coverage}\\  
&&&&&Model/Boot& Model/Boot\\
\\ 
{3}  & \multirow{8}{*}{AIPW}    &  \red{Cox}/\red{Cox}-logit & \multicolumn{1}{c}{~0.146~} & \multicolumn{1}{c}{~0.106~} & \multicolumn{1}{c}{0.108/0.109} & \multicolumn{1}{c}{0.79/0.78} \\  
 &   &  \red{Cox}/\red{Cox}-GBM & \multicolumn{1}{c}{~0.146~} & \multicolumn{1}{c}{~0.108~} & \multicolumn{1}{c}{0.114/0.120} & \multicolumn{1}{c}{0.81/0.83} \\  
 &   &  \red{Cox}/RSF-logit & \multicolumn{1}{c}{~0.015~} & \multicolumn{1}{c}{~0.103~} & \multicolumn{1}{c}{0.122/0.118} & \multicolumn{1}{c}{0.98/0.97} \\  
 &   &  \red{Cox}/RSF-GBM & \multicolumn{1}{c}{~0.014~} & \multicolumn{1}{c}{~0.104~} & \multicolumn{1}{c}{0.126/0.125} & \multicolumn{1}{c}{0.98/0.98} \\  
 &   &  RSF/\red{Cox}-logit & \multicolumn{1}{c}{~0.007~} & \multicolumn{1}{c}{~0.092~} & \multicolumn{1}{c}{0.098/0.097} & \multicolumn{1}{c}{0.95/0.94} \\  
 &   &  RSF/\red{Cox}-GBM & \multicolumn{1}{c}{~0.001~} & \multicolumn{1}{c}{~0.095~} & \multicolumn{1}{c}{0.101/0.104} & \multicolumn{1}{c}{0.95/0.95} \\  
 &   &  RSF/RSF-logit & \multicolumn{1}{c}{~0.019~} & \multicolumn{1}{c}{~0.097~} & \multicolumn{1}{c}{0.102/0.104} & \multicolumn{1}{c}{0.96/0.95} \\  
 &   &  RSF/RSF-GBM & \multicolumn{1}{c}{~0.026~} & \multicolumn{1}{c}{~0.098~} & \multicolumn{1}{c}{0.103/0.111} & \multicolumn{1}{c}{0.96/0.97} \\ 
 \\ 
 &  \multirow{4}{*}{IPW} &   ~~~~\red{Cox}-logit & \multicolumn{1}{c}{~0.305~} & \multicolumn{1}{c}{~0.112~} & \multicolumn{1}{c}{~~~-~/0.103} & \multicolumn{1}{c}{~~-~/0.21} \\  
 &   &  ~~~~\red{Cox}-GBM & \multicolumn{1}{c}{~0.335~} & \multicolumn{1}{c}{~0.115~} & \multicolumn{1}{c}{~~~-~/0.104} & \multicolumn{1}{c}{~~-~/0.16} \\  
 &   &  ~~~~RSF-logit & \multicolumn{1}{c}{~0.077~} & \multicolumn{1}{c}{~0.078~} & \multicolumn{1}{c}{~~~-~/0.079} & \multicolumn{1}{c}{~~-~/0.84} \\  
 &   &  ~~~~RSF-GBM & \multicolumn{1}{c}{~0.109~} & \multicolumn{1}{c}{~0.082~} & \multicolumn{1}{c}{~~~-~/0.082} & \multicolumn{1}{c}{~~-~/0.73} \\ 
 \\ 
 &  \red{Naive Cox} &  & \multicolumn{1}{c}{~0.895~} & \multicolumn{1}{c}{~0.108~} & \multicolumn{1}{c}{0.110/0.110} & \multicolumn{1}{c}{0.00/0.00} \\  
 &  Full Data &  & \multicolumn{1}{c}{~0.001~} & \multicolumn{1}{c}{~0.029~} & \multicolumn{1}{c}{0.028/~~-~~} & \multicolumn{1}{c}{0.95/~~-~} \\  
\\ 
{4}  & \multirow{8}{*}
{AIPW}    &  \red{Cox}/\red{Cox}-\red{logit} & \multicolumn{1}{c}{~0.347~} & \multicolumn{1}{c}{~0.120~} & \multicolumn{1}{c}{0.138/0.143} & \multicolumn{1}{c}{0.14/0.24} \\  
 &   &  \red{Cox}/\red{Cox}-GBM & \multicolumn{1}{c}{~0.135~} & \multicolumn{1}{c}{~0.131~} & \multicolumn{1}{c}{0.116/0.129} & \multicolumn{1}{c}{0.72/0.78} \\  
 &   &  \red{Cox}/RSF-\red{logit} & \multicolumn{1}{c}{~0.361~} & \multicolumn{1}{c}{~0.201~} & \multicolumn{1}{c}{0.343/0.224} & \multicolumn{1}{c}{0.50/0.71} \\  
 &   &  \red{Cox}/RSF-GBM & \multicolumn{1}{c}{~0.084~} & \multicolumn{1}{c}{~0.140~} & \multicolumn{1}{c}{0.149/0.157} & \multicolumn{1}{c}{0.88/0.90} \\  
 &   &  RSF/\red{Cox}-\red{logit} & \multicolumn{1}{c}{~0.075~} & \multicolumn{1}{c}{~0.107~} & \multicolumn{1}{c}{0.101/0.105} & \multicolumn{1}{c}{0.89/0.89} \\  
 &   &  RSF/\red{Cox}-GBM & \multicolumn{1}{c}{~0.026~} & \multicolumn{1}{c}{~0.105~} & \multicolumn{1}{c}{0.096/0.106} & \multicolumn{1}{c}{0.90/0.92} \\  
 &   &  RSF/RSF-\red{logit} & \multicolumn{1}{c}{~0.143~} & \multicolumn{1}{c}{~0.137~} & \multicolumn{1}{c}{0.174/0.155} & \multicolumn{1}{c}{0.88/0.93} \\  
 &   &  RSF/RSF-GBM & \multicolumn{1}{c}{~0.020~} & \multicolumn{1}{c}{~0.124~} & \multicolumn{1}{c}{0.119/0.132} & \multicolumn{1}{c}{0.94/0.94} \\ 
 \\ 
 &  \multirow{4}{*}{IPW} &   ~~~~\red{Cox}-\red{logit} & \multicolumn{1}{c}{~0.063~} & \multicolumn{1}{c}{~0.130~} & \multicolumn{1}{c}{~~~-~/0.124} & \multicolumn{1}{c}{~~-~/0.90} \\  
 &   &  ~~~~\red{Cox}-GBM & \multicolumn{1}{c}{~0.249~} & \multicolumn{1}{c}{~0.130~} & \multicolumn{1}{c}{~~~-~/0.117} & \multicolumn{1}{c}{~~-~/0.41} \\  
 &   &  ~~~~RSF-\red{logit} & \multicolumn{1}{c}{~0.163~} & \multicolumn{1}{c}{~0.112~} & \multicolumn{1}{c}{~~~-~/0.104} & \multicolumn{1}{c}{~~-~/0.66} \\  
 &   &  ~~~~RSF-GBM & \multicolumn{1}{c}{~0.005~} & \multicolumn{1}{c}{~0.129~} & \multicolumn{1}{c}{~~~-~/0.112} & \multicolumn{1}{c}{~~-~/0.91} \\ 
 \\ 
 &  \red{Naive Cox} &  & \multicolumn{1}{c}{~0.562~} & \multicolumn{1}{c}{~0.120~} & \multicolumn{1}{c}{0.132/0.132} & \multicolumn{1}{c}{0.00/0.00} \\  
 &  Full Data &  & \multicolumn{1}{c}{~0.001~} & \multicolumn{1}{c}{~0.028~} & \multicolumn{1}{c}{0.028/~~-~~} & \multicolumn{1}{c}{0.95/~~-~} 
\end{tabular} \end{center}  \end{table}

\subsection{Honolulu-Asia Aging Study}

We analyze data from the Honolulu-Asia Aging Study (HAAS), which began in 1991 as an extension of the Honolulu Heart Program project (HHP, 1965-1974). The objective is to examine the impact of various mid- and late-life  exposures,  on  outcomes including cognitive and motor impairment, stroke, other common chronic conditions, and mortality. Both studies followed a cohort of Japanese men born between 1900 and 1919, residing in Hawaii.
The dataset comprises 2079 participants, with 552 classified as heavy drinkers ($>2$ drinks/day) at some point during mid-life. We consider covariates such as age, systolic blood pressure, heart rate at baseline, and years of education attained. The maximum follow-up duration is set at 13 years. In total, 47\% of participants died, 21\% were censored due to loss to follow-up, and the remaining 32\% were alive at the 13-year mark. Our specific focus here is the impact of mid-life alcohol exposure on late-life mortality.

\begin{figure}  
\begin{center}
\includegraphics[width=0.5\textwidth]{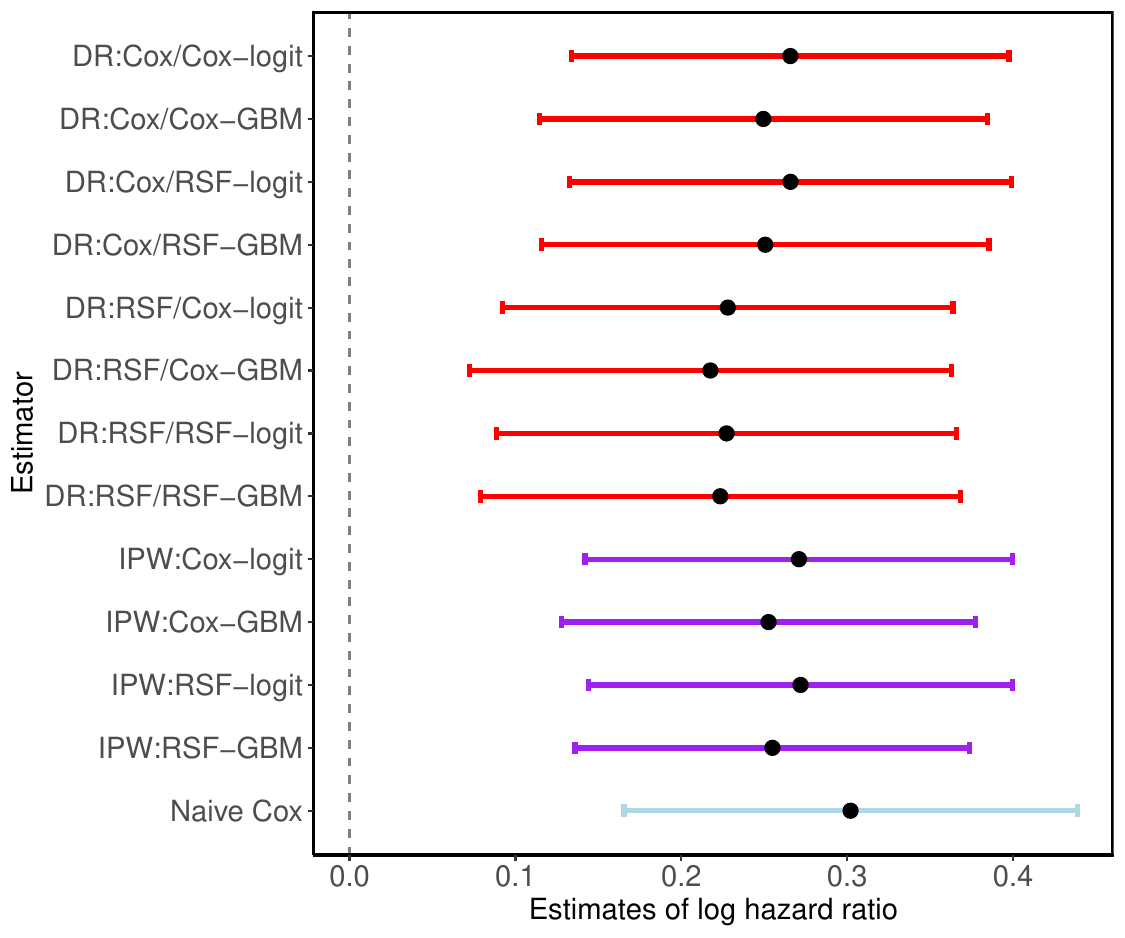} 
\end{center}
\caption{Forest plot of the (time-averaged) log hazard ratio estimates of the effect of mid-life alcohol exposure (heavy versus not heavy) on overall survival for the HAAS data.} \label{AIPW:haas_survival_forest}
\end{figure}

  Figure~\ref{AIPW:haas_survival_forest}  shows the  point estimates and the corresponding $95\%$ bootstrap-based confidence intervals for all 13 estimators; a corresponding table is also provided in the Supplementary Material.   
The naive Cox estimate yields the largest point estimate at 0.30, whereas the IPW estimates are smaller after adjusting for confounding and informative censoring. Notably, the AIPW estimates, particularly those employing machine learning methods, show considerably smaller log hazard ratio estimates compared to the IPW estimates.
All estimates confirm a statistically significant increased risk associated with mid-life alcohol exposure on mortality, although the precision of these estimates varies.  
We note also that the lack of validity of the IPW based inference renders the confidence intervals questionable. 
Additionally, our approach provides estimates of  the risk difference $P(T(1)\le t) -P(T(0)\le t)$ and the risk ratio $P(T(1) \le t)/P(T(0) \le t)$; see  a figure in the Supplementary Material   for $t = 3, 4, \ldots,12$ years.

We would like to remind the reader that when we refer to the log hazard ratio above, we assume that the proportional hazards assumption holds. To date, there is a notable absence of diagnostic tools designed for assessing the marginal structural Cox model. Nevertheless, as expounded upon in the subsequent section, the development of such tools appears feasible with the aid of the AIPW estimators introduced in this paper, which would hold significant practical significance. 

If there are concerns about the proportional hazards assumption, interpreting the estimand as the time-averaged log hazard ratio (Section 3) is a useful approach. This interpretation  is always useful, as discussed in \cite{buja:etal:2019c}, where it is noted that testing the assumption on the same data set can lead to post-selection inference issues.

\section{Discussion}


We have derived an AIPW estimator for the  log hazard ratio under the marginal structural Cox model, both with the proportional hazard assumption and allowing it to be violated. The estimator possesses rate doubly robust property 
which provides a solution to the challenge otherwise posed by the non-collapsibility of the Cox model. 
In this way it becomes  possible 
 to protect against misspecification of the propensity score model which would otherwise invalidate the Cox-IPW approach. We believe that in place of the Cox-IPW estimator commonly used in practice, the AIPW estimator should be routinely used. In addition, as indicated in \cite{L2023} the explicit expression of the AIPW estimator can lead to diagnostics of the proportional hazard assumption in the marginal structural Cox model, and this would be of interest for future work. 
 
 The full data estimating equations involving martingale type increments considered in this paper are more broadly applicable to other semiparametric survival models, such as the additive hazards model \citep{ly} or the transformation cure model \citep{LY2004}. We expect that the joint augmentation technique can be similarly applied, so that doubly robust approaches may be developed for these other marginal structural models besides the Cox one, for analyzing observational studies with survival outcomes. 

In the context of model misspecification 
we could not agree more with  \cite{cox:95} that models are simplifications and idealizations of reality. While it is important to carefully examine any estimation procedure under the model assumptions, it is also very useful to know, in the (often inevitable) situations when a model is misspecified, what we are estimating. Our development in this paper clearly established that the estimand $\beta^*$ (and $\Lambda^*$), outside of the marginal structural Cox model, meets the criteria of {well-specification}
\citep{buja:etal:2019b, buja:etal:2019c}. That is, it is a functional of the potential outcome distributions only, and does not depend on the treatment assignment mechanism, or the covariate distribution, or the censoring mechanism. This is imperative for such an estimand, so that it remains interpretable in the absence of randomization, as well as not affected by specific recruitment patterns of any studies which translate to censoring mechanisms.

We note that among the discussion surrounding \cite{buja:etal:2019a, buja:etal:2019b}, \cite{whit:etal:2019} advocated for  an estimand closely related to ours: $\int \beta(t) f(t) dt$. 
We speculate that it is possible to incorporate different weights into \eqref{eq:ave} through  deconstruction of Lemma \ref{TATE:lem:full data} together with the expression of $\omega(t)$ following it, although this is beyond the scope of the current paper. 
\cite{vansteelandt2022assumption} also focused on assumption-lean Cox regression. They considered an averaged  log cumulative hazard ratio as the estimand, but the time interval on which the average is taken has to be bounded away from time zero. On the other hand, the proportional hazards assumption, if at all, is more likely satisfied before too much time has elapsed. In this sense the approach seems more like an alternative instead of assumption-lean. 
 
Our AIPW approach with the saturated model may also offer  a novel perspective for handling situations with a large number of covariates or effect modifiers, addressing both precise parameter identification and high-dimensionality. The concept of an average, also referred to as least-false, parameter introduces a fresh viewpoint with the potential for more efficient utilization of high-dimensional structures than precise models. While significant and promising, this extension falls beyond our study's scope; nevertheless, our joint augmentation framework has laid the groundwork for future exploration.

\section*{Acknowledgement}
This research was partially supported by 
NIH/NIA grant R03 AG062432 as well as NSF-DMS grant 1712481.

\bibliographystyle{apalike}
\bibliography{causalpropoddsmodel}

\newpage
{\textbf{\Large Supplementary Materials} 

\appendix

\begin{center} {\large Summary} \end{center}
~~~~~~~~The Supplementary Material contains the following:
\begin{enumerate}
\item Notation and expressions;
\item Some results under misspecified Cox model;
\item Proof of identifiability and double robustness;
\item Additional plots;
\item Simulation under the $\beta(t)$ model;
\item Proof of asymptotic results.
\end{enumerate}

\newpage
\section{Notation and Expressions}  \label{AIPW:appendix:notations}

For any random quantities $a$ and $b$, we will use $a \lesssim b$ to denote that $a$ is less than or equal to $b$ up to a constant factor. 

For $a = 0,1$, $l = 0,1$ and  $i= 1,\ldots, n$,
\ban{
    &N_{Ti}(t) = I(T_i \le t),\quad\quad\quad \quad\quad\quad \quad\quad\quad\quad Y_{Ti}(t) = I(T_i \ge t), \\
    &M_{Ti}(t; \beta, \Lambda) =N_{Ti}(t) - \int_0^t Y_{Ti}(u)e^{\beta A_i}d\Lambda(u), \\
    &N_{ci}(t) = I(X_i \le t, \Delta_i = 0), \quad\quad\quad \quad\quad\quad~~ Y_i(t) = I(X_i \ge t), \\
    &M_{ci}(t;a,S_c) = N_{ci}(t) - \int_0^t Y_i(u) d\Lambda_c(u;a,Z_i), \\
    & N^a(t) = I\{X(a) \le t, T(a) \le C(a)\}, \quad\quad\quad Y^a(t) = I\{X(a) \ge t\} \\
    & N_c^a(t) = I\{X(a) \le t, T(a) > C(a) \}, \quad\quad\quad \Delta^a(t) = I\{\min(T(a),t) \leq C(a)\},\\
    &J_i(t;a,S,S_c) = \int_0^t \frac{dM_{ci}(u;a,S_c)}{S(u;a,Z_i)S_c(u;a,Z_i)}, \\ 
    & D_{1i}^w(t;\beta, \Lambda, \pi, S_c) = \frac{dM_i(t;\beta, \Lambda)}{\pi(Z_i)^A_i\{1-\pi(Z_i)\}^{1-A_i}S_c(t;A_i,Z_i)} \\
    &d\mathcal{N}_i^{(l)}(t;\pi,S,S_c) = \frac{A_i^l dN_i(t)}{\pi(Z_i)^{A_i}\{1-\pi(Z_i)\}^{1-A_i}S_c(t;A_i,Z_i)} + \frac{A^l_i dS(t;A_i,Z_i)}{\pi(Z_i)^{A_i}\{1-\pi(Z_i)\}^{1-A_i}} \\
    &\quad\quad\quad\quad\quad\quad\quad\quad- \sum_{a=0,1} a^l \left\{1 + \frac{A_i^a (1-A_i)^{1-a}}{\pi(Z_i)^a\{1-\pi(Z_i)\}^{1-a}}J_i(t;a,S,S_c) \right\} dS(t;a,Z_i), \\
    &\Gamma_i^{(l)}(t;\beta,\pi,S, S_c) = \frac{A^l_i Y_i(t)e^{\beta A_i}}{\pi(Z_i)^{A_i}\{1-\pi(Z_i)\}^{1-A_i}S_c(t;A_i,Z_i)} - \frac{A^l_i S(t;A_i,Z_i)e^{\beta A_i}}{\pi(Z_i)^{A_i}\{1-\pi(Z_i)\}^{1-A_i}} \\
    &\quad\quad\quad\quad\quad\quad\quad\quad+ \sum_{a=0,1} a^l \left\{1 + \frac{A_i^a (1-A_i)^{1-a}}{\pi(Z_i)^a \{1 - \pi(Z_i)\}^{1-a}}J_i(t;a,S,S_c)\right\}S(t;a,Z_i)e^{\beta a}, \\
    &D_{1i}(t; \beta, \Lambda, \pi, S, S_c) = d\mathcal{N}_{i}^{(0)}(t;\pi, S, S_c) - \Gamma_i^{(0)}(t;\beta,\pi,S, S_c) d\Lambda(t), \\
    &D_{2i}(\beta, \Lambda, \pi, S, S_c) = \int_0^\tau d\mathcal{N}_{i}^{(1)}(t;\pi, S, S_c) - \Gamma_i^{(1)}(t;\beta,\pi,S, S_c) d\Lambda(t), \\
    &\mathpzc{s}^{(1)}(t;\beta,\pi, S, S_c) = \frac{\partial}{\partial \beta} \mathpzc{s}^{(0)}(t;\beta,\pi, S, S_c) = \frac{\partial ^2}{\partial \beta^2} \mathpzc{s}^{(0)}(t;\beta,\pi, S, S_c), \\
    &\bar{\alpha}(t;\beta,\pi, S, S_c) = \frac{\mathpzc{s}^{(1)}(t;\beta,\pi, S, S_c)}{\mathpzc{s}^{(0)}(t;\beta,\pi, S, S_c)}, \\
    & v(t;\beta,\pi, S, S_c) = \bar{\alpha}(t;\beta,\pi, S, S_c) - \bar{\alpha}(t;\beta,\pi, S, S_c)^2, \\
    &\nu(\beta, \pi, S, S_c) = \int_0^\tau v(t;\beta,\pi, S, S_c)\mathpzc{s}^{(0)}(t;\beta^*,\pi, S, S_c) d\Lambda^*(t) \\
    &\mu(\beta, \pi, S, S_c) = \int_0^\tau \{\bar{\alpha}(t; \beta^*, \pi, S, S_c) - \bar{\alpha}(t; \beta, \pi, S, S_c) \} \mathpzc{s}^{(0)}(t; \beta^*, \pi, S, S_c) d\Lambda^*(t).
}
Note that the quantities in the last 4 lines are  defined in the Additional Assumptions Section \ref{appendix:assumption} for the Proof of Asymptotics Results later.

Next, we define quantities evaluated over the entire sample of $n$ observations:
\ban{
    &\mathcal{S}^{(l)}(t;\beta,\pi,S, S_c) = \frac{1}{n} \sum_{i = 1}^n  \Gamma_i^{(l)}(t;\beta,\pi,S,S_c), 
    \quad \mathpzc{s}^{(l)}(t;\beta,\pi,S, S_c) = E \{ \mathcal{S}^{(l)}(t;\beta,\pi,S, S_c) \}\\
    &\bar{A}(t;\beta,\pi,S, S_c) = \frac{\mathcal{S}^{(1)}(t;\beta,\pi,S, S_c) }{\mathcal{S}^{(0)}(t;\beta,\pi,S, S_c) }, \\
   &V(t;\beta,\pi,S, S_c) = \bar{A}(t;\beta,\pi,S, S_c) - \bar{A}(t;\beta,\pi,S, S_c)^2, \\
    &\wt{\Lambda}(t;\beta, \pi, S, S_c) = \frac{1}{n} \sum_{i = 1}^n \int_0^t \frac{ d\mathcal{N}_{i}(u;\pi, S, S_c)}{\mathcal{S}^{(0)}(u; \beta,\pi, S, S_c)},  \\
    &U(\beta, \pi, S, S_c) =  \frac{1}{n} \sum_{i = 1}^n \int_0^\tau d\mathcal{N}_i^{(1)}(t;\pi, S, S_c) - \bar{A}(t;\beta,\pi, S, S_c)d\mathcal{N}_i^{(0)}(t;\pi, S, S_c).
}
Analogous to the quantities above, for each fold $m$, we define the fold-specific quantities:
\ban{
    &\mathcal{S}^{(l)}_m(t;\beta, \pi, S, S_c) = \frac{1}{|\mathcal{I}_m|} \sum_{i \in \mathcal{I}_m}  \Gamma_i^{(l)}(t;\beta,\pi, S,S_c), \\
    &\bar{A}_m(t;\beta,\pi, S, S_c) = \frac{\mathcal{S}^{(1)}_m(t;\beta,\pi, S, S_c) }{\mathcal{S}^{(0)}_m(t;\beta,\pi, S, S_c) }, \\
   &V_m(t;\beta,\pi, S, S_c) = \bar{A}_m(t;\beta,\pi, S, S_c) - \bar{A}_m(t;\beta,\pi, S, S_c)^2, \\
    &\wt{\Lambda}_m(t;\beta, \pi,  S, S_c) = \frac{1}{|\mathcal{I}_m|}\sum_{i \in \mathcal{I}_m}\int_0^t \frac{ d\mathcal{N}_{i}^{(0)}(u; \pi, S, S_c)}{\mathcal{S}_m^{(0)}(u; \beta,\pi,  S, S_c)}, \\
    & \wt{\psi}_{m,i}(\beta^*, \Lambda, \pi, S, S_c) =   D_{2i}(t;\beta, \Lambda, \pi, S, S_c) - \int_0^\tau \bar{A}_m(t;\beta, \pi,S,S_c)D_{1i}(t;\beta, \Lambda, \pi, S, S_c).\\
}


The asymptotic variance of $\whb$ in Theorems~\ref{thm:AN} is 
\ba{
\sigma^2 = E\{\psi(\beta^*, \Lambda^*, \pi^o, S^o, S_c^o)^2\}/\nu^2(\beta^*, \pi^o, S^o, S_c^o),
}
where
\ba{
\psi(\beta^*, \Lambda, \pi, S, S_c) =  D_2(\beta, \Lambda, \pi, S, S_c) - \int_0^\tau \bar{\alpha}(t;\beta, \pi,S,S_c)D_1(t;\beta, \Lambda, \pi, S, S_c). \label{AIPW:psi}
}
The asymptotic variance $\sigma^2$ can be consistently estimated using
\ba{
    \wh{\sigma}^2(\whb) = \frac{ {n} \sum_{m=1}^k \sum_{i \in \mathcal{I}_m}  \wt{\psi}_{m,i}(\whb, \wt{\Lambda}_m(\cdot;\whb,\wh{\pi}^{(-m)}, \wh{S}^{(-m)}, \wh{S}_c^{(-m)}), \wh{\pi}^{(-m)}, \wh{S}^{(-m)} , \wh{S}_c^{(-m)})^2}{\left\{ \sum_{m=1}^k  \sum_{i \in \mathcal{I}_m} \int_0^\tau V_m(t;\whb, \wh{\pi}^{(-m)}, \wh{S}^{(-m)}, \wh{S}_c^{(-m)}) d\mathcal{N}_i^{(0)}(t; \wh{\pi}^{(-m)}, \wh{S}^{(-m)}, \wh{S}^{(-m)}_c)\right\}^2}. \label{AIPW:var.estimator}
}

Without cross-fitting, and when all the working models are correctly specified (semi)parametrically,  the asymptotic variance can be consistently estimated using
\ba{
    & \tilde{\sigma}^2(\hat\beta) = \frac{ {n} \sum_{i=1}^n  \wt{\psi}_i(\hat\beta, \wt{\Lambda}(\cdot;\beta,\wh{\pi}, \wh{S}, \wh{S}_c), \wh{\pi}, \wh{S} , \wh{S}_c)^2}{\left\{ \sum_{i=1}^n  \int_0^\tau V(t; \hat\beta, \wh{\pi}, \wh{S}, \wh{S}_c) d\mathcal{N}_i^{(0)}(t; \wh{\pi}, \wh{S}, \wh{S}_c)\right\}^2}, \label{TATE:var.est}
}
where $\wh{\pi}$, $\wh{S}$ and $\wh{S}_c$ are estimated using the same sample of $n$ observations.

\section{Some results under misspecified Cox model}


\noindent{\it Proof of Lemma~\ref{TATE:lem:full data}.} 
First, we solve $E\{D_1^f(t; \beta, \Lambda) \} = 0$. 
By Leibniz integral rule, we may exchange the order of differentiation and integral and have 
\ba{
d\Lambda(t) =& \frac{\sum_{a=0,1} dE\{I\{T(a)<t\} \}}{\sum_{a=0,1}  e^{\beta a} E\left[ I\{T(a) \ge t\} \right ]}
            = \frac{\sum_{a=0,1} dF_a(t)}{\sum_{a=0,1}  e^{\beta a} S_a(t) }. 
     \label{TATE:lem2.1}
}
Plugging it into $E\{D_2^f(\beta, \Lambda) \} = 0$, we have
\ban{
0 =& \int_0^\tau \sum_{a=0,1}  a \cdot dE[I\{T(a)<t\}] -   \frac{\sum_{a=0,1} a e^{\beta a }E\left[I\{T(a) \ge t\}\right] \cdot \sum_{a=0,1} dF_a(t)}{\sum_{a=0,1}  e^{\beta a} S_a(t) }  \\
    =& \int_0^\tau \sum_{a=0,1}  a \cdot f_a(t) dt - \frac{\sum_{a=0,1} a e^{\beta a } S_a(t)}{\sum_{a=0,1}  e^{\beta a} S_a(t)} \sum_{a=0,1} f_a(t) dt, \\
    =& \int_0^\tau \left\{  \frac{\sum_{a=0,1} a \cdot f_a(t) }{\sum_{a=0,1} f_a(t)}   - \frac{\sum_{a=0,1} a e^{\beta a} S_a(t) }{\sum_{a=0,1} e^{\beta a} S_a(t)} \right \} \sum_{a=0,1}f_a(t) dt \\
    =& \int_0^\tau \left\{  \frac{\sum_{a=0,1} a \cdot \Lambda(t) e^{\beta(t) a} S_a(t) }{\sum_{a=0,1} \Lambda(t) e^{\beta(t) a} S_a(t)  }   - \frac{\sum_{a=0,1} a e^{\beta a} S_a(t) }{\sum_{a=0,1} e^{\beta a} S_a(t)} \right \} \sum_{a=0,1}f_a(t) dt \\
    =& \int_0^\tau \left\{  \frac{\sum_{a=0,1} a e^{\beta(t) a} S_a(t) }{\sum_{a=0,1} e^{\beta(t) a} S_a(t)}   - \frac{\sum_{a=0,1} a e^{\beta a} S_a(t) }{\sum_{a=0,1} e^{\beta a} S_a(t)} \right \} \sum_{a=0,1}f_a(t) dt,
}
which is equivalent to the definition of $\beta^*$ defined in \eqref{eq:soln1}. 
In addition $v(\beta, t)>0$. 
Therefore $\beta^*$ is the unique solution to $\beta$ in the full data estimating functions. Plugging $\beta^*$ into \eqref{TATE:lem2.1}, we also see that $\Lambda^*(t)$ as defined in \eqref{eq:soln2} 
is also the solution to $\Lambda(t)$ in the full data estimating functions.
\qed

\bigskip
We next show that under 1:1 randomization, equation \eqref{eq:beta*} is equivalent to \eqref{eq:soln1}.

By  consistency we have
\ban{
f(t) = P(A=1)f(t;A=1) + P(A=0)f(t;A=0) = \frac{1}{2}\{f_1(t) + f_0(t)\}.
}
Bayes' rule implies that for two random variables $X$ and $Y$,
\ban{
    E(X|Y=y) = \frac{E\{Xf_{Y|X}(y|X)\}}{f_Y(y)}.
}
Applying this to $E_{\beta(t)}(A|T=t)$
we have
\ban{
E_{\beta(t)}(A|T=t) =& \frac{E\{Af(t;A)\}}{f(t)} \\
=& \frac{ \sum_{a=0,1} af_a(t) }{
\sum_{a=0,1} f_a(t)} \\
=& \frac{\sum_{a=0,1} a\lambda_{T(a)}(t)S_a(t) }{\sum_{a=0,1} \lambda_{T(a)}(t)S_a(t)} \\
=& \frac{\sum_{a=0,1} a e^{\beta(t) a} S_a(t) }{\sum_{a=0,1} e^{\beta(t) a} S_a(t)}. 
}
Replacing $\beta(t)$ in the above with a constant $\beta$ 
we have
\ban{
E_{\beta}(A|T=t) = \frac{\sum_{a=0,1} a e^{\beta a} S_a(t) }{\sum_{a=0,1} e^{\beta a} S_a(t)}. 
}
Substituting these two quantities into \eqref{eq:beta*} we  have \eqref{eq:soln1}. 
\qed

\section{Proof of Identifiability and Double Robustness}

We first state and prove some lemmas that will be used. 

\begin{lemma} \label{TATE:lem:AZ}
For any real-valued functions $g$ and $h$, we have
\ba{
    E\{g(A,Z)h(T,C,A,Z) \} = \sum_{a=0,1} E \left[ g(a,Z) \pi^o(Z)^a \{(1-\pi^o(Z)\}^{1-a} E \{h(T,C,A,Z)|A=a, Z \} \right]
}
\end{lemma}

{\it Proof.}
By the law of total expectation we have
\ba{
  &E\{g(A,Z)h(T,C,A,Z) \}     \\
  =& E [ E\{g(A,Z)h(T,C,A,Z)|A,Z \} ] \\
  =& E \left[  \sum_{a=0,1} E \{g(A,Z)h(T,C,A,Z)|A=a,Z \} \pi^o(Z)^a\{1-\pi^o(Z)\}^{1-a}   \right]  \\
  =& \sum_{a=0,1} E \left[ g(a,Z) \pi^o(Z)^a \{(1-\pi^o(Z)\}^{1-a} E \{h(T,C,A,Z)|A=a,Z \}    \right].
}
\qed

\begin{lemma} \label{TATE:lem:observed_M}
Denote  $\Delta^a(t) = I\{\min(T(a),t) \leq C(a)\}$ for $a=0,1$.
Then 
\ba{
    M(t;\beta, \Lambda)  = A\Delta^1(t)M_T^1(t;\beta, \Lambda) + (1-A)\Delta^0(t) M_T^0(t;\beta, \Lambda).
}
\end{lemma}

{\it Proof.}
By definition $N^a(t) = I\{T(a) \le C(a)\} I\{T(a) \le t\}$.
Meanwhile 
$$N^a_T(t)\Delta^a(t) = I\{T(a) \le t\} I\{ \min(T(a), t) \le C(a)\}
= I\{T(a) \le t\} I\{T(a) \le C(a)\}.$$ 
Therefore $N^a(t) = N^a_T(t)\Delta^a(t) $. 

In addition, 
\be
Y^a_T(t)\Delta^a(t)  
    &=& I(T(a) \ge t) I\{ \min(T(a),t) \le C(a)\} \\
    &=& I(T(a) \ge t) I\{ C(a) \ge t\} 
    = I(X(a) \ge t) = Y^a(t).
 \ee
    
Then by the consistency Assumption~\ref{assump2}, we have
\ba{
    N(t) &= AN^1(t) + (1 - A)N^0(t) \\
    &= AN^1_T(t)\Delta^1(t) + (1-A)N^0_T(t)\Delta^0(t). \label{TATE:lem1.3}
}
Similarly,
\ba{
    Y(t) = AY^1_T(t)\Delta^1(t) + (1-A)Y^0_T(t)\Delta^0(t). \label{TATE:lem1.4}
}
Combining \eqref{TATE:lem1.3} and \eqref{TATE:lem1.4} completes the proof.
\qed

\begin{lemma}\label{TATE:lem:dMc to Deltac}
For $a = 0,1$, 
\ba{
    \frac{\Delta^a(t) dM_T^a(t; \beta, \Lambda)}{S_c(t;a,Z)} = dM_T^a(t; \beta, \Lambda) - dM_T^a(t; \beta, \Lambda)\int_0^t \frac{dM_c(u;a,S_c)}{S_c(u;a,Z)}.
}
\end{lemma}

{\it Proof.}
We prove the result for $a=1$, and the same arguments can be made for $a=0$.  

The following is a potential outcome version of Lemma 1 from \cite{LX2022}. 
Note that 
\ba{
\int_0^t \frac{ dN_c^1(u) }{S_c(u;1,Z)} = \frac{N_c^1(t-)}{S_c(X;1,Z)}.
\label{TATE:l3.1}  
}
Because $ \Lambda_c(u;1,Z) = - \log \{S_c(u;1,Z)\}$, we have
\ba{
    &\int_0^t  \frac {- Y^1(u) d\Lambda_c(u;1,Z) }{S_c(u;1,Z) }  \\
    =& I\{X(1) \ge t\}\int_0^t \frac{dS_c(u;1,Z) }{S_c(u;1,Z)^2} +I\{X(1)
    < t\}\int_0^{X(1)} \frac{dS_c(u;1,Z) }{S_c(u;1,Z)^2}  \\
    =& I\{X(1) \ge t\}\{ - S_c(u;1,Z)^{-1} \}|_{u=0}^{u=t} + I\{X(1) < t\}\{ - S_c(u;1,Z)^{-1} \}|_{u=0}^{u=X(1)}   \\
    =& \frac{I\{X(1) \ge t\}}{S_c(0;1,Z)} + \frac{I\{X(1) < t\}}{S_c(0;1,Z)} - \frac{I\{X(1) \ge t\}}{ S_c(t;1,Z)}  - \frac{ I(X(1) < t) }{S_c(X(1);1,Z)},  \\
    =& 1 - \frac{ Y^1(t)}{ S_c(t;1,Z)}  - \frac{ I(X(1) < t) }{S_c(X(1);1,Z)}. \label{TATE:l3.2}
}
Since $I(X(1) < t) = N^1(t-) + N_c^1(t-) $, \eqref{TATE:l3.1}  $+$ \eqref{TATE:l3.2}  gives
\ba{
\int_0^t \frac{dM_c(u;1,S_c)}{ S_c(u;1,Z)}  = 1 - \frac{Y^1(t)}{ S_c (t;1,Z)} - \frac{N^1(t-)}{ S_c(X(1);1,Z)}. \label{TATE:l3.3}
}
The rest of the proof is analogous to part (b) of the proof of Theorem 1 from \cite{LX2022}.  
Note that 
$Y^1(t) dN_T^1(t)= dN^1(t) $,  and $dN_T^1(t)N^1(t-) = Y_T^1(t)N^1(t-) = 0$.  
Multiplying \eqref{TATE:l3.3} by $dM_T^1(t; \beta, \Lambda) = dN_T^1(t) - Y_T^1(t) e^{\beta} d\Lambda(t)$ we have
\ban{
& dM_T^1(t; \beta, \Lambda)\int_0^t \frac{dM_c(u;1,S_c)}{S_c(u;1,Z)} \\
=&     dN_T^1(t)\int_0^t \frac{dM_c(u;1,S_c)}{S_c(u;1,Z)}
- Y_T^1(t)e^{\beta}d\Lambda(t)\int_0^t \frac{dM_c(u;1,S_c)}{S_c(u;1,Z)} \\
=& dN_T^1(t) - \frac{dN_T^1(t) Y^1(t)}{S_c(t;1,Z)} - \frac{dN_T^1(t) N^1(t-)}{S_c(X(1);1,Z)} 
- Y_T^1(t)e^{\beta}d\Lambda(t) + \frac{Y^1(t)e^{\beta}d\Lambda(t) }{S_c(t;1,Z)} + \frac{ Y_T^1(t)N^1(t-)e^{\beta^*}d\Lambda^*(t) }{S_c(X(1);1,Z)}. \\
=&  dN_T^1(t) - Y_T^1(t)e^{\beta}d\Lambda(t) - \frac{dN^1(t)}{S_c(t;1,Z)} +  \frac{Y^1(t)e^{\beta}d\Lambda(t)}{S_c(t;1,Z)} \\
=& dM_T^1(t; \beta, \Lambda) - \frac{dN^1(t) - Y^1(t)e^{\beta}d\Lambda(t)}{S_c(t;1,Z)} \\
=& dM_T^1(t; \beta, \Lambda) -  \frac{\Delta^1(t) dN_T^1(t) - \Delta^1(t) Y_T^1(t)e^{\beta}d\Lambda(t)}{S_c(t;1,Z)} \\
=& dM_T^1(t; \beta, \Lambda) -  \frac{\Delta^1(t) dM_T^1(t; \beta, \Lambda)}{S_c(t;1,Z)} .
}

\qed

\subsection{Identifiability via IPW}

\begin{lemma} \label{TATE:lem:identifiability}
Under Assumptions~\ref{assump1}-\ref{assump5}, for $t \in [0,\tau]$,
\be
E\{D_1^w(t; \beta^*,\Lambda^*, \pi, S_c)\} = 0~~~~\text{and}~~~~E\{D_2^w(\beta^*,\Lambda^*, \pi, S_c)\} = 0. 
\ee
\end{lemma}

{\it Proof. } 
Using Lemma~\ref{TATE:lem:AZ} and Lemma~\ref{TATE:lem:observed_M}, we have 
\ba{
& E\left[ \frac{A^ldM(t;\beta^*, \Lambda^*)}{\pi(Z)^A\{1-\pi(Z)\}^{1-A}S_c(t;A,Z)}   \right] \\
      =&\sum_{a=0,1} E\left[\frac{a^l}{S_c(t;a,Z)} E\left\{dM(t;\beta^*, \Lambda^*)|A=a, Z  \right\}       \right]    \\
     =& \sum_{a=0,1} E\left[\frac{a^l}{S_c(t;a,Z)} E\left\{\Delta^a(t)  dM_T^a(t;\beta^*, \Lambda^*)| Z  \right\}       \right]   \\
     =& \sum_{a=0,1} E\left(\frac{a^l}{S_c(t;a,Z)} E\left[ E\left\{ \Delta^a(t)  dM_T^a(t;\beta^*, \Lambda^*)|T(a)=t, Z \right\}|Z \right]  \right)  \\
     =&  \sum_{a=0,1} E\left\{\frac{a^l}{S_c(t;a,Z)} E\left( E\left[\{ dN_T^a(t) I(C(a) \ge t) - Y_T^a(t) I(C(a) \ge t) e^{\beta^* a} d\Lambda^*(t) \}|T(a)=t, Z \right] \Big|Z \right)  \right\}  \\
     =&  \sum_{a=0,1} E\left\{\frac{a^l E\{I(C(a) \ge t) |Z\} }{S_c(t;a,Z)} E\left( E\left[\{ dN_T^a(t) - Y_T^a(t) e^{\beta^* a} d\Lambda^*(t) \}|T(a)=t, Z \right] \Big|Z \right)  \right\}\\ \label{TATE:thm1.16} \\
     =& \sum_{a=0,1} a^l dE\{ M_T^a(t;\beta^*, \Lambda^*) \}    \label{TATE:thm1.11}\\
     =& 0,
}
where \eqref{TATE:thm1.16} makes use of the informative censoring Assumption~\ref{assump5}, and \eqref{TATE:thm1.11} uses the consistency Assumption~\ref{assump2} and the tower property.
This then gives both $E\{D_1^w(t;\beta^*, \Lambda^*, \pi, S_c)\}=0$ and $E\{D_2^w(\beta^*, \Lambda^*, \pi, S_c)\}=0$. 

\subsection{\texorpdfstring{Proof of Theorem~1$'$ (double robustness)}{Proof of Theorem 1' (double robustness)}}
Note that 
\ban{
    &D_1(t; \beta^*, \Lambda^*, \pi, S, S_c) = d\mathcal{N}_{i}^{(0)}(t;\pi, S, S_c) - \Gamma_i^{(0)}(t;\beta^*,\pi,S, S_c) d\Lambda^*(t), \\
    &D_2(\beta^*, \Lambda^*, \pi, S, S_c) = \int_0^\tau d\mathcal{N}_{i}^{(1)}(t;\pi, S, S_c) - \Gamma_i^{(1)}(t;\beta^*,\pi,S, S_c) d\Lambda^*(t).
}

By Fubini's theorem, in obvious short-hand notation 
it suffices to show that $E\{d\mathcal{N}_{i}^{(l)}(t) - \Gamma_i^{(l)}(t)d\Lambda^*(t)\} = 0$ for $l = 0,1$ and  any $t \in [0,\tau]$.

a) \underline{Assume $(\pi, S_c) = (\pi^o, S^o_c)$}. 
We can  write $E\{d\mathcal{N}_{i}^{(l)}(t) - \Gamma_i^{(l)}(t)d\Lambda^*(t)\} = R_1 + R_2 - R_3$, where 
\ba{
  R_1&=  E\left[ \frac{A^ldM(t;\beta^*, \Lambda^*)}{\pi^o(Z)^A\{1-\pi^o(Z)\}^{1-A}S_c^o(t;A,Z)}   \right] , \\
  R_2&= E\left[ \frac{A^l\{ dS(t;A,Z) + S(t;A,Z)e^{\beta^* A} d\Lambda^*(t)\}}{\pi^o(Z)^A\{1-\pi^o(Z)\}^{1-A}}  - \sum_{a=0,1} a^l \{dS(t;a,Z) + S(t;a,Z)e^{\beta^* a} d\Lambda^*(t) \} \right],    \\
  R_3&= E\left[ \sum_{a=0,1} a^l\frac{A^a(1-A)^{1-a}}{\pi^o(Z)^a\{1-\pi^o(Z)\}^{1-a}} J(t;a,S,S_c^o)\{ dS(t;a,Z) + S(t;a,Z)e^{\beta^* a} d\Lambda^*(t) \}  \right].
}
$R_1 = 0$ follows directly from the identifiability Lemma~\ref{TATE:lem:identifiability}.

Applying Lemma~\ref{TATE:lem:AZ} to $R_2$, we have
\ba{
  R_2=&  E\left[ \sum_{a=0,1} a^l\{dS(t;a,Z) + S(t;a,Z)e^{\beta^* a} d\Lambda^*(t) \} - \sum_{a=0,1}a^l \{dS(t;a,Z) + S(t;a,Z)e^{\beta^* a} d\Lambda^*(t) \} \right]  \\
     =& 0.
}
Finally, again applying Lemma~\ref{TATE:lem:AZ}, we have
\ba{
  R_3=& \sum_{a=0,1} \sum_{\alpha = 0,1}  a^l E\bigg[ \frac{\alpha^a(1-\alpha)^{1-a}\pi^o(Z)^{\alpha} \{(1-\pi^o(Z)\}^{1-\alpha}}{\pi^o(Z)^a\{1-\pi^o(Z)\}^{1-a}} \{ dS(t;a,Z) + S(t;a,Z)e^{\beta^* a} d\Lambda^*(t) \} \\
  &\times E\{J(t;a,S,S_c^o) |A=\alpha, Z\} \bigg]    \\
  =& \sum_{a=0,1} a^l E\left[ \{ dS(t;a,Z) + S(t;a,Z)e^{\beta^* a} d\Lambda^*(t) \} E\{J(t;a,S,S_c^o) |A=a, Z\} \right] \label{TATE:thm1.14}   \\
 =& \sum_{a=0,1} a^l E\bigg[ \{ dS(t;a,Z) + S(t;a,Z)e^{\beta^* a} d\Lambda^*(t) \}   \int_0^t \frac{dE\{M_c(u;a,S_c^o)|A=a,Z\}}{S(u;a,Z)S_c^o(u;a,Z)} \bigg] \\
 =& 0, \label{TATE:thm1.15}
}
where \eqref{TATE:thm1.14} comes from $\alpha^a(1-\alpha)^{1-a} = I(a = \alpha)$, and \eqref{TATE:thm1.15} uses the fact that for each $A=a$, $M_c(t;a,S_c^o)$ given $Z$ is a martingale when $S_c = S_c^o$.

b) \underline{Assume $S = S^o$}. We have $E\{d\mathcal{N}_{i}^{(l)}(t) - \Gamma_i^{(l)}(t)d\Lambda^*(t)\} = R_4 + R_5 + R_6$, where 
\ba{
  R_4=& E\bigg[ \frac{A^l dM(t;\beta^*, \Lambda^*)}{\pi(Z)^A\{1-\pi(Z)\}^{1-A}S_c(t;A,Z)} \\
  &-\sum_{a=0,1} a^l\frac{A^a(1-A)^{1-a}}{\pi(Z)^a\{1-\pi(Z)\}^{1-a}} J(t;a,S^o,S_c)\{ dS^o(t;a,Z) + S^o(t;a,Z)e^{\beta^* a} d\Lambda^*(t)\} \bigg],    \\
  R_5=&  E\left[ \frac{A^l\{ dS^o(t;A,Z) + S^o(t;A,Z)e^{\beta^* A} d\Lambda^*(t) \}}{\pi(Z)^A\{1-\pi(Z)\}^{1-A}} \right],    \\
  R_6=& - \sum_{a=0,1}a^l E\{dS^o(t;a,Z) + S^o(t;a,Z)e^{\beta^* a} d\Lambda^*(t) \},
}

We first make use of the fact that under $S = S^o$, 
$$ 
E\{dM_T(t;\beta,\Lambda)|T \ge u, A,  Z\} =- \frac{dS(t;A,Z) + S(t;A,Z)e^{\beta A}d\Lambda(t)} 
{S(u|A,Z)}. 
$$
Therefore  
\ba{
  R_4  =& E\bigg[ \frac{A^ldM(t;\beta^*, \Lambda^*)}{\pi(Z)^A\{1-\pi(Z)\}^{1-A}S_c(t;A,Z)} \label{TATE:thm1.12} \\
  &+\sum_{a=0,1} a^l\frac{A^a(1-A)^{1-a}}{\pi(Z)^a\{1-\pi(Z)\}^{1-a}} \int_0^t \frac{dM_c(u;a,S_c)}{S_c(u;a,Z)} E\{ dM_T(t;\beta^*,\Lambda^*)|T\ge u, A=a, Z \} \bigg]. \label{TATE:thm1.13}
}
Applying Lemma~\ref{TATE:lem:AZ} to both \eqref{TATE:thm1.12} and \eqref{TATE:thm1.13}, we have
\ba{
  R_4 =& \sum_{a=0,1} a^l E\left[ \frac{\pi^o(Z)^a\{1-\pi^o(Z)\}^{1-a}}{\pi(Z)^a\{1-\pi(Z)\}^{1-a}} \frac{E\{ dM(t;\beta^*, \Lambda^*)|A=a, Z\}}{S_c(t;a,Z)} \right] \\
  &+ \sum_{a=0,1} \sum_{\alpha=0,1} a^lE\bigg( \frac{\alpha^a(1-\alpha)^{1-a}\pi^o(Z)^\alpha\{1-\pi^o(Z)\}^{1-\alpha}}{\pi(Z)^a\{1-\pi(Z)\}^{1-a}} \\
  &\quad\times E\left[ \int_0^t \frac{dM_c(u;a,S_c)}{S_c(u;a,Z)} E\{ dM_T(t;\beta^*,\Lambda^*)|T\ge u,A=a, Z \} \Big|A=\alpha, Z \right] \bigg) \\
  =& \sum_{a=0,1} a^l E\left[ \frac{\pi^o(Z)^a\{1-\pi^o(Z)\}^{1-a}}{\pi(Z)^a\{1-\pi(Z)\}^{1-a}} \frac{\Delta^a(t) dM_T^a(t;\beta^*, \Lambda^*)}{S_c(t;a,Z)} \right] \label{TATE:thm1.21} \\
  &+ \sum_{a=0,1} a^lE\bigg( \frac{\pi^o(Z)^a\{1-\pi^o(Z)\}^{1-a}}{\pi(Z)^a\{1-\pi(Z)\}^{1-a}} \\
  &\quad\times E\left[\int_0^t \frac{dM_c(u;a,S_c)}{S_c(u;a,Z)} E\{ dM_T^a(t;\beta^*,\Lambda^*)|T(a)\ge u,A=a, Z \}\bigg|A=a, Z \right] \bigg) \\
  \label{TATE:thm1.17} \\
  =& \sum_{a=0,1} a^l E\bigg[ \frac{\pi^o(Z)^a\{1-\pi^o(Z)\}^{1-a}}{\pi(Z)^a\{1-\pi(Z)\}^{1-a}} \bigg( \frac{\Delta^a(t) dM_T^a(t;\beta^*, \Lambda^*)}{S_c(t;a,Z)} \\
  &+ E\left[\int_0^t \frac{dM_c(u;a,S_c)}{S_c(u;a,Z)} E\{ dM_T^a(t;\beta^*,\Lambda^*)|T(a)\ge u,A=a, Z \}\bigg|A=a, Z \right] \bigg ) \bigg]   \\
  =& \sum_{a=0,1} a^l E\left[ \frac{\pi^o(Z)^a\{1-\pi^o(Z)\}^{1-a}}{\pi(Z)^a\{1-\pi(Z)\}^{1-a}} dM_T^a(t;\beta^*, \Lambda^*) \right] \label{TATE:thm1.18} 
  + R_7, 
}
where \eqref{TATE:thm1.21} uses Lemma~\ref{TATE:lem:observed_M} and the tower property, \eqref{TATE:thm1.17}  makes use of the fact that $\alpha^a(1-\alpha)^{1-a} = I(a = \alpha)$ and the consistency Assumption~\ref{assump2},  \eqref{TATE:thm1.18} makes use of Lemma~\ref{TATE:lem:dMc to Deltac}, and 
\ban{
     R_7 
    =& \sum_{a=0,1} a^l E\bigg[ \frac{\pi^o(Z)^a\{1-\pi^o(Z)\}^{1-a}}{\pi(Z)^a\{1-\pi(Z)\}^{1-a}} \\ 
     & \times\bigg\{ E\Big[ \int_0^t \frac{dM_c(u;a,S_c)}{S_c(u;a,Z)} E\{ dM_T^a(t;\beta^*,\Lambda^*)|T(a)\ge u, A=a, Z\} \bigg|A=a, Z \Big] \\
     &\quad -  \int_0^t \frac{dM_c(u;a,S_c)}{S_c(u;a,Z)} dM_T^a(t;\beta^*,\Lambda^*)  \bigg \} \bigg] \\
     =& \sum_{a=0,1} a^l E\bigg[ \frac{\pi^o(Z)^a\{1-\pi^o(Z)\}^{1-a}}{\pi(Z)^a\{1-\pi(Z)\}^{1-a}} \\ 
     & \times E\Big\{ \int_0^t \frac{dM_c(u;a,S_c)}{S_c(u;a,Z)} [E\{ dM_T^a(t;\beta^*,\Lambda^*)|T(a)\ge u, A=a, Z\} - dM_T^a(t;\beta^*,\Lambda^*) ] \bigg|A=a, Z \Big\} \bigg].
}
We show that $R_7=0$ by showing that the inner conditional expectation is zero:
\ba{
    &  E\Big\{ \int_0^t \frac{dM_c(u;a,S_c)}{S_c(u;a,Z)} [E\{ dM_T^a(t;\beta^*,\Lambda^*)|T(a)\ge u, A=a,Z\} - dM_T^a(t;\beta^*,\Lambda^*) ] \bigg|A=a, Z \Big\} \\
    =&  E\bigg[ E\Big\{ \int_0^t \frac{dN_c(u)}{S_c(u;a,Z)} \\
    &\quad \times [E\{ dM_T^a(t;\beta^*,\Lambda^*)|T(a)\ge u, A=a, Z\} - dM_T^a(t;\beta^*,\Lambda^*) ] \bigg|A=a, Z, T \ge u, C =u \Big\} \bigg|A=a,Z \bigg] \\
    &- E\bigg[ E\Big\{ \int_0^t \frac{Y(u)d\Lambda_c(u;a,Z)}{S_c(u;a,Z)} \\
    &\quad\times [E\{ dM_T^a(t;\beta^*,\Lambda^*)|T(a)\ge u,A=a, Z\} - dM_T^a(t;\beta^*,\Lambda^*) ] \bigg|A=a, Z, T \ge u, C =u \Big\} \bigg| A=a,Z \bigg] \\
     =& E \bigg \{ \int_0^t \frac{dN_c^a(u) }{S_c(u;a,Z)} \\
     &\quad\times \Big[E\{dM_T^a(t;\beta^*,\Lambda^*)|T(a)\ge u,A=a, Z\} - E\{dM_T^a(t;\beta^*,\Lambda^*)|T(a)\ge u,A=a, Z\} \Big] \bigg| A=a, Z \bigg\} \label{TATE:thm1.19}\\
     &-  E \bigg \{ \int_0^t \frac{Y^a(u)d\Lambda_c(u;a,Z)}{S_c(u;a,Z)} \\
     &\quad\times \Big[E\{dM_T^a(t;\beta^*,\Lambda^*)|T(a)\ge u, A=a, Z\} - E\{dM_T^a(t;\beta^*,\Lambda^*)|T(a)\ge u,A=a, Z\} \Big]  \bigg| A=a, Z \bigg\} \label{TATE:thm1.20} \\
    =& 0,
}
where \eqref{TATE:thm1.19} and \eqref{TATE:thm1.20} uses consistency and informative censoring from Assumptions~\ref{assump2} and \ref{assump5}.

Next, using Lemma~\ref{TATE:lem:AZ}, we have 
\ba{
  R_5=& \sum_{a=0,1} a^l E\left[ \frac{\pi^o(Z)^a\{1-\pi^o(Z)\}^{1-a}}{\pi(Z)^a\{1-\pi(Z)\}^{1-a}} \{dS^o(t;a,Z) + S^o(t;a,Z)e^{\beta^* a} d\Lambda^*(t)\} \right]    \\
     =& \sum_{a=0,1} a^l E\left[ \frac{\pi^o(Z)^a\{1-\pi^o(Z)\}^{1-a}}{\pi(Z)^a\{1-\pi(Z)\}^{1-a}} E\{-dN_T^a(t) + Y_T^a(t)e^{\beta^* a} d\Lambda^*(t) |Z\} \right]      \\
     =& - \sum_{a=0,1} a^l E\left[ E \left \{\frac{\pi^o(Z)^a\{1-\pi^o(Z)\}^{1-a}}{\pi(Z)^a\{1-\pi(Z)\}^{1-a}} dM_T^a(t;\beta^*, \Lambda^*) \Big |Z \right \} \right]      \\
     =& - \sum_{a=0,1} a^l E\left[ \frac{\pi^o(Z)^a\{1-\pi^o(Z)\}^{1-a}}{\pi(Z)^a\{1-\pi(Z)\}^{1-a}} dM_T^a(t;\beta^*, \Lambda^*) \right].    
}
Lastly,
\ba{
  R_6=& - \sum_{a=0,1}a^l E(E[\{dS^o(t;a,Z) + S^o(t;a,Z)e^{\beta^* a} d\Lambda^*(t) \}|Z])    \\
     =& - \sum_{a=0,1}a^l E(E[\{-dN_T^a(t) + Y_T^a(t)e^{\beta^* a} d\Lambda^*(t) \}|Z])   \\
     =& \sum_{a=0,1}a^l E \{ dM_T^a(t;\beta^*, \Lambda^*)\}  \\
     =& 0.
}
The above gives $R_4 + R_5 + R_6 = 0$ as desired.
\qed

\section{Additional plots and tables}

\begin{figure}[htbp]
\begin{center}
\begin{tabular}{cc}
\includegraphics[angle = 90, height = 92mm, width=0.46\linewidth]{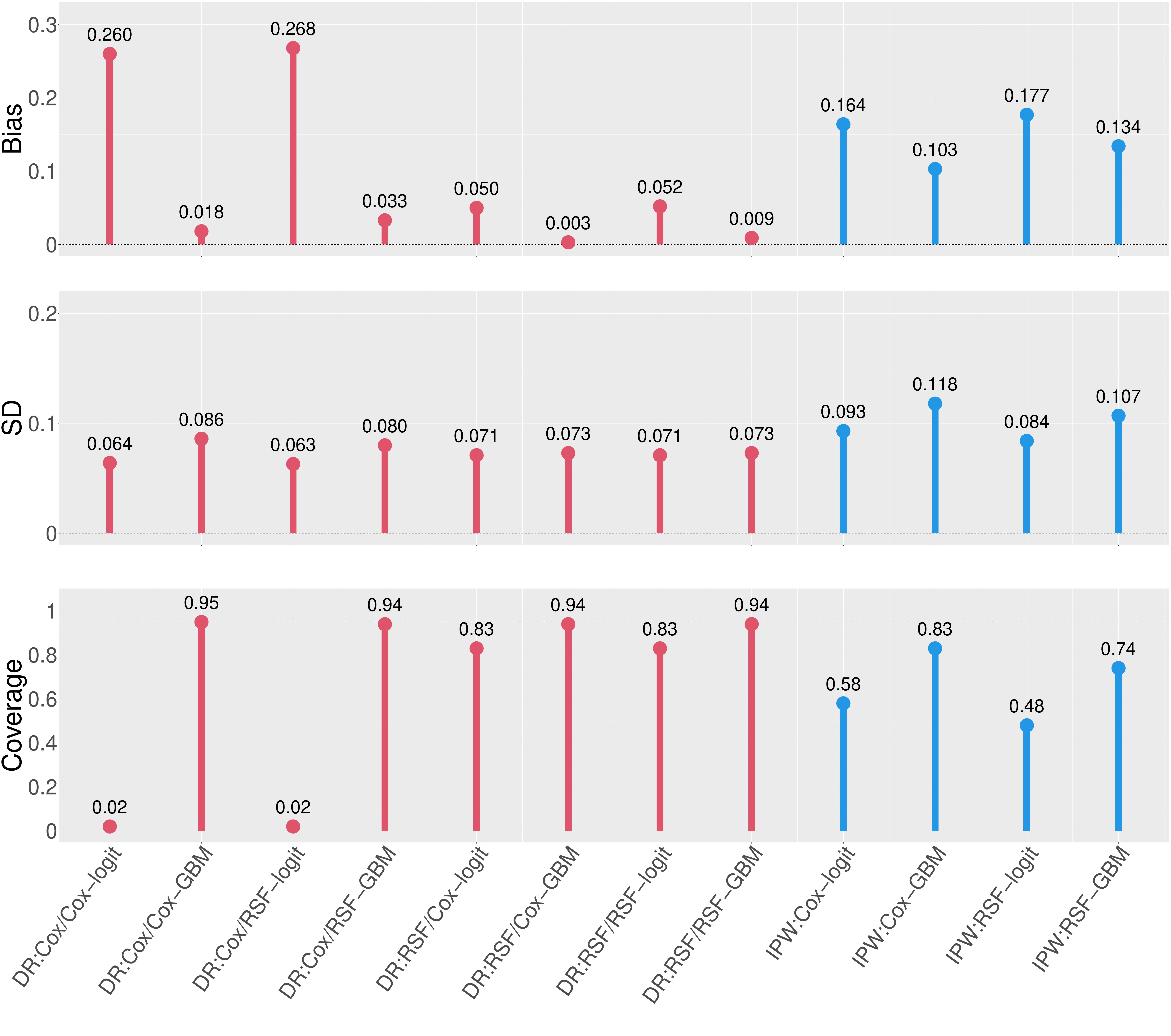} &   \includegraphics[angle = 90, height = 92mm, width=0.46\linewidth]{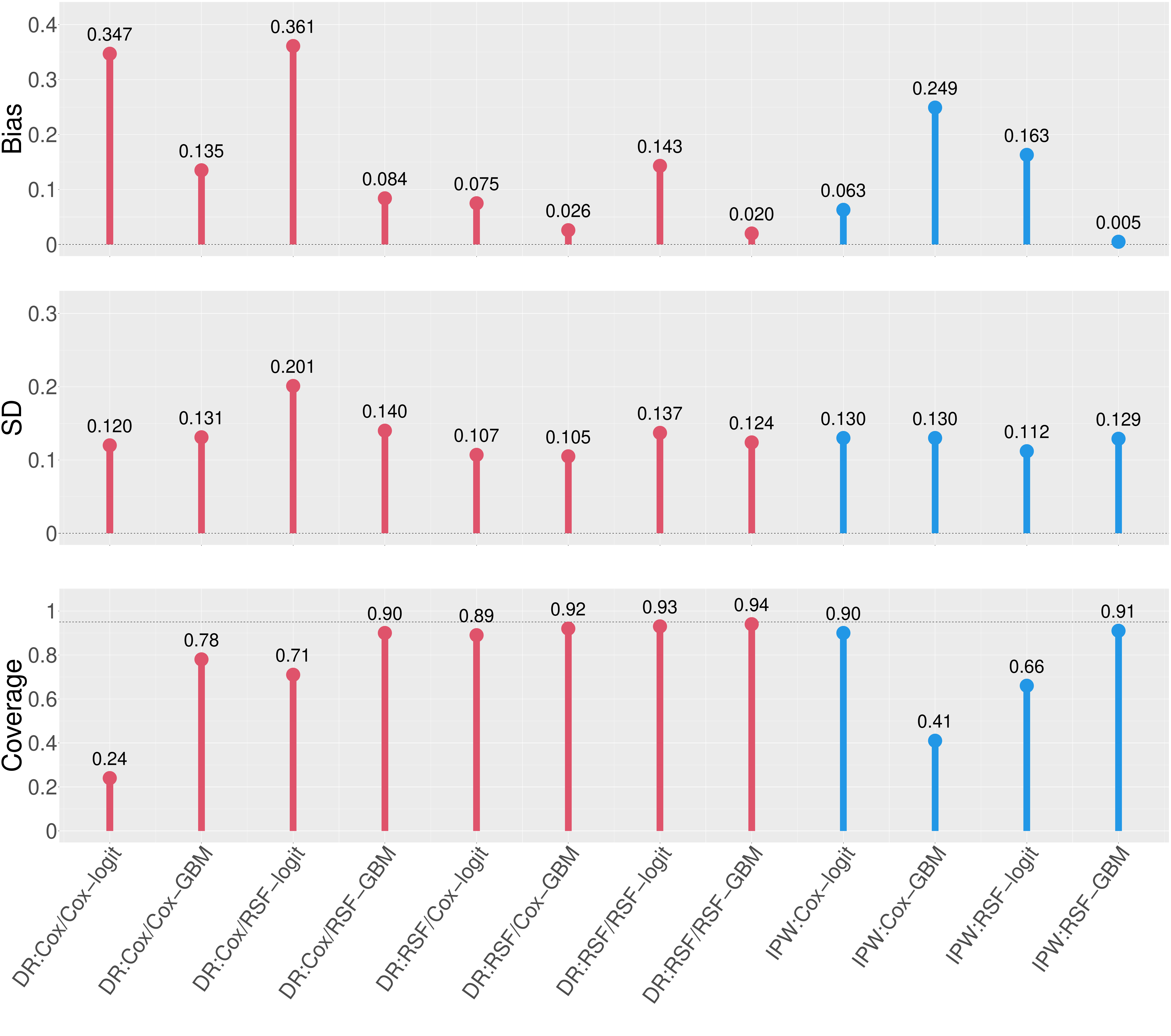} \\
\includegraphics[angle = 90,height = 92mm,  width=0.46\linewidth]{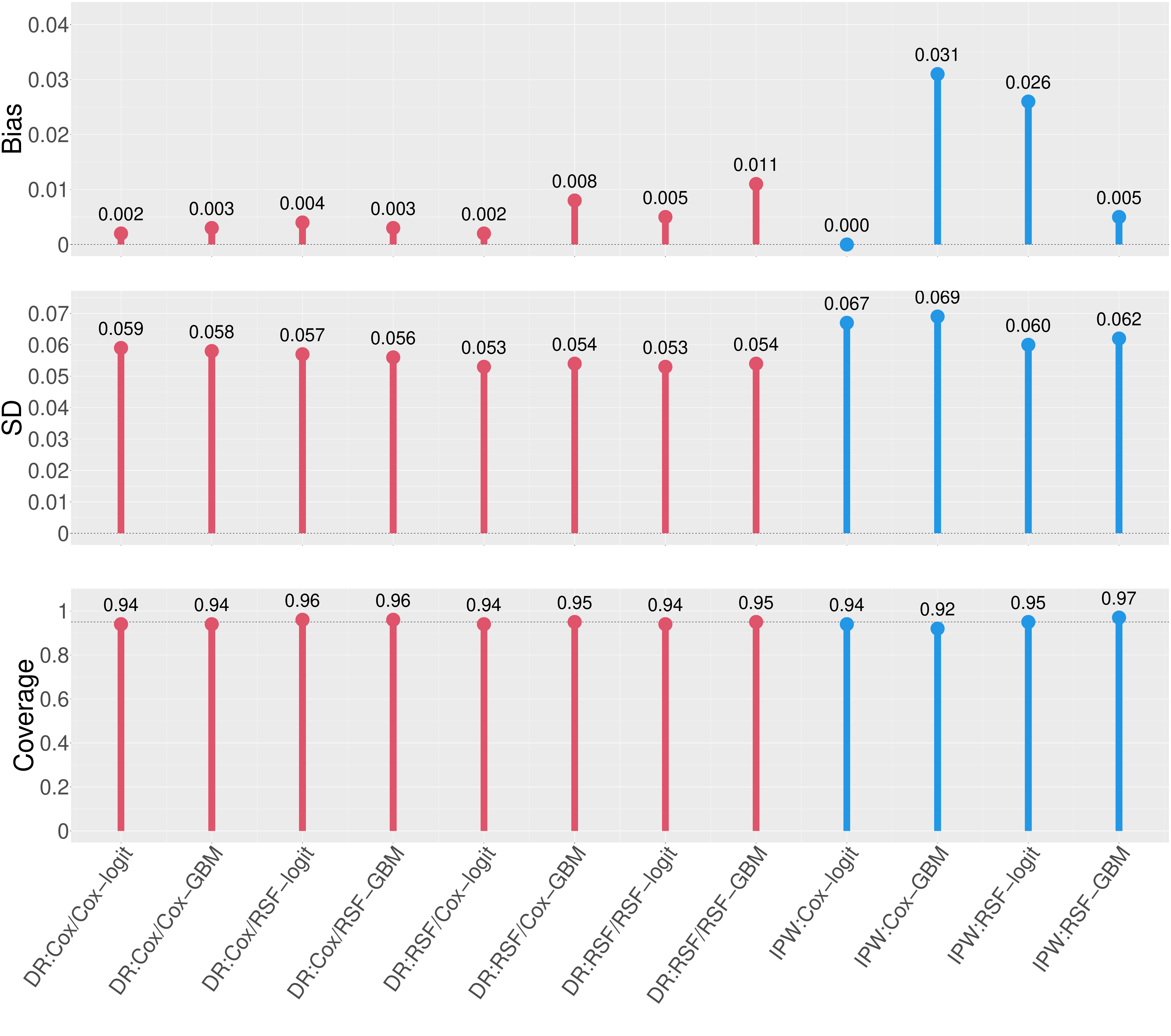} &   \includegraphics[angle = 90, height = 92mm, width=0.46\linewidth]{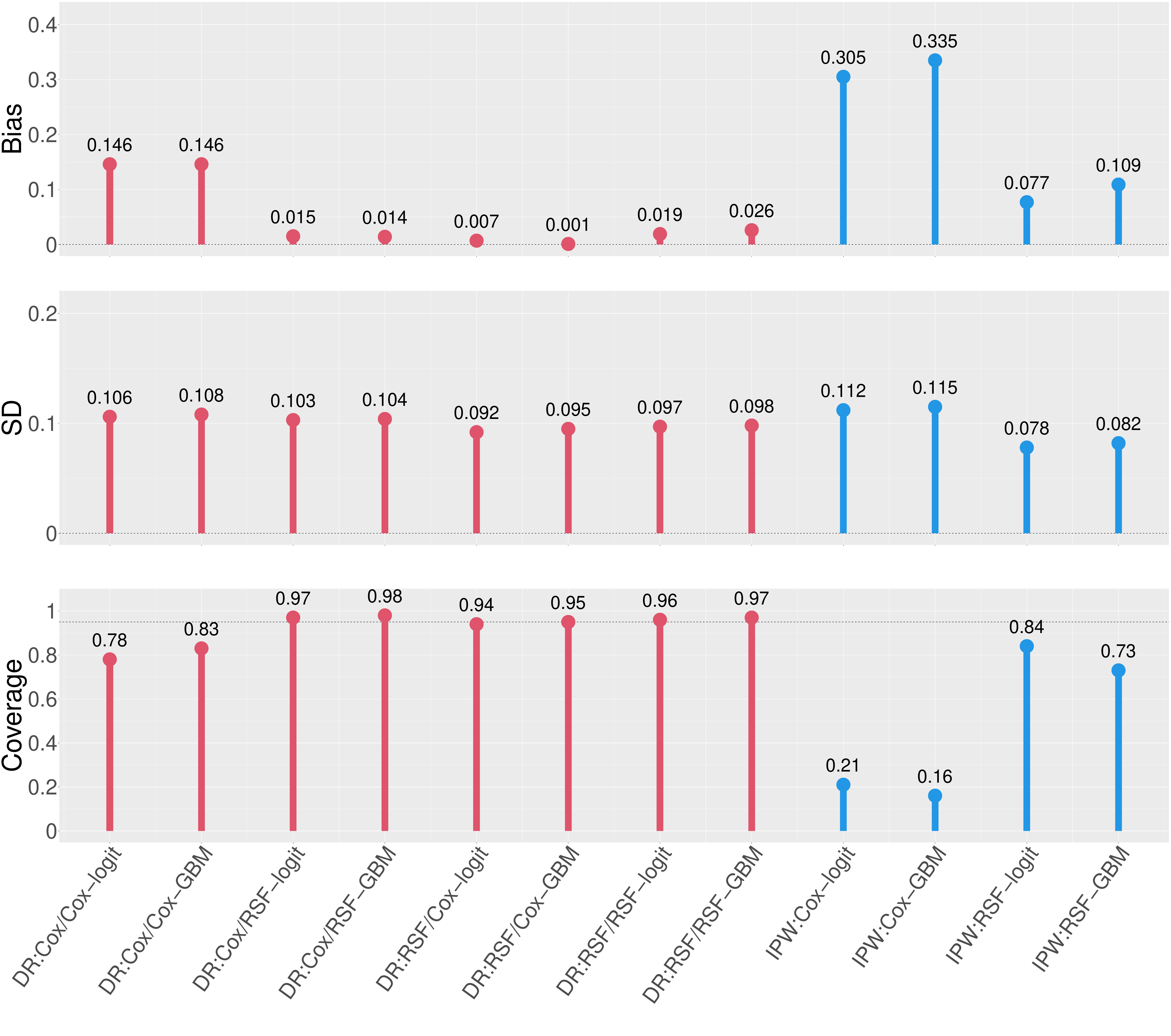} \\
\end{tabular}
\caption{Plots of bias, SD, and bootstrap coverage for each of the four Scenarios considered in the main paper Simulation under the marginal structural Cox model. Top-left, top-right, bottom-left, and bottom-right in landscape view correspond to Scenario 1 to Scenario 4, respectively.}  \label{AIPW:Scenarios.plot}
\end{center}
\end{figure}

\begin{table}[htbp] \small \begin{center} 
\caption{The (time-averaged) log hazard ratio estimated for the HAAS data as in Fig.~\ref{AIPW:haas_survival_forest} 
} \label{AIPW:haas_survival_estimates} 
\begin{tabular}{cllll} 
\multicolumn{1}{c}{Estimator} & \multicolumn{1}{c}{T/C-PS Models} & \multicolumn{1}{c}{Estimate} & \multicolumn{1}{c}{Boot SE} & \multicolumn{1}{c}{95$\%$ Boot CI} \\ 
\\ 
  \multirow{8}{*}{AIPW}  & Cox/Cox-logit & \multicolumn{1}{c}{0.27} & \multicolumn{1}{c}{0.07} & \multicolumn{1}{c}{(0.13, 0.40)} \\
& Cox/Cox-GBM & \multicolumn{1}{c}{0.25} & \multicolumn{1}{c}{0.07} & \multicolumn{1}{c}{(0.11, 0.38)} \\
& Cox/RSF-logit & \multicolumn{1}{c}{0.27} & \multicolumn{1}{c}{0.07} & \multicolumn{1}{c}{(0.13, 0.40)} \\
& Cox/RSF-GBM & \multicolumn{1}{c}{0.25} & \multicolumn{1}{c}{0.07} & \multicolumn{1}{c}{(0.12, 0.39)} \\
& RSF/Cox-logit & \multicolumn{1}{c}{0.23} & \multicolumn{1}{c}{0.07} & \multicolumn{1}{c}{(0.09, 0.36)} \\
& RSF/Cox-GBM & \multicolumn{1}{c}{0.22} & \multicolumn{1}{c}{0.07} & \multicolumn{1}{c}{(0.07, 0.36)} \\
& RSF/RSF-logit & \multicolumn{1}{c}{0.23} & \multicolumn{1}{c}{0.07} & \multicolumn{1}{c}{(0.09, 0.37)} \\
& RSF/RSF-GBM & \multicolumn{1}{c}{0.22} & \multicolumn{1}{c}{0.07} & \multicolumn{1}{c}{(0.08, 0.37)} \\
\\ 
\multirow{4}{*}{IPW}  & ~~~~~~~Cox-logit & \multicolumn{1}{c}{0.27} & \multicolumn{1}{c}{0.07} & \multicolumn{1}{c}{(0.14, 0.40)} \\
& ~~~~~~~Cox-GBM & \multicolumn{1}{c}{0.25} & \multicolumn{1}{c}{0.06} & \multicolumn{1}{c}{(0.13, 0.38)} \\
& ~~~~~~~RSF-logit & \multicolumn{1}{c}{0.27} & \multicolumn{1}{c}{0.07} & \multicolumn{1}{c}{(0.14, 0.40)} \\
& ~~~~~~~RSF-GBM & \multicolumn{1}{c}{0.25} & \multicolumn{1}{c}{0.06} & \multicolumn{1}{c}{(0.14, 0.37)} \\
\\ 
Naive Cox  &  & \multicolumn{1}{c}{0.30} & \multicolumn{1}{c}{0.07} & \multicolumn{1}{c}{(0.17, 0.44)} \\
\end{tabular} \end{center}  \end{table}

\begin{table}[htbp] \small \begin{center}  
\caption{Estimated risk difference and risk ratio for mortality at a given year since 1991 between the mid-life heavy and the not-heavy drinkers using the RSF/RSF-GBM estimator, in () are the bootstrapped $95\%$ confidence intervals.  
} \label{AIPW:HAAS_risk_contrasts_table} 
\begin{tabular}{rll} 
\multicolumn{1}{c}{Year} & \multicolumn{1}{c}{Risk Difference} & \multicolumn{1}{c}{Risk Ratio} \\ 
\\ 
3 & 0.001 (-0.0001,0.003) & 1.249 (1.070,1.429) \\
4 & 0.012 (0.004,0.021) & 1.242 (1.068,1.417) \\
5 & 0.022 (0.007,0.036) & 1.236 (1.067,1.404) \\
6 & 0.033 (0.011,0.055) & 1.227 (1.065,1.388) \\
7 & 0.044 (0.015,0.073) & 1.217 (1.063,1.371) \\
8 & 0.054 (0.018,0.089) & 1.207 (1.061,1.352) \\
9 & 0.062 (0.021,0.103) & 1.196 (1.059,1.334) \\
10 & 0.069 (0.024,0.115) & 1.186 (1.057,1.315) \\
11 & 0.074 (0.025,0.122) & 1.177 (1.055,1.298) \\
12 & 0.078 (0.027,0.129) & 1.166 (1.052,1.280) \\
\end{tabular} \end{center} \end{table}

\clearpage
\section{\texorpdfstring{Simulation under the $\beta(t)$ model}{Simulation under the b(t) model} }\label{TATE:sec:simulation}

In this section, similar to the simulation under the marginal structural Cox model \eqref{msm}, we simulated under the general $\beta(t)$ model \eqref{eq:betat}. Since the model is saturated, the non-collapsibility of the Cox model is no longer an issue, and it is possible that the conditional $T$ model given $A$ and $Z$ is correctly specified parametrically or semiparametrically. 
 
The data generation process is summarized in Fig.~\ref{TATE:dag}. 
We simulate 1000 datasets with $n=1000$, and 5-fold cross-fitting is used. 
We set $\tau = 1$ and first simulate  the covariate $Z \sim$
 Unif$(-1, 1)$. We then simulate $T(a)$, $C(a)$ and $ A$ given $Z$ under four different scenarios described in Table~\ref{TATE:table.dgp}. 
After simulating the potential outcomes and $A$, we obtain  $T = AT(1) + (1-A)T(0)$ and $C = AC(1) + (1-A)C(0)$.
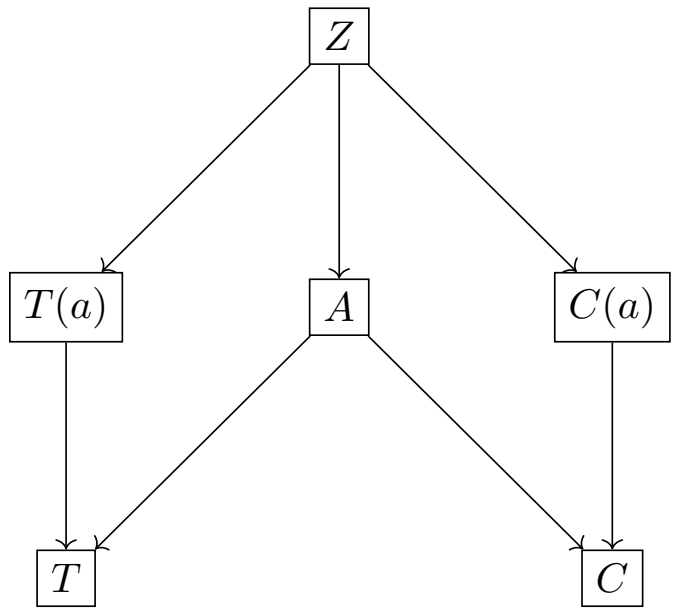
\begin{figure}[ht]
\centering
\resizebox{0.5\textwidth}{!}{
\begin{tikzpicture} 
    \node[draw] (v0) at (-2.3cm,0cm) {$T(a)$};
    \node[draw] (v1) at (2.3cm, 0cm) {$C(a)$};
    \node[draw] (v2) at (-0cm,0cm) {$A$};
    \node[draw] (v3) at (0cm,2.3cm) {$Z$};
    \node[draw] (v6) at (-2.3cm,-2.3cm) {$T$};
    \node[draw] (v7) at (2.3cm,-2.3cm) {$C$};
    \draw [->] (v3) edge (v1);
    \draw [->] (v3) edge (v0);
    \draw [->] (v3) edge (v2);
    \draw [->] (v2) edge (v6);
    \draw [->] (v2) edge (v7);
    \draw [->] (v0) edge (v6);
    \draw [->] (v1) edge (v7);
\end{tikzpicture}}
\caption{DAG for simulation} \label{TATE:dag}
\end{figure}

\begin{table}[ht]
\centering
\small
\caption{Data-generating mechanisms for $T(a)$, $C(s)$ and $A$  in the simulation} \label{TATE:table.dgp}
\setlength{\tabcolsep}{2pt}
\begin{tabular}{@{\extracolsep{5pt}} cll} 
\multicolumn{2}{c}{Scenario} 
 & Details 
 \\ 
\\
\multirow{2}{*}{1}& $T(a)$: Cox  & $\lambda_{T(a)}(t;Z) = \exp(2 - 1.12a - 2Z)$.\\  
& $C(a)$: Cox  & $\lambda_{C(a)}(t;Z) = \exp(3.5 - 2a - 2.5Z)$.  \\ 
& $A$: Logistic & $\logit \{\pi(Z) \} = 2Z$  \\
\\
\multirow{3}{*}{2}& $T(a)$: Cox  & $\lambda_{T(a)}(t;Z) = \exp(2 - 1.12a - 2Z)$. \\ 
& \multirow{2}{*}{$C(a)$: Mixture} & $Z \leq 0$: $\lambda_{C(a)}(t;Z) = \exp(3.5 - 3a - 0.5Z)$,  \\
&& $Z > 0$: $C(a) \sim$ Unif$(0, 1.05)$. \\ 
& $A$: Soft Partition & $\logit \{\pi(Z) \} = 2 \cdot \mathbf{1}\{Z < -1/3\} - 2 \cdot \mathbf{1}\{-1/3 \leq Z < 1/3\}$ \\
&&\quad\quad\quad\quad\quad\quad+ $2 \cdot \mathbf{1}\{Z \geq 1/3\}$  \\
\\
\multirow{3}{*}{3}& \multirow{2}{*}{$T(a)$: Mixture} & $Z \leq 0$: $\lambda_{T(a)}(t;Z) = \exp(5 - 3.4a + 2.5Z)$,  \\
&& $Z > 0$: $T \sim$ Unif$(0, 1.05)$. \\ 
& $C(a)$: Cox  & $\lambda_{C(a)}(t;Z) = \exp(3.5 - 2a - 2.5Z)$. \\ 
& $A$: Logistic & $\logit \{\pi(Z) \} = 2Z$  \\
\\
\multirow{4}{*}{4}& \multirow{2}{*}{$T(a)$: Mixture} & $Z \leq 0$: $\lambda_{T(a)}(t;Z) = \exp(5 - 3.4a + 2.5Z)$,  \\
&& $Z > 0$: $T(a) \sim$ Unif$(0, 1.05)$. \\ 
& \multirow{2}{*}{$C(a)$: Mixture} & $Z \leq 0$: $\lambda_{C(a)}(t;Z) = \exp(3.5 - 3a - 0.5Z)$,  \\
&& $Z > 0$: $C(a) \sim$ Unif$(0, 1.05)$. \\ 
& $A$: Soft Partition & $\logit \{\pi(Z) \} = 2 \cdot \mathbf{1}\{Z < -1/3\} - 2 \cdot \mathbf{1}\{-1/3 \leq Z < 1/3\}$ \\
&&\quad\quad\quad\quad\quad\quad + $2 \cdot \mathbf{1}\{Z \geq 1/3\}$  \\
\end{tabular}
\end{table}

\begin{figure}[ht]
\begin{center}
\includegraphics[width=.41\linewidth]{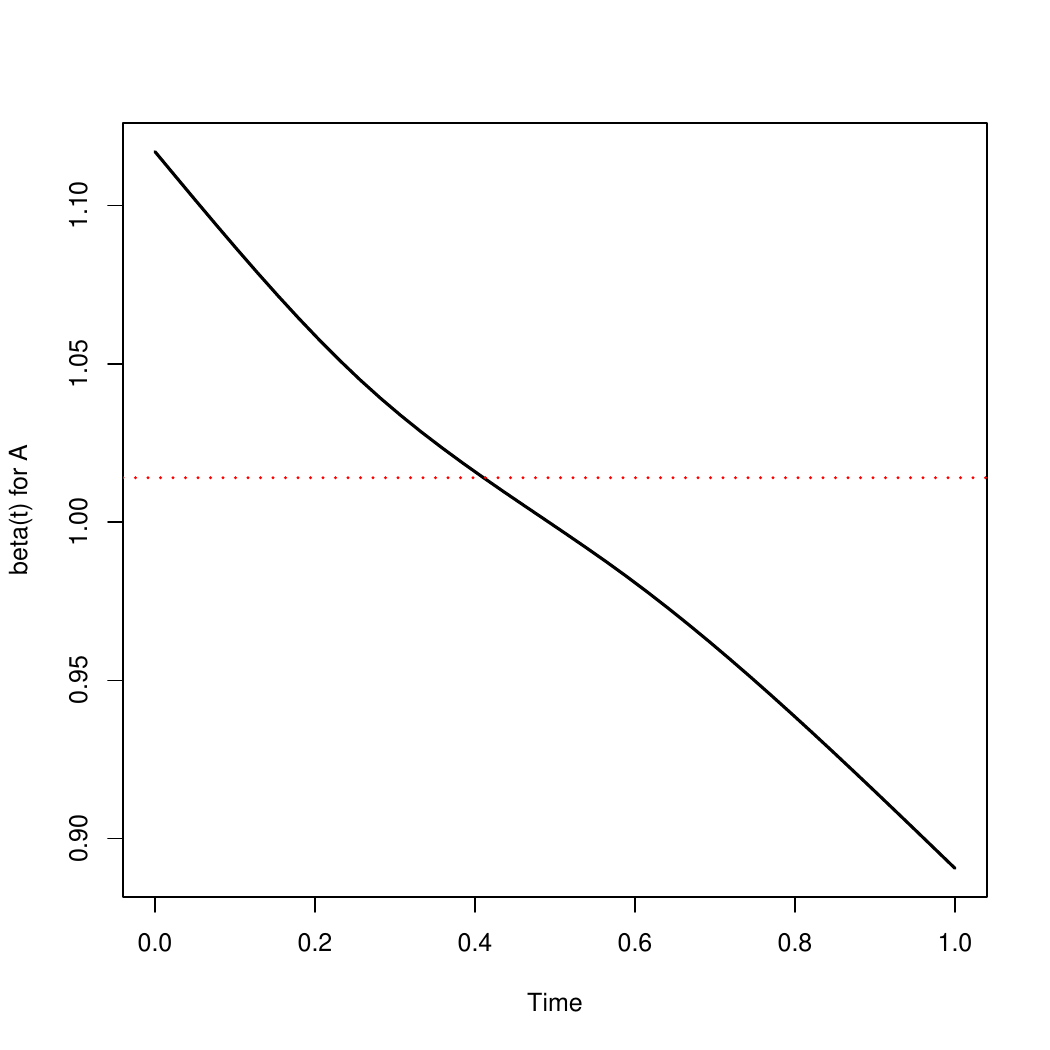}\\
\includegraphics[width=.41\linewidth]{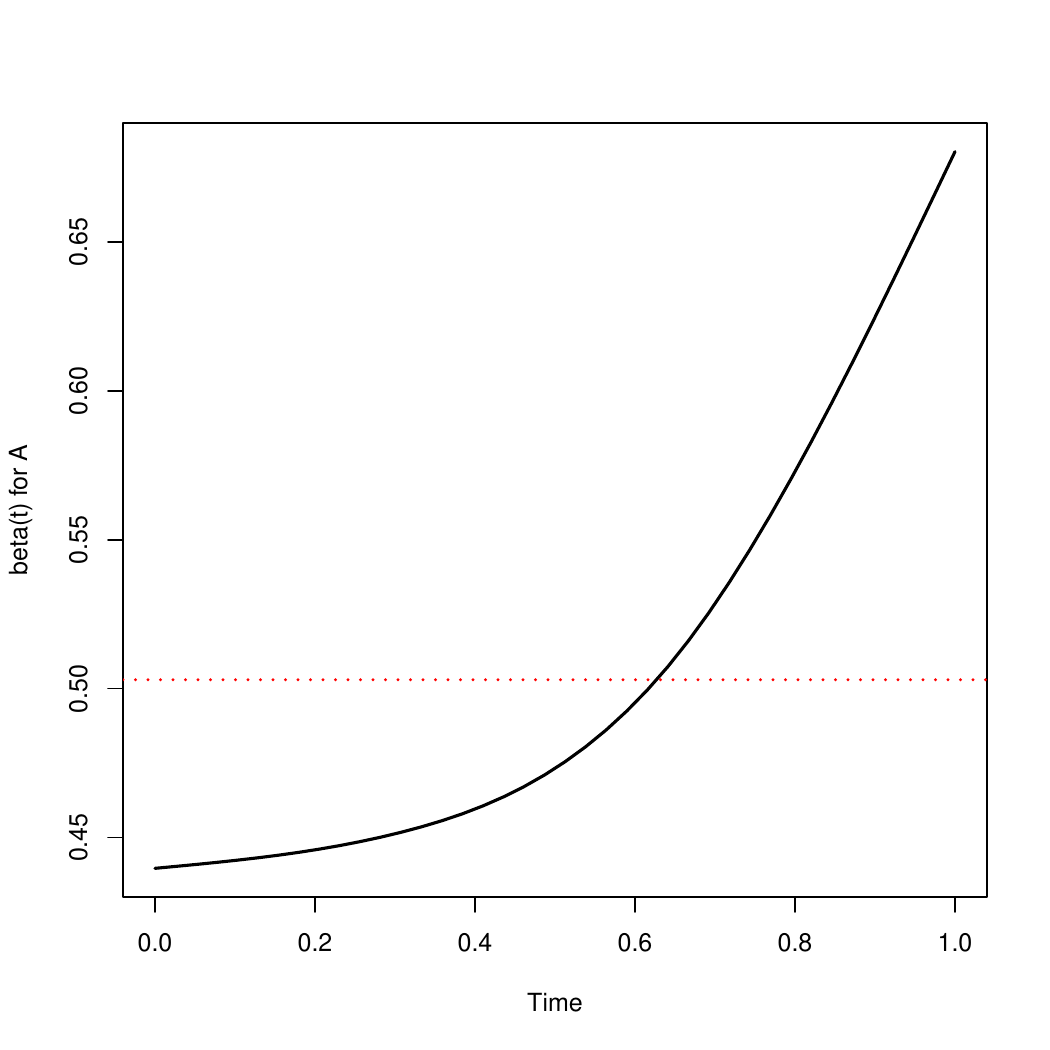}
\caption{True $\beta(t)$ plot for Scenarios 1 and 2 (top), and 3 and 4 (bottom). 
True $\beta^*$ is also shown in \textcolor{red}{red}.}  \label{TATE:sim.beta.t.plot}
\end{center}
\end{figure}

All four scenarios have an event rate between $30 - 50\%$,  censoring rate between$30 - 50\%$ and administrate censoring rate ($X > \tau$) between $10 - 30\%$. 
True $\beta^*=1.014$ and $0.503$ when $T(a)$ follows the Cox model and the Mixture setting, respectively. 
Since there is no analytical solution for $\beta^*$, the true value is calculated using a simulated sample of one million full data points. 
Fig.~\ref{TATE:sim.beta.t.plot} shows a smoothed plot of how $\beta(t)$ changes over time under the Mixture setting for $T(a)$, 
created using the 'cox.zph' function from the 'survival' package applied to the simulated full data. 

We consider the same estimators as in the main paper, with the exception that 
the Cox/Cox-logit estimator is not cross-fitted, and is discussed in more details below.

Tables~\ref{TATE:sim.table.1} and \ref{TATE:sim.table.2} show the bias, standard deviation (SD), and bootstrap-based coverage probability of the  estimators under Scenarios 1 -- 4 respectively. Additional visualization are plotted in Figures \ref{TATE:four.scenarios.plots}. 
We see that the Cox/Cox-logit estimator performs well as long as one of the working models is correctly specified (Scenarios 1, 2, 3).  This is the classical  model doubly robust behavior, with (semi)parametric working models. 
There is no theoretical guarantee for the AIPW estimators when some of the (semi)parametric working models is wrong and machine learning methods are used for the rest of the working models; in Scenario 3 for example, there appears to be slight over coverage of the confidence intervals. 
The RSF/RSF-GBM estimator continues to perform well under all four scenarios.
 along with excellent model coverage and bootstrap coverage. 
 
The IPW Cox-logit estimator performs well under Scenarios 1 and 3 when both the $C$ and $A$ models are correctly specified,  but performs poorly otherwise (Scenarios 2 and 4). 
The other IPW estimators with machine learning has no theoretical guarantees and performs less satisfactorily even when no model is `wrong'. 

Finally, it's worth noting that in all four scenarios, the naive Cox estimator lies outside the whole range of  $\beta(t)$ (Fig.~\ref{TATE:sim.beta.t.plot}). This is different from the regression setting considered in \cite{XO2000}, where 
neither confounding nor informative censoring was present.

\begin{table}[ht] \begin{center} 
\caption{Simulation based on 1000 data sets for Scenarios 1 and 2, each with 1000 observations. True $\beta^* = 1.014$. \textcolor{red}{Red} indicates that the working model or the approach is invalid.}   \label{TATE:sim.table.1} 
\renewcommand{\arraystretch}{0.82} 
\begin{tabular}{cclllll} 
\multicolumn{1}{c}{Scenario} & \multicolumn{1}{c}{Estimator} & \multicolumn{1}{c}{T/C-PS Models}  & \multicolumn{1}{c}{Bias} & \multicolumn{1}{c}{SD} & \multicolumn{1}{c}{SE} & \multicolumn{1}{c}{Coverage}\\  &&&&&Model/Boot& Model/Boot\\
\\
{1}  & \multirow{8}{*}{AIPW}    &  Cox/Cox-logit & \multicolumn{1}{c}{~0.002~} & \multicolumn{1}{c}{~0.151~} & \multicolumn{1}{c}{0.151/0.151} & \multicolumn{1}{c}{0.95/0.95} \\  
 &   &  Cox/Cox-GBM & \multicolumn{1}{c}{~0.002~} & \multicolumn{1}{c}{~0.157~} & \multicolumn{1}{c}{0.164/0.168} & \multicolumn{1}{c}{0.95/0.96} \\  
 &   &  Cox/RSF-logit & \multicolumn{1}{c}{~0.001~} & \multicolumn{1}{c}{~0.151~} & \multicolumn{1}{c}{0.154/0.157} & \multicolumn{1}{c}{0.95/0.95} \\  
 &   &  Cox/RSF-GBM & \multicolumn{1}{c}{~0.003~} & \multicolumn{1}{c}{~0.156~} & \multicolumn{1}{c}{0.165/0.167} & \multicolumn{1}{c}{0.96/0.96} \\  
 &   &  RSF/Cox-logit & \multicolumn{1}{c}{~0.000~} & \multicolumn{1}{c}{~0.154~} & \multicolumn{1}{c}{0.155/0.159} & \multicolumn{1}{c}{0.95/0.96} \\  
 &   &  RSF/Cox-GBM & \multicolumn{1}{c}{~0.000~} & \multicolumn{1}{c}{~0.156~} & \multicolumn{1}{c}{0.166/0.175} & \multicolumn{1}{c}{0.96/0.97} \\  
 &   &  RSF/RSF-logit & \multicolumn{1}{c}{~0.004~} & \multicolumn{1}{c}{~0.153~} & \multicolumn{1}{c}{0.156/0.159} & \multicolumn{1}{c}{0.95/0.95} \\  
 &   &  RSF/RSF-GBM & \multicolumn{1}{c}{~0.004~} & \multicolumn{1}{c}{~0.155~} & \multicolumn{1}{c}{0.167/0.175} & \multicolumn{1}{c}{0.96/0.97} \\ 
 \\
 &  \multirow{4}{*}{IPW} &   ~~~~~~~~Cox-logit & \multicolumn{1}{c}{~0.002~} & \multicolumn{1}{c}{~0.155~} & \multicolumn{1}{c}{~~~-~~~~/0.153} & \multicolumn{1}{c}{~~~-~~/0.95} \\  
 &   &  ~~~~~~~~Cox-GBM & \multicolumn{1}{c}{~0.040~} & \multicolumn{1}{c}{~0.155~} & \multicolumn{1}{c}{~~~-~~~~/0.144} & \multicolumn{1}{c}{~~~-~~/0.92} \\  
 &   &  ~~~~~~~~RSF-logit & \multicolumn{1}{c}{~0.001~} & \multicolumn{1}{c}{~0.154~} & \multicolumn{1}{c}{~~~-~~~~/0.153} & \multicolumn{1}{c}{~~~-~~/0.94} \\  
 &   &  ~~~~~~~~RSF-GBM & \multicolumn{1}{c}{~0.039~} & \multicolumn{1}{c}{~0.154~} & \multicolumn{1}{c}{~~~-~~~~/0.144} & \multicolumn{1}{c}{~~~-~~/0.92} \\ 
 \\
 &  \red{Naive Cox} &  & \multicolumn{1}{c}{~0.470~} & \multicolumn{1}{c}{~0.152~} & \multicolumn{1}{c}{0.151/0.151} & \multicolumn{1}{c}{0.11/0.11} \\  
 &  Full Data &  & \multicolumn{1}{c}{~0.001~} & \multicolumn{1}{c}{~0.061~} & \multicolumn{1}{c}{0.063/0.063} & \multicolumn{1}{c}{0.96/0.96} \\  
\\
{2}  & \multirow{8}{*}{AIPW}    &  Cox/\red{Cox}-\red{logit} & \multicolumn{1}{c}{~0.009~} & \multicolumn{1}{c}{~0.188~} & \multicolumn{1}{c}{0.175/0.189} & \multicolumn{1}{c}{0.94/0.95} \\  
 &   &  Cox/\red{Cox}-GBM & \multicolumn{1}{c}{~0.023~} & \multicolumn{1}{c}{~0.338~} & \multicolumn{1}{c}{0.314/0.338} & \multicolumn{1}{c}{0.94/0.96} \\  
 &   &  Cox/RSF-\red{logit} & \multicolumn{1}{c}{~0.008~} & \multicolumn{1}{c}{~0.205~} & \multicolumn{1}{c}{0.189/0.209} & \multicolumn{1}{c}{0.93/0.96} \\  
 &   &  Cox/RSF-GBM & \multicolumn{1}{c}{~0.017~} & \multicolumn{1}{c}{~0.320~} & \multicolumn{1}{c}{0.308/0.318} & \multicolumn{1}{c}{0.96/0.95} \\  
 &   &  RSF/\red{Cox}-\red{logit} & \multicolumn{1}{c}{~0.145~} & \multicolumn{1}{c}{~0.250~} & \multicolumn{1}{c}{0.176/0.208} & \multicolumn{1}{c}{0.74/0.81} \\  
 &   &  RSF/\red{Cox}-GBM & \multicolumn{1}{c}{~0.028~} & \multicolumn{1}{c}{~0.366~} & \multicolumn{1}{c}{0.337/0.373} & \multicolumn{1}{c}{0.91/0.94} \\  
 &   &  RSF/RSF-\red{logit} & \multicolumn{1}{c}{~0.152~} & \multicolumn{1}{c}{~0.258~} & \multicolumn{1}{c}{0.189/0.219} & \multicolumn{1}{c}{0.76/0.83} \\  
 &   &  RSF/RSF-GBM & \multicolumn{1}{c}{~0.040~} & \multicolumn{1}{c}{~0.365~} & \multicolumn{1}{c}{0.352/0.390} & \multicolumn{1}{c}{0.94/0.95} \\ 
 \\
 &  \multirow{4}{*}{IPW} &   ~~~~~~~~\red{Cox}-\red{logit} & \multicolumn{1}{c}{~0.494~} & \multicolumn{1}{c}{~0.181~} & \multicolumn{1}{c}{~~~-~~~~/0.180} & \multicolumn{1}{c}{~~~-~~/0.20} \\  
 &   &  ~~~~~~~~\red{Cox}-GBM & \multicolumn{1}{c}{~0.170~} & \multicolumn{1}{c}{~0.302~} & \multicolumn{1}{c}{~~~-~~~~/0.227} & \multicolumn{1}{c}{~~~-~~/0.79} \\  
 &   &  ~~~~~~~~RSF-\red{logit} & \multicolumn{1}{c}{~0.269~} & \multicolumn{1}{c}{~0.184~} & \multicolumn{1}{c}{~~~-~~~~/0.181} & \multicolumn{1}{c}{~~~-~~/0.67} \\  
 &   &  ~~~~~~~~RSF-GBM & \multicolumn{1}{c}{~0.052~} & \multicolumn{1}{c}{~0.280~} & \multicolumn{1}{c}{~~~-~~~~/0.223} & \multicolumn{1}{c}{~~~-~~/0.89} \\ 
 \\
 &  \red{Naive Cox} &  & \multicolumn{1}{c}{~0.245~} & \multicolumn{1}{c}{~0.164~} & \multicolumn{1}{c}{0.167/0.166} & \multicolumn{1}{c}{0.71/0.70} \\  
 &  Full Data &  & \multicolumn{1}{c}{~0.001~} & \multicolumn{1}{c}{~0.061~} & \multicolumn{1}{c}{0.063/0.063} & \multicolumn{1}{c}{0.96/0.96} \\  
\end{tabular} \end{center} \end{table}

\begin{table}[ht] \begin{center} 
\caption{Simulation based on 1000 data sets for Scenarios 3 and 4, each with 1000 observations. True $\beta^* = 0.503$. \textcolor{red}{Red} indicates that the working model or the approach is invalid.} \label{TATE:sim.table.2}  \renewcommand{\arraystretch}{0.82} 
\begin{tabular}{cclllll} 
\multicolumn{1}{c}{Scenario} & \multicolumn{1}{c}{Estimator} & \multicolumn{1}{c}{T/C-PS Models}  & \multicolumn{1}{c}{Bias} & \multicolumn{1}{c}{SD} & \multicolumn{1}{c}{SE} & \multicolumn{1}{c}{Coverage}\\  &&&&&Model/Boot& Model/Boot\\
\\
{3}  & \multirow{8}{*}{AIPW}    &  \red{Cox}/Cox-logit & \multicolumn{1}{c}{~0.001~} & \multicolumn{1}{c}{~0.082~} & \multicolumn{1}{c}{0.086/0.083} & \multicolumn{1}{c}{0.96/0.96} \\  
 &   &  \red{Cox}/Cox-GBM & \multicolumn{1}{c}{~0.012~} & \multicolumn{1}{c}{~0.077~} & \multicolumn{1}{c}{0.094/0.088} & \multicolumn{1}{c}{0.98/0.98} \\  
 &   &  \red{Cox}/RSF-logit & \multicolumn{1}{c}{~0.005~} & \multicolumn{1}{c}{~0.085~} & \multicolumn{1}{c}{0.089/0.091} & \multicolumn{1}{c}{0.96/0.96} \\  
 &   &  \red{Cox}/RSF-GBM & \multicolumn{1}{c}{~0.010~} & \multicolumn{1}{c}{~0.078~} & \multicolumn{1}{c}{0.095/0.089} & \multicolumn{1}{c}{0.98/0.97} \\  
 &   &  RSF/Cox-logit & \multicolumn{1}{c}{~0.004~} & \multicolumn{1}{c}{~0.071~} & \multicolumn{1}{c}{0.072/0.074} & \multicolumn{1}{c}{0.95/0.95} \\  
 &   &  RSF/Cox-GBM & \multicolumn{1}{c}{~0.010~} & \multicolumn{1}{c}{~0.073~} & \multicolumn{1}{c}{0.077/0.081} & \multicolumn{1}{c}{0.96/0.96} \\  
 &   &  RSF/RSF-logit & \multicolumn{1}{c}{~0.007~} & \multicolumn{1}{c}{~0.072~} & \multicolumn{1}{c}{0.074/0.076} & \multicolumn{1}{c}{0.95/0.96} \\  
 &   &  RSF/RSF-GBM & \multicolumn{1}{c}{~0.013~} & \multicolumn{1}{c}{~0.075~} & \multicolumn{1}{c}{0.079/0.083} & \multicolumn{1}{c}{0.96/0.97} \\ 
 \\
 &  \multirow{4}{*}{IPW} &   ~~~~~~~~Cox-logit & \multicolumn{1}{c}{~0.001~} & \multicolumn{1}{c}{~0.081~} & \multicolumn{1}{c}{~~~-~~~~/0.082} & \multicolumn{1}{c}{~~~-~~/0.95} \\  
 &   &  ~~~~~~~~Cox-GBM & \multicolumn{1}{c}{~0.021~} & \multicolumn{1}{c}{~0.074~} & \multicolumn{1}{c}{~~~-~~~~/0.070} & \multicolumn{1}{c}{~~~-~~/0.93} \\  
 &   &  ~~~~~~~~RSF-logit & \multicolumn{1}{c}{~0.022~} & \multicolumn{1}{c}{~0.088~} & \multicolumn{1}{c}{~~~-~~~~/0.085} & \multicolumn{1}{c}{~~~-~~/0.93} \\  
 &   &  ~~~~~~~~RSF-GBM & \multicolumn{1}{c}{~0.001~} & \multicolumn{1}{c}{~0.081~} & \multicolumn{1}{c}{~~~-~~~~/0.073} & \multicolumn{1}{c}{~~~-~~/0.93} \\ 
 \\
 &  \red{Naive Cox} &  & \multicolumn{1}{c}{~0.518~} & \multicolumn{1}{c}{~0.097~} & \multicolumn{1}{c}{0.099/0.099} & \multicolumn{1}{c}{0.00/0.00} \\  
 &  Full Data &  & \multicolumn{1}{c}{~0.001~} & \multicolumn{1}{c}{~0.035~} & \multicolumn{1}{c}{0.034/0.034} & \multicolumn{1}{c}{0.95/0.94} \\  
\\
{4}  & \multirow{8}{*}{AIPW}    &  \red{Cox}/\red{Cox}-\red{logit} & \multicolumn{1}{c}{~0.469~} & \multicolumn{1}{c}{~0.125~} & \multicolumn{1}{c}{0.110/0.117} & \multicolumn{1}{c}{0.02/0.03} \\  
 &   &  \red{Cox}/\red{Cox}-GBM & \multicolumn{1}{c}{~0.224~} & \multicolumn{1}{c}{~0.157~} & \multicolumn{1}{c}{0.154/0.153} & \multicolumn{1}{c}{0.70/0.70} \\  
 &   &  \red{Cox}/RSF-\red{logit} & \multicolumn{1}{c}{~0.180~} & \multicolumn{1}{c}{~0.138~} & \multicolumn{1}{c}{0.180/0.183} & \multicolumn{1}{c}{0.83/0.92} \\  
 &   &  \red{Cox}/RSF-GBM & \multicolumn{1}{c}{~0.010~} & \multicolumn{1}{c}{~0.210~} & \multicolumn{1}{c}{0.276/0.218} & \multicolumn{1}{c}{0.97/0.97} \\  
 &   &  RSF/\red{Cox}-\red{logit} & \multicolumn{1}{c}{~0.044~} & \multicolumn{1}{c}{~0.085~} & \multicolumn{1}{c}{0.077/0.085} & \multicolumn{1}{c}{0.88/0.92} \\  
 &   &  RSF/\red{Cox}-GBM & \multicolumn{1}{c}{~0.010~} & \multicolumn{1}{c}{~0.113~} & \multicolumn{1}{c}{0.104/0.112} & \multicolumn{1}{c}{0.94/0.95} \\  
 &   &  RSF/RSF-\red{logit} & \multicolumn{1}{c}{~0.036~} & \multicolumn{1}{c}{~0.116~} & \multicolumn{1}{c}{0.116/0.130} & \multicolumn{1}{c}{0.94/0.97} \\  
 &   &  RSF/RSF-GBM & \multicolumn{1}{c}{~0.008~} & \multicolumn{1}{c}{~0.156~} & \multicolumn{1}{c}{0.160/0.177} & \multicolumn{1}{c}{0.94/0.97} \\ 
 \\
 &  \multirow{4}{*}{IPW} &   ~~~~~~~~\red{Cox}-\red{logit} & \multicolumn{1}{c}{~0.592~} & \multicolumn{1}{c}{~0.119~} & \multicolumn{1}{c}{~~~-~~~~/0.112} & \multicolumn{1}{c}{~~~-~~/0.00} \\  
 &   &  ~~~~~~~~\red{Cox}-GBM & \multicolumn{1}{c}{~0.308~} & \multicolumn{1}{c}{~0.159~} & \multicolumn{1}{c}{~~~-~~~~/0.131} & \multicolumn{1}{c}{~~~-~~/0.37} \\  
 &   &  ~~~~~~~~RSF-\red{logit} & \multicolumn{1}{c}{~0.218~} & \multicolumn{1}{c}{~0.105~} & \multicolumn{1}{c}{~~~-~~~~/0.105} & \multicolumn{1}{c}{~~~-~~/0.44} \\  
 &   &  ~~~~~~~~RSF-GBM & \multicolumn{1}{c}{~0.047~} & \multicolumn{1}{c}{~0.152~} & \multicolumn{1}{c}{~~~-~~~~/0.127} & \multicolumn{1}{c}{~~~-~~/0.89} \\ 
 \\
 &  \red{Naive Cox} &  & \multicolumn{1}{c}{~0.431~} & \multicolumn{1}{c}{~0.117~} & \multicolumn{1}{c}{0.111/0.112} & \multicolumn{1}{c}{0.03/0.03} \\  
 &  Full Data &  & \multicolumn{1}{c}{~0.001~} & \multicolumn{1}{c}{~0.035~} & \multicolumn{1}{c}{0.035/0.035} & \multicolumn{1}{c}{0.94/0.94} \\  
\end{tabular} \end{center}  \end{table}

\begin{figure}[htbp]
\begin{center}
\begin{tabular}{cc}
\includegraphics[angle = 90, height = 92mm, width=0.47\linewidth]{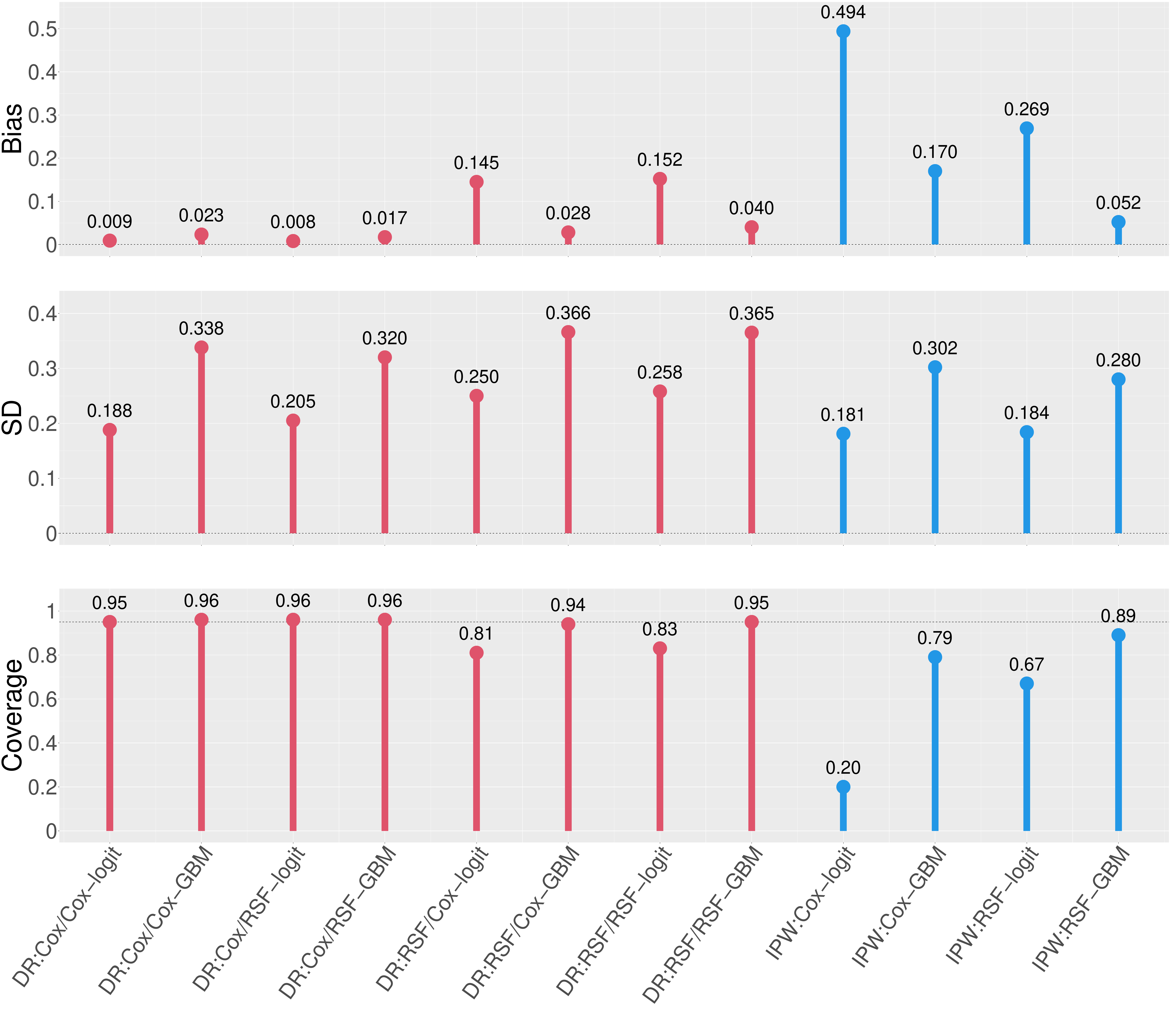} &   \includegraphics[angle = 90, height = 92mm, width=0.47\linewidth]{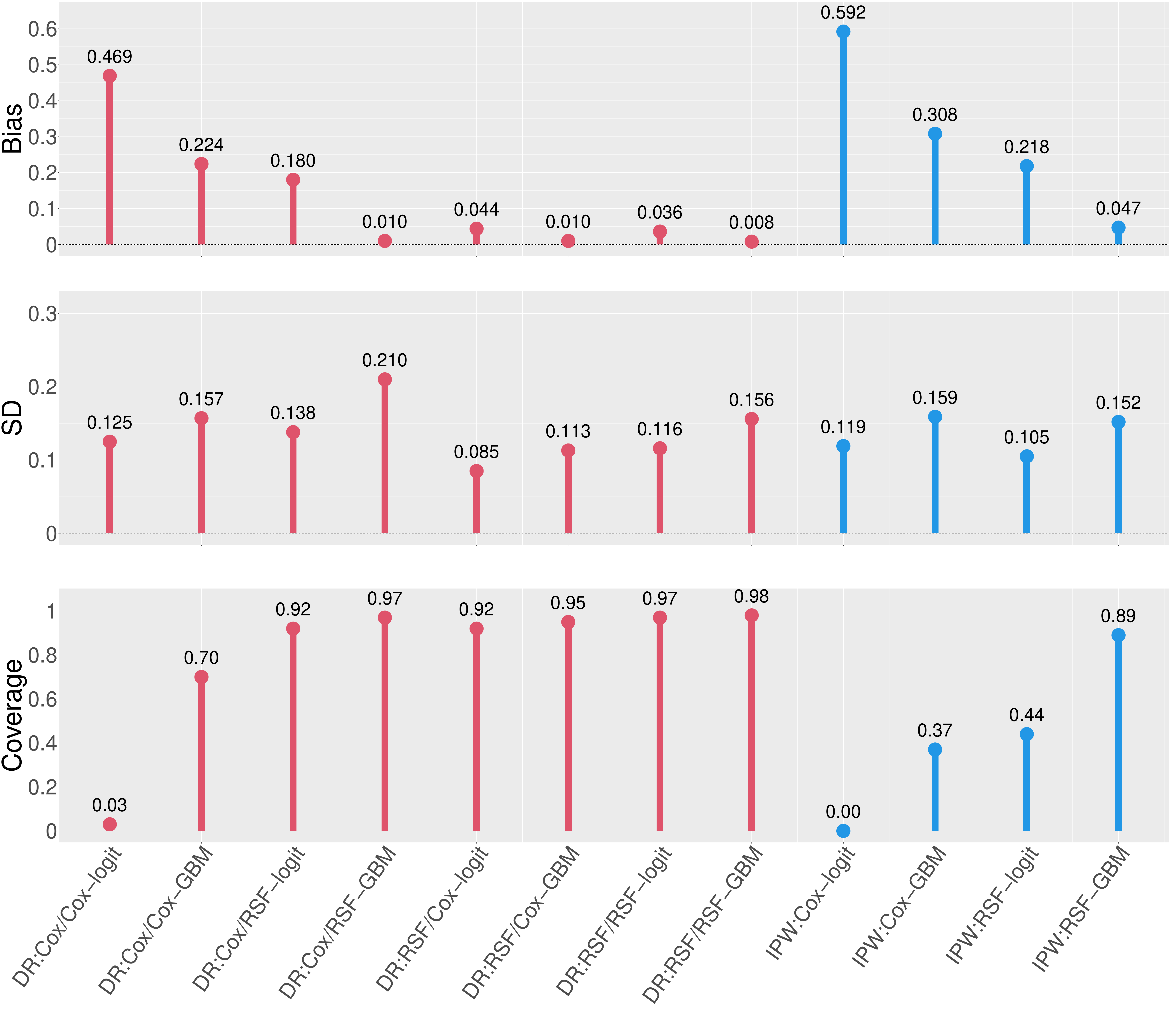} \\
\includegraphics[angle = 90,height = 92mm,  width=0.47\linewidth]{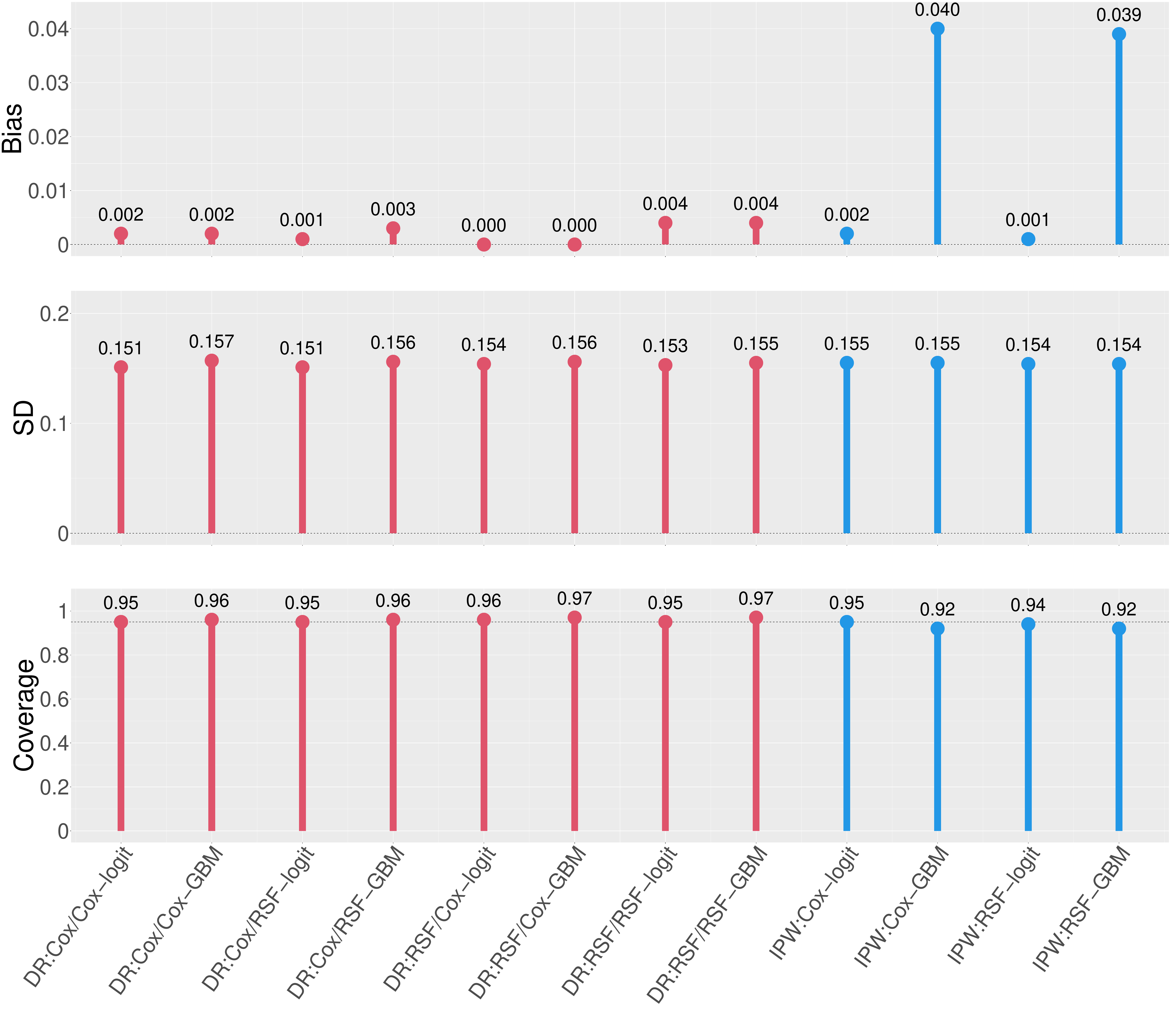} &   \includegraphics[angle = 90, height = 92mm, width=0.47\linewidth]{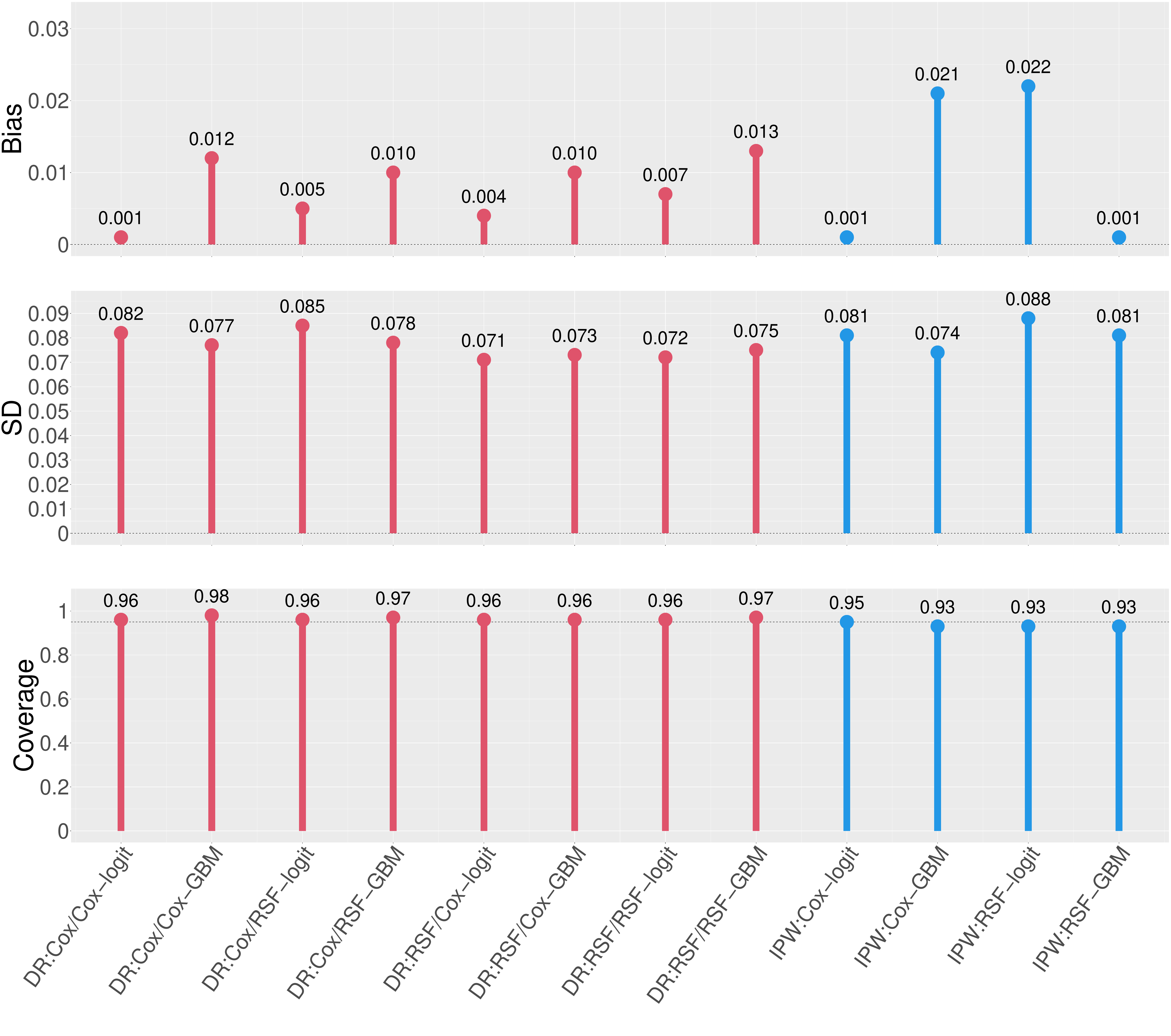} \\
\end{tabular}
\caption{Plots of bias, bootstrap SD and bootstrap coverage for all four scenarios under the $\beta(t)$ simulation. Top-left, top-right, bottom-left, and bottom-right in the landscape view correspond to Scenario 1 to Scenario 4, respectively.}  \label{TATE:four.scenarios.plots}
\end{center}
\end{figure}

\clearpage
\section{Proof of Asymptotic Results}

\subsection{Additional Assumptions} \label{appendix:assumption}

We remind the reader that the notations below can be found at the start of the Supplementary Material. 
We assume that the nuisance function estimates $\wh{\pi}, \wh{S}, \wh{S}_c$ and their limits $\pi^*, S^*, S_c^*$ only take values in $[0,1]$. 
In addition, $\wh{S}$ and $ \wh{S}_c$ are non-increasing in $t$.
\begin{assumption}
There exists a neighbourhood $\mathcal{B}$ of $\beta^*$ such that $\sup_{t \in [0,\tau], \beta \in \mathcal{B}} |\mathcal{S}^{(l)}(t;\beta,\pi^*, S^*, S_c^*) - \mathpzc{s}^{(l)}(t;\beta,\pi^*, S^*, S^*_c)|$ $= o_p(1)$. 
\label{assumpA1}
\end{assumption}

\begin{assumption}  \label{assumpA2}
    For $l = 0,1$, $\mathpzc{s}^{(l)}(t;\beta,\pi^*, S^*, S_c^*)$ are continuous functions of $\beta \in \mathcal{B}$, uniformly in $t \in [0,\tau]$  and are bounded on $\mathcal{B} \times [0,\tau]$. $\mathpzc{s}^{(0)}(t;\beta,\pi^*, S^*, S_c^*)$ is bounded away from zero on $\mathcal{B} \times [0,\tau]$.
    For all $\beta \in \mathcal{B}$, $t \in [0,\tau]$:
    \ba{
    \mathpzc{s}^{(1)}(t;\beta,\pi, S, S_c) = \frac{\partial}{\partial \beta} \mathpzc{s}^{(0)}(t;\beta,\pi, S, S_c) = \frac{\partial ^2}{\partial \beta^2} \mathpzc{s}^{(0)}(t;\beta,\pi, S, S_c) .   
    }
    In addition, let $\bar{\alpha} = \mathpzc{s}^{(1)}/ \mathpzc{s}^{(0)}$ and $v = \bar{\alpha} - \bar{\alpha}^2$. We have
    \ba{
    \nu(\beta^*, \pi^*, S^*, S_c^*) = \int_0^\tau v(t;\beta^*,\pi^*, S^*, S_c^*)\mathpzc{s}^{(0)}(t;\beta^*,\pi^*, S^*, S_c^*) d\Lambda^*(t) > 0.
    }
\end{assumption}

Assumptions \ref{assumpA1} and \ref{assumpA2} are the typical regularity assumptions that are made under the Cox type models similar to those in  \cite{andersen1982cox}.

\begin{assumption} \label{assumpA3}\
There exist unique zeros to both the estimating equation $U_{cf}(\beta) = 0$ and the equation $\mu(\beta) = 0$.
\end{assumption}

This assumption is needed for showing the consistency of $\whb$ since our estimating function $U_{cf}(\beta)$ is not monotone in general. The uniqueness also agrees with what we observe in the simulation.

\begin{assumption} \label{assumpA4}\
For $\pi=\wh{\pi}$ or $\pi^*$, $S =\wh{S}$ or $S^*$, and $S_c= \wh{S}_c$ or $S_c^*$ below, where $\wh{\pi}$, $\wh{S}$ and $\wh{S}_c$ are estimated using an independent sample, we have
\ba{ \label{C3first}
    E \left\{ \left[ \sup_{t\in[0,\tau]} \bigg|\mathpzc{s}^{(l)}(t;\beta, \pi, S, S_c) - \mathpzc{s}^{(l)}(t;\beta, \pi^*, S^*, S_c^*)\bigg| \right]^2  \right\}= o(1),
}
and
\ba{ \label{C3second}
\sup_{t \in [0,\tau]} |\mathcal{S}^{(l)}(t;\beta,\pi, S, S_c) - \mathpzc{s}^{(l)}(t;\beta,\pi, S, S_c)| = O_p(n^{-1/2}),
}
for $\beta \in \mathcal{B}$ and $l = 0,1$. Moreover, 
\ba{ 
    \int_0^\tau \{\bar{A}(t;\beta^*, \pi, S, S_c) - \bar{\alpha}(t;\beta^*, \pi, S, S_c)\}\cdot \frac{1}{\sqrt{n}}\sum_{i =1}^n  D_{1i}(t;\beta^*, \Lambda^*, \pi^o, S^o, S_c^o) = o_p(1). \nn\\
    \label{C3third}
}
\end{assumption}

Assumption~\ref{assumpA4} is required due to the involvement of the time-dependent nuisance functions as well as the risk sets that are specific to the possibly misspecified Cox MSM. Condition \eqref{C3first} simply states that the convergence of $\wh{\pi}, \wh{S}, \wh{S}_c$ carries over to $\mathpzc{s}^{(l)}(t;\beta^*, \pi, S, S_c)$; an example might be: $E \left\{ \left[ \sup_{t\in[0,\tau]} \bigg|\mathpzc{s}^{(l)}(t;\beta, \wh{\pi}, S^*, \wh{S}_c) - \mathpzc{s}^{(l)}(t;\beta, \pi^*, S^*, S_c^*)\bigg| \right]^2  \right\}= o(1) $. An example of \eqref{C3second} might be: $ \sup_{t \in [0,\tau]} |\mathcal{S}^{(l)}(t;\beta, \wh{\pi}, S^*, \wh{S}_c) - \mathpzc{s}^{(l)}(t;\beta, \wh{\pi}, S^*, \wh{S}_c)| = O_p(n^{-1/2})$. 
Condition \eqref{C3second} should hold for most functions with simple structures even though the estimates of the nuisance function may converge at a slower than root-$n$ rate.
For example, if we have $G(t; h) = n^{-1} \sum_{i=1}^n A_i/h(t)$ and its limit $g(t;h) = E(A)/h(t)$, then
\ba{
\sup_{t \in [0,\tau]} |G(t;\wh{h}) - g(t;\wh{h})| \le \left| \frac{1}{n}\sum_{i=1}^n A_i - E(A) \right| \cdot \sup_{t \in [0,\tau]}\left| \frac{1}{\wh{h}(t)} \right| = O_p(n^{-1/2})
}
for any out-of-sample estimates $\wh{h}(t)$ that are bounded away from zero. 
Condition \eqref{C3third} is required for the same reason the integral terms $\mathcal{D}^\dag_1$ and $\mathcal{D}^\dag_2$ in Assumption \ref{assump7} are required.
Although we have $\sqrt{n}\{\bar{A}(t;\beta^*, \pi, S, S_c) - \bar{\alpha}(t;\beta^*, \pi, S, S_c)\} = O_p(1)$ from \eqref{C3second}, and $n^{-1}\sum_{i =1}^n  \int_0^\tau D_{1i}(t;\beta^*, \Lambda^*, \pi^o, S^o, S_c^o) = o(1) $ from Theorem~\ref{thm:dr:avebeta} and the law of large numbers, no existing tools allow us to generalize this product rate to increments within an integral, which is specific to our problem.

\subsection{Proof of Main Results}

We prove in this section the consistency and asymptotic normality of the cross-fitted AIPW estimator $\whb$. The proof of the main results is intentionally kept short and easy to follow, while the tedious details are put into the Lemmas~\ref{AIPW:lem:U_to_mu and dU_to_nu} and \ref{AIPW:lem:U_root_n}. The proof of Lemma~\ref{AIPW:lem:U_to_mu and dU_to_nu} involves standard convergence in probability arguments, regardless of whether we use cross-fitting or not. On the other hand, the proof of Lemma \ref{AIPW:lem:U_root_n} makes use of the independence induced by cross-fitting and the rate condition Assumption~\ref{assump7}, which we will elaborate on in more details later.


Here, we first state Lemma 5.10 from \cite{V1998}, which will be used in the consistency proof.
\begin{lemma}\label{AIPW:lem:van der vaart}
Let $\Theta$ be a subset of the real line and let $\Psi_n$ be random functions and $\Psi$ a fixed function of $\theta$ such that $\Psi_n(\theta) \to \Psi(\theta)$ in probability for every $\theta$. Assume that each map $\theta \to \Psi_n(\theta)$ is continuous and has exactly one zero $\wh{\theta}_n$, or is non-decreasing with $\Psi_n(\wh{\theta}_n) = o_p(1)$. Let $\theta_0$ be a point such that $\Psi(\theta_0 - \epsilon) < 0 < \Psi(\theta_0 + \epsilon)$  for every $\epsilon > 0$. Then $\wh{\theta}_n \overset{p}{\to} \theta_0$.
\end{lemma}

\begin{lemma}\label{AIPW:lem:U_to_mu and dU_to_nu}
Under Assumptions~\ref{assump4}, \ref{assump6} and \ref{assumpA1}-\ref{assumpA4}, if either $S^* = S^o$ or $(\pi^*, S_c^*) = (\pi^o, S_c^o)$, then for $\beta \in \mathcal{B}$,
\ba{
U_{cf}(\beta) &\overset{p}{\to} \mu(\beta, \pi^*, S^*, S_c^*), \label{AIPW:lem2:part1}\\
\frac{\partial}{\partial \beta}U_{cf}(\beta) &\overset{p}{\to} -\nu(\beta, \pi^*, S^*, S_c^*) \label{AIPW:lem2:part2},
}
where 
\ba{
\mu(\beta, \pi, S, S_c) &= \int_0^\tau \{\bar{\alpha}(t; \beta^*, \pi, S, S_c) - \bar{\alpha}(t; \beta, \pi, S, S_c) \} \mathpzc{s}^{(0)}(t; \beta^*, \pi, S, S_c) d\Lambda^*(t), \\
\nu(\beta, \pi, S, S_c) &= \int_0^\tau v(t;\beta, \pi, S, S_c)\mathpzc{s}^{(0)}(t; \beta^*, \pi, S, S_c) d\Lambda^*(t).
}
\end{lemma}

\begin{lemma}\label{AIPW:lem:U_root_n}
Under Assumptions~\ref{assump4}, \ref{assump6}-\ref{assump7} and \ref{assumpA1}-\ref{assumpA4}, 
\ba{
    \sqrt{n} U_{cf}(\beta^*) = \frac{1}{\sqrt{n}} \sum_{i=1}^n \psi_i(\beta^*, \Lambda^*, \pi^o, S^o,S_c^o) + o_p(1).
}
\end{lemma}

\noindent{\bf Proof of Theorem~\ref{thm:consistency}}

To show consistency, we make use of Lemma~\ref{AIPW:lem:van der vaart}. Equation \eqref{AIPW:lem2:part1} of Lemma \ref{AIPW:lem:U_to_mu and dU_to_nu} states that
\ba{
    U_{cf}(\beta) \overset{p}{\to}  \mu(\beta,\pi^*, S^*, S_c^*), 
}
for $\beta$ in a neighbourhood $\mathcal{B}$ of $\beta^*$.

Since $U_{cf}(\wh{\beta}) = 0$ and $\mu(\beta^*) = 0$, it follows from Assumption~\ref{assumpA3} that $\wh{\beta}$ is a unique zero of $U_{cf}(\beta)$ and $\beta^*$ is a unique zero of $\mu(\beta,\pi^*, S^*, S_c^*)$. Using Assumption \ref{assumpA2}, we have $\partial \mu(\beta, \pi^*, S^*, S_c^*)/\partial \beta|_{\beta = \beta^*} = -\nu(\beta^*, \pi^*, S^*, S_c^*) < 0$, and that $\mu(\beta, \pi^*, S^*, S_c^*)$ is continuous for $\beta \in \mathcal{B}$. These conditions together imply that
\ba{
\mu(\beta - \epsilon, \pi^*, S^*, S_c^*) > 0 > \mu(\beta + \epsilon, \pi^*, S^*, S_c^*)
}
for any $\epsilon > 0$.

Lastly, by noting that $ U_{cf}(\beta)$ is also continuous in $\beta$, we have $\whb \overset{p}{\to} \beta^*$ from applying Lemma~\ref{AIPW:lem:van der vaart}.
\qed

\noindent {\bf Proof of Theorem~\ref{thm:AN}}
Applying the mean value theorem to $U_{cf}(\beta^*)$ around $\beta^*$, we have 
\ba{
    \sqrt{n}(\whb - \beta^*) = \frac{-\sqrt{n}U_{cf}(\beta^*)}{
    U_{cf}(\wt{\beta}) / {\partial \beta}},   
}
where $\wt{\beta}$ is some value between $\whb$ and $\beta^*$. From Theorem \ref{thm:consistency} then $\wt{\beta} \overset{p}{\to} \beta^* $.

By \eqref{AIPW:lem2:part2} of Lemma \ref{AIPW:lem:U_to_mu and dU_to_nu}, we have $\partial U_{cf}(\beta^*)/ \partial  \beta \overset{p}{\to} -\nu(\beta^*, \pi^o, S^o, S_c^o)$. Using the same arguments as those in the proof of Lemma~\ref{AIPW:lem:U_to_mu and dU_to_nu}, we also have $\partial U_{cf}(\wt{\beta})/ \partial  \beta - \partial U_{cf}(\beta^*)/ \partial  \beta = o_p(1)$, so 
\ba{
\frac{\partial}{\partial \beta}U_{cf}(\wt{\beta}) \overset{p}{\to} -\nu(\beta^*, \pi^o, S^o, S_c^o).
}

The asymptotic expansion of $\sqrt{n}U_{cf}(\beta^*)$ is derived in Lemma \ref{AIPW:lem:U_root_n}:
\ba{
    \sqrt{n}U_{cf}(\beta^*) = \frac{1}{\sqrt{n}}\sum_{i=1}^n \psi_i(\beta^*, \Lambda^*, \pi^o, S^o, S_c^o)  + o_p(1). 
}
By Assumptions~\ref{assump4} and \ref{assumpA2}, it's easy to see that $|\psi(\beta^*, \Lambda^*, \pi^o, S^o, S_c^o)|$ is bounded a.s., so by the central limit theorem,
\ba{
    \sqrt{n}U_{cf}(\beta^*) \overset{d}{\to} N(0,E\{\psi(\beta^*, \Lambda^*, \pi^o, S^o, S_c^o)^2\}).
}
Applying Slutsky's Theorem, we therefore have
\ba{
    \sqrt{n}(\whb - \beta^*) \overset{d}{\to} N(0,\sigma^2),
}
where $\sigma^2 = E\{\psi(\beta^*, \Lambda^*, \pi^o, S^o, S_c^o)^2\}/\nu^2(\beta^*, \pi^o, S^o, S_c^o)$.

Lastly, to show that $\wh{\sigma}^2$ is a consistent estimator of $\sigma^2$, we show separately the convergence of its numerator and its denominator in probability:
\ban{
& \frac{1}{n} \sum_{m=1}^k \sum_{i \in \mathcal{I}_k}  \wt{\psi}_{m,i}(\whb, \wt{\Lambda}_m(\cdot; \whb, \wh{\pi}^{(-m)}, \wh{S}^{(-m)}, \wh{S}_c^{(-m)}), \wh{\pi}^{(-m)}, \wh{S}^{(-m)} , \wh{S}_c^{(-m)})
^2 \overset{p}{\to} E\{\psi(\beta^*, \Lambda^*, \pi^o, S^o, S_c^o)^2\}, \\
& \left\{\frac{1}{n} \sum_{m=1}^k  \sum_{i \in \mathcal{I}_k} \int_0^\tau V_m(t; \whb, \wh{\pi}^{(-m)},  \wh{S}^{(-m)}, \wh{S}_c^{(-m)}) d\mathcal{N}_i(t; \wh{\pi}^{(-m)}, \wh{S}^{(-m)}, \wh{S}^{(-m)}_c)\right\}^2 \overset{p}{\to} \nu^2(\beta^*, \pi^o, S^o, S_c^o).
}
These can be shown using the same arguments as used in Lemma~\ref{AIPW:lem:U_to_mu and dU_to_nu}, so we omit the proof here.
Applying Slutsky's theorem again, we have
\ba{
    \wh{\sigma}^{-1} \sqrt{n}(\whb - \beta^*) \overset{d}{\to} N(0,1). 
}
\qed

\subsection{Proof of lemmas} \label{AIPW:proof of lemmas}

Since the number of folds $k$ is fixed as $n\rightarrow\infty$, to show that results in Lemma~\ref{AIPW:lem:U_to_mu and dU_to_nu} hold for the cross-fitted estimating equations $U_{cf}$, it is sufficient to show that they hold for sample-splitting. Therefore, in the proof of Lemma~\ref{AIPW:lem:U_to_mu and dU_to_nu} below, we will show that
\ba{
U(\beta, \wh{\pi}, \wh{S}, \wh{S}_c) &\overset{p}{\to} \mu(\beta, \pi^*, S^*, S_c^*), \\
\frac{\partial}{\partial \beta}U(\beta, \wh{\pi}, \wh{S}, \wh{S}_c) &\overset{p}{\to} -\nu(\beta, \pi^*, S^*, S_c^*), \label{AIPW:l2.second}
}
where with a slight abuse of notation, we let $\wh{\pi}, \wh{S}, \wh{S}_c$ denote nuisance functions estimated using a different set of data independent from but with the same distribution as the dataset that $U$ is evaluated on.
Similarly, in the proof of Lemma~\ref{AIPW:lem:U_root_n} below, we will show that
\ba{
    \sqrt{n} U(\beta^*, \wh{\pi}, \wh{S}, \wh{S}_c) = \frac{1}{\sqrt{n}} \sum_{i=1}^n \psi_i(\beta^*, \Lambda^*, \pi^o, S^o,S_c^o) + o_p(1).
}

Before we begin the proof of the lemmas, note from the strict positivity Assumption~\ref{assump4} that $S^*(t;a,z)$ is bounded away from zero. 
By the uniform convergence Assumption~\ref{assump6}, $\wh{S}(t;a,z)$ converges to $S^*(t;a,z)$ in probability uniformly in $t$, so the probability that $\wh{S}(t;a,z)$ is bounded away from zero goes to one. Same argument also applies to $\wh{S}_c(t;a,z)$, $\wh{\pi}(z)$, and $1 - \wh{\pi}(z)$. We can also derive from \eqref{C3first} and \eqref{C3second} of Assumption~\ref{assumpA4} that for nuisance functions $\pi, S, S_c$ that are either the estimates $\wh{\pi}, \wh{S}, \wh{S}_c$ or their limits, and for $\beta \in \mathcal{B}$, $\mathcal{S}^{(l)}(t;\beta,\pi, S, S_c)$ converges to $\mathpzc{s}^{(l)}(t;\beta,\pi^*, S^*, S_c^*)$ in probability uniformly in $t$. Since assumption~\ref{assumpA2} states that $\mathpzc{s}^{(l)}(t;\beta,\pi^*, S^*, S_c^*)$ and $1/\mathpzc{s}^{(0)}(t;\beta,\pi^*, S^*, S_c^*)$ are bounded, so $\mathcal{S}^{(l)}(t;\beta,\pi, S, S_c)$ and $1/\mathcal{S}^{(0)}(t;\beta,\pi, S, S_c)$ are bounded with probability going to one. 
In the following to simplify the proofs, we will assume WLOG that the quantities are bounded almost surely, and this is due to the conditioning event argument below.


Both Lemmas~\ref{AIPW:lem:U_to_mu and dU_to_nu} and \ref{AIPW:lem:U_root_n} claim convergence in probability results. To prove them, we want to show that for some random quantity (i.e.~remainder term) $X_n$ and for any $\epsilon > 0$, $P(|X_n| < \epsilon) \to 1$ as $n \to \infty$. 
Let $\mathcal{G}_n$
denote the event that all those terms above are bounded. From Assumptions~\ref{assump4}, \ref{assump6}, \ref{assumpA2}, and \ref{assumpA4}, we showed earlier that $P(\mathcal{G}_n) \to 1$ as $n \to \infty$. 
In our approach we first show that $E(|X_n|\;|\mathcal{G}_n) \to 0$ as $n \to \infty$, which by Markov' inequality implies that
\ba{
P(|X_n| < \epsilon\;| \mathcal{G}_n) > 1 - \frac{E(|X_n|\;| \mathcal{G}_n)}{\epsilon} \to 1
}
as $n \to \infty$. This  leads to 
\ba{
P(|X_n| < \epsilon) = P(|X_n| < \epsilon \cap \mathcal{G}_n) + P(|X_n| < \epsilon \cap \mathcal{G}_n^c) \ge P(|X_n| < \epsilon \cap \mathcal{G}_n) =P(|X_n| < \epsilon \;| \mathcal{G}_n) P(\mathcal{G}_n) \to 1
}
as $n \to \infty$. 




\noindent {\bf Proof of Lemma~\ref{AIPW:lem:U_to_mu and dU_to_nu}}\label{AIPW:pf:lem2}
First, we have
\ban{
    U(\beta, \wh{\pi}, \wh{S}, \wh{S}_c) = U(\beta, \pi^*, S^*, S_c^*)  + Q_1 + Q_2 + Q_3,
}
where
\ba{
    Q_1 =&U(\beta, \wh{\pi}, \wh{S}, \wh{S}_c) -  U(\beta, \pi^*, \wh{S}, \wh{S}_c) \\
    Q_2 =&U(\beta, \pi^*, \wh{S}, \wh{S}_c) -   U(\beta, \pi^*, S^*, \wh{S}_c) \\
    Q_3 =&U(\beta, \pi^*, S^*, \wh{S}_c) -  U(\beta, \pi^*, S^*, S^*_c).
}
We now show that $Q_1, Q_2$, and $Q_3$ are $o_p(1)$. 

Consider $Q_1$. We write
\ba{
    Q_1 = Q_{11} - Q_{12} - Q_{13} \label{AIPW:l2.A1}
}
\ba{
    Q_{11} =& \frac{1}{n} \sum_{i=1}^n \int_0^{\tau}   d\mathcal{N}_i^{(1)}(t;\wh{\pi},\wh{S}, \wh{S}_c) - d\mathcal{N}_i^{(1)}(t;\pi^*,\wh{S}, \wh{S}_c)    \\
    Q_{12} =& \frac{1}{n} \sum_{i=1}^n \int_0^{\tau} \left\{ \bar{A}(t;\beta,\wh{\pi},\wh{S}, \wh{S}_c) - \bar{A}(t;\beta,\pi^*,\wh{S}, \wh{S}_c) \right \} d\mathcal{N}_i^{(0)}(t;\wh{\pi},\wh{S}, \wh{S}_c) \\
    Q_{13} =& \frac{1}{n} \sum_{i=1}^n \int_0^{\tau} \bar{A}(t;\beta,\pi^*,\wh{S}, \wh{S}_c) \left\{ d\mathcal{N}_i^{(0)}(t;\wh{\pi},\wh{S}, \wh{S}_c) - d\mathcal{N}_i^{(0)}(t;\pi^*,\wh{S}, \wh{S}_c) \right \}.
}

First, we note that $d\mathcal{N}_i^{(1)}(t;\pi, S, S_c)$ is a sum of several terms, each term is a product of a term that is bounded a.s. and an increment of a monotone function. Specifically, we have
\ba{
    Q_{11} =& \frac{1}{n} \sum_{i=1}^n \left [  \frac{1}{\wh{\pi}(Z_i)^{A_i}\{1-\wh{\pi}(Z_i)\}^{1-A_i}} - \frac{1}{\pi^*(Z_i)^{A_i}\{1-\pi^*(Z_i)\}^{1-A_i}}  \right ] \int_0^\tau \frac{A_i}{\wh{S}_c(t;A_i,Z_i)} dN_i(t) \\
    &+ \frac{1}{n} \sum_{i=1}^n \left [  \frac{1}{\wh{\pi}(Z_i)^{A_i}\{1-\wh{\pi}(Z_i)\}^{1-A_i}} - \frac{1}{\pi^*(Z_i)^{A_i}\{1-\pi^*(Z_i)\}^{1-A_i}}  \right ] \int_0^\tau A_i d\wh{S}(t;A_i,Z_i) \\
    &- \frac{1}{n} \sum_{i=1}^n  A_i \left\{\frac{1}{\wh{\pi}(Z_i)} - \frac{1}{\pi^*(Z_i)} \right\} \int_0^\tau  J_i(t;1,\wh{S},\wh{S}_c)d\wh{S}(t;1,Z_i).
}
This allows us to make use of the following property: for any function $f(t)$, and any monotone function $G(t)$ defined on $[a,b]$, we have 
\ba{
   \left |\int_a^b f(t) dG(t) \right| \le \sup_{t \in [a,b]} |f(t)| \cdot |G(b) - G(a)|. \label{AIPW:sup ineq}
}
Since $N(t)$ and $\wh{S}(t;a,z)$
are monotone in $t$, we apply \eqref{AIPW:sup ineq} to each of the 3 terms in $Q_{11}$ above and have
\ba{
&|Q_{11}| \\
\le& \frac{1}{n} \sum_{i=1}^n \left |  \frac{1}{\wh{\pi}(Z_i)^{A_i}\{1-\wh{\pi}(Z_i)\}^{1-A_i}} - \frac{1}{\pi^*(Z_i)^{A_i}\{1-\pi^*(Z_i)\}^{1-A_i}}  \right | \cdot \sup_{t \in [0,\tau]} \left|\frac{A_i}{\wh{S}_c(t;A_i,Z_i)} \right| \cdot \left|N_i(\tau) - N_i(0) \right| \\
&+ \frac{1}{n} \sum_{i=1}^n \left |  \frac{1}{\wh{\pi}(Z_i)^{A_i}\{1-\wh{\pi}(Z_i)\}^{1-A_i}} - \frac{1}{\pi^*(Z_i)^{A_i}\{1-\pi^*(Z_i)\}^{1-A_i}}  \right | \cdot | A_i | \cdot \left|\wh{S}(\tau;A_i,Z_i) - \wh{S}(0;A_i,Z_i)  \right| \\
&+ \frac{1}{n} \sum_{i=1}^n | A_i | \left | \frac{1}{\wh{\pi}(Z_i)} - \frac{1}{\pi^*(Z_i)}  \right | \cdot  \sup_{t \in [0,\tau]} \left|  J_i(t;1,\wh{S},\wh{S}_c)  \right| \cdot \left|\wh{S}(\tau;1,Z_i) - \wh{S}(0;1,Z_i) \right|. \\
\le& \frac{1}{n} \sum_{i=1}^n  \frac{|\wh{\pi}(Z_i) - \pi^*(Z_i)|}{\left| \{\wh{\pi}(Z_i)\pi^*(Z_i)\}^{A_i}[\{1-\wh{\pi}(Z_i)\}\{1-\pi^*(Z_i)\} ]^{1-A_i}\right|} \cdot \sup_{t \in [0,\tau]} \left|\frac{A_i}{\wh{S}_c(t;A_i,Z_i)} \right| \cdot \left|N_i(\tau) - N_i(0) \right| \\
&+ \frac{1}{n} \sum_{i=1}^n  \frac{|\wh{\pi}(Z_i) - \pi^*(Z_i)|}{\left| \{\wh{\pi}(Z_i)\pi^*(Z_i)\}^{A_i}[\{1-\wh{\pi}(Z_i)\}\{1-\pi^*(Z_i)\} ]^{1-A_i}\right|} \cdot | A_i | \cdot \left|\wh{S}(\tau;A_i,Z_i) - \wh{S}(0;A_i,Z_i)  \right| \\
&+ \frac{1}{n} \sum_{i=1}^n | A_i | \frac{|\wh{\pi}(Z_i) - \pi^*(Z_i)|}{|\wh{\pi}(Z_i)\pi^*(Z_i)|} \cdot  \sup_{t \in [0,\tau]} \left|  J_i(t;1,\wh{S},\wh{S}_c)  \right| \cdot \left|\wh{S}(\tau;1,Z_i) - \wh{S}(0;1,Z_i) \right|.
}
Since 
$\wh{S}(t;a,z)$ and $\wh{S}_c(t;a,z)$ are bounded away from zero a.s., we can again apply \eqref{AIPW:sup ineq} to 
\ba{
    J_i(t;1,\wh{S},\wh{S}_c) =\int_0^t \frac{dN_{ci}(u) + Y_i(u) d\log\{ \wh{S}_c(u;1,Z_i)\} }{\wh{S}(u;1,Z_i)\wh{S}_c(u;1,Z_i)},
}
and have 
\ba{
\sup_{t \in [0,\tau]} \left|  J_i(t;1,\wh{S},\wh{S}_c)  \right| 
\le& \sup_{t \in [0,\tau]} \Bigg\{\sup_{u \in [0,t]} \bigg|\frac{1}{\wh{S}(u;1,Z_i)\wh{S}_c(u;1,Z_i)} \bigg| \cdot \left|N_{ci}(t) - N_{ci}(0)  \right|  \Bigg\} \\
&+ \sup_{t \in [0,\tau]} \Bigg\{\sup_{u \in [0,t]} \bigg|\frac{Y_i(u)}{\wh{S}(u;1,Z_i)\wh{S}_c(u;1,Z_i)} \bigg| \cdot \left| \log\{ \wh{S}_c(t;1,Z_i)\} - \log\{ \wh{S}_c(0;1,Z_i)\} \right |   \Bigg\}\\
\lesssim& 1 \label{AIPW:l3.4}.
}
In addition, since $\wh{\pi}(z)$ and $ 1 - \wh{\pi}(z)$ are bounded away from zero a.s., we have 
\ba{ 
|Q_{11}| \lesssim \frac{1}{n} \sum_{i=1}^n |\wh{\pi}(Z_i) - \pi^*(Z_i)|.
}

As a reminder, $E^\dag$ denotes expectations taken with respect to a sample $O^\dag$ of $n$ observations, and $E$ denotes expectations taken with respect to an independent data $O$ conditional on $O^\dag$. $O$ is used for constructing $U$, 
while $O^\dag$ is used to estimate $(\wh{\pi}, \wh{S}, \wh{S}_c)$. Using this notation, we have 
\ba{
E(|Q_{11}|) \lesssim& E^\dag\left [ E \left\{  | \wh{\pi}(Z) - \pi^*(Z) |   \right\} \right] \leq \| \wh{\pi} - \pi^* \|_\dag   = o(1),
}
where the last inequality follows from Jensen's inequality, and the last equality follows from Assumption~\ref{assump6}. So we have $Q_{11} = o_p(1)$ from Markov's inequality.

Consider $Q_{12}$. We again break $d\mathcal{N}_i^{(0)}(t;\pi, S, S_c)$ into a sum of terms, each being a product of a term that is bounded a.s. and an increment of a monotone function.
\ba{
Q_{12} =& \frac{1}{n} \sum_{i=1}^n  \int_0^{\tau} \left\{ \bar{A}(t;\beta,\wh{\pi},\wh{S}, \wh{S}_c) - \bar{A}(t;\beta,\pi^*,\wh{S}, \wh{S}_c) \right \} \cdot \frac{1}{\wh{\pi}(Z_i)^{A_i}\{1-\wh{\pi}(Z_i)\}^{1-A_i}\wh{S}_c(t;A_i,Z_i)} dN_i(t) \\
&+ \frac{1}{n} \sum_{i=1}^n  \int_0^{\tau} \left\{ \bar{A}(t;\beta,\wh{\pi},\wh{S}, \wh{S}_c) - \bar{A}(t;\beta,\pi^*,\wh{S}, \wh{S}_c) \right \} \cdot \frac{1}{\wh{\pi}(Z_i)^{A_i}\{1-\wh{\pi}(Z_i)\}^{1-A_i}} d\wh{S}(t;A_i,Z_i) \\
&- \frac{1}{n} \sum_{i=1}^n  \int_0^{\tau} \left\{ \bar{A}(t;\beta,\wh{\pi},\wh{S}, \wh{S}_c) - \bar{A}(t;\beta,\pi^*,\wh{S}, \wh{S}_c) \right \} \\
&\quad \cdot \sum_{a=0,1} \left\{1 + \frac{A_i^a (1-A_i)^{1-a}}{\wh{\pi}(Z_i)^a\{1-\wh{\pi}(Z_i)\}^{1-a}}J_i(t;a,\wh{S},\wh{S}_c) \right\} d\wh{S}(t;a,Z_i).
}
Applying \eqref{AIPW:sup ineq} and similar arguments as the above, 
also recall that $\mathcal{S}^{(0)}(t;\beta,\wh{\pi},\wh{S}, \wh{S}_c)$ and $\mathcal{S}^{(0)}(t;\beta,\pi^*,\wh{S}, \wh{S}_c)$ are bounded away from zero a.s. and $\mathcal{S}^{(l)}(t;\beta,\wh{\pi},\wh{S}, \wh{S}_c)$ is bounded a.s.,
we have
\ba{
|Q_{12}| \lesssim& 
\sup_{t \in [0,\tau]}\left| \bar{A}(t;\beta,\wh{\pi},\wh{S}, \wh{S}_c) - \bar{A}(t;\beta,\pi^*,\wh{S}, \wh{S}_c) \right | \\
=& \sup_{t \in [0,\tau]}\left| \frac{\mathcal{S}^{(0)}(t;\beta,\wh{\pi},\wh{S}, \wh{S}_c)\mathcal{S}^{(1)}(t;\beta,\pi^*,\wh{S}, \wh{S}_c) - \mathcal{S}^{(1)}(t;\beta,\wh{\pi},\wh{S}, \wh{S}_c)\mathcal{S}^{(0)}(t;\beta,\pi^*,\wh{S}, \wh{S}_c)}{\mathcal{S}^{(0)}(t;\beta,\wh{\pi},\wh{S}, \wh{S}_c)\mathcal{S}^{(0)}(t;\beta,\pi^*,\wh{S}, \wh{S}_c)} \right| \\
\lesssim&  \sup_{t \in [0,\tau]}\left|  \mathcal{S}^{(0)}(t;\beta,\wh{\pi},\wh{S}, \wh{S}_c)\mathcal{S}^{(1)}(t;\beta,\pi^*,\wh{S}, \wh{S}_c) - \mathcal{S}^{(1)}(t;\beta,\wh{\pi},\wh{S}, \wh{S}_c)\mathcal{S}^{(0)}(t;\beta,\pi^*,\wh{S}, \wh{S}_c) \right| \label{AIPW:l2.19}\\
\le& \sup_{t \in [0,\tau]}\left| \mathcal{S}^{(0)}(t;\beta,\wh{\pi},\wh{S}, \wh{S}_c)\{\mathcal{S}^{(1)}(t;\beta,\pi^*,\wh{S}, \wh{S}_c) - \mathcal{S}^{(1)}(t;\beta,\wh{\pi},\wh{S}, \wh{S}_c)\}    \right| \\
&+ \sup_{t \in [0,\tau]}\left|  \mathcal{S}^{(1)}(t;\beta,\wh{\pi},\wh{S}, \wh{S}_c)\{\mathcal{S}^{(0)}(t;\beta,\pi^*,\wh{S}, \wh{S}_c) - \mathcal{S}^{(0)}(t;\beta,\wh{\pi},\wh{S}, \wh{S}_c)\}  \right| \\
\lesssim& \sum_{l=0,1} \sup_{t \in [0,\tau]} \left| \mathcal{S}^{(l)}(t;\beta,\wh{\pi},\wh{S}, \wh{S}_c) - \mathcal{S}^{(l)}(t;\beta,\pi^*,\wh{S}, \wh{S}_c) \right| \label{AIPW:l2.20}\\
\le & \sum_{l=0,1} \cdot \frac{1}{n} \sum_{i=1}^n \sup_{t \in [0,\tau]} \left|  \Gamma_i^{(l)}(t;\beta,\wh{\pi},\wh{S}, \wh{S}_c) - \Gamma_i^{(l)}(t;\beta,\pi^*,\wh{S}, \wh{S}_c) \right| \\
\lesssim& \sum_{l=0,1} \cdot \frac{1}{n} \sum_{i=1}^n \left|\frac{1}{\wh{\pi}(Z_i)^{A_i}\{1 - \wh{\pi}(Z_i)\}^{1 - A_i}} - \frac{1}{\pi^*(Z_i)^{A_i}\{1 - \pi^*(Z_i)\}^{1 - A_i}} \right| \\
& + \sum_{l=0,1} \cdot \frac{1}{n}  \sum_{i=1}^n \sum_{a=0,1} a^l \left|\frac{1}{\wh{\pi}(Z_i)^a\{1 - \wh{\pi}(Z_i)\}^{1 - a}} - \frac{1}{\pi^*(Z_i)^a\{1 - \pi^*(Z_i)\}^{1 - a}} \right|  \label{AIPW:l2.21} \\
\lesssim& \frac{1}{n} \sum_{i=1}^n |\wh{\pi}(Z_i) - \pi^*(Z_i)|,
}
where \eqref{AIPW:l2.21} follows since $\wh{S}_c(t;A_i,Z_i)$ is bounded away from zero a.s. and $J_i(t;a,\wh{S},\wh{S}_c)$ is bounded a.s. following  \eqref{AIPW:l3.4}.
Therefore, we again have $E(|Q_{12}|) = o(1)$ from Assumption~\ref{assump6}, so $Q_{12} = o_p(1)$ by Markov's inequality.

$Q_{13} = o_p(1)$ can be shown using exactly the same arguments. We therefore have $Q_1 = o_p(1)$.

Next, we show $Q_2 = o_p(1)$.
First, we write
\ba{
    Q_2 = Q_{21} - Q_{22} - Q_{23} \label{AIPW:l2.A2}
}
where
\ba{
    Q_{21} =& \frac{1}{n} \sum_{i=1}^n \int_0^{\tau}   d\mathcal{N}_i^{(1)}(t;\pi^*,\wh{S}, \wh{S}_c) - d\mathcal{N}_i^{(1)}(t;\pi^*,S^*, \wh{S}_c)    \\
    Q_{22} =& \frac{1}{n} \sum_{i=1}^n \int_0^{\tau} \left\{ \bar{A}(t;\beta,\pi^*,\wh{S}, \wh{S}_c) - \bar{A}(t;\beta,\pi^*,S^*, \wh{S}_c) \right \} d\mathcal{N}_i^{(0)}(t;\pi^*,\wh{S}, \wh{S}_c) \\
    Q_{23} =& \frac{1}{n} \sum_{i=1}^n \int_0^{\tau} \bar{A}(t;\beta,\pi^*,S^*, \wh{S}_c) \left\{ d\mathcal{N}_i^{(0)}(t;\pi^*,\wh{S}, \wh{S}_c) - d\mathcal{N}_i^{(0)}(t;\pi^*,S^*, \wh{S}_c) \right \}.
}

Consider $Q_{21}$. We have 
\ba{
    Q_{21} = Q_{211} - Q_{212} - Q_{213}, 
}
where
\ban{
    Q_{211} =& \frac{1}{n} \sum_{i=1}^n  \frac{A_i \{\wh{S}(\tau;A_i,Z_i) - S^*(\tau;A_i,Z_i)\}}{\pi^*(Z_i)^{A_i}\{1-\pi^*(Z_i)\}^{1-A_i}} + \frac{1}{n} \sum_{i=1}^n \sum_{a=0,1} a \{\wh{S}(\tau;a,Z_i) - S^*(\tau;a,Z_i)\}, \\
    Q_{212} =& \frac{1}{n} \sum_{i=1}^n \sum_{a=0,1} \frac{aA_i^a (1-A_i)^{1-a}}{\pi^*(Z_i)^a\{1-\pi^*(Z_i)\}^{1-a}} \int_0^{\tau} J_i(t;a,\wh{S},\wh{S}_c) \{d\wh{S}(t;a,Z_i) - dS^*(t;a,Z_i)\}, \\
    Q_{213} =& \frac{1}{n} \sum_{i=1}^n \sum_{a=0,1} \frac{aA_i^a (1-A_i)^{1-a}}{\pi^*(Z_i)^a\{1-\pi^*(Z_i)\}^{1-a}} \\
    &\times \int_0^{\tau} \left[ \int_0^t \left\{\frac{1}{\wh{S}(u;a,Z_i)} - \frac{1}{S^*(u;a,Z_i)}\right\}  \frac{dM_{ci}(u;a,\wh{S}_c)}{\wh{S}_c(u;a,Z_i)} \right] dS^*(t;a,Z_i).
}
For $Q_{211}$, we can easily see that
\ba{
|Q_{211}| \lesssim \frac{1}{n} \sum_{i=1}^n \sum_{a=0,1} |\{\wh{S}(\tau;a,Z_i) - S^*(\tau;a,Z_i)\}| \lesssim \frac{1}{n}\sum_{i=1}^n \sup_{t \in [0,\tau], a \in \{0,1\}}\left|\{\wh{S}(t;a,Z_i) - S^*(t;a,Z_i)\} \right|,
}
so $E(|Q_{211}|)=o(1)$ by Assumption~\ref{assump6} and $Q_{211} = o_p(1)$ by Markov's inequality. 

Term $Q_{212}$ involves a difference in increments $d\wh{S}(t;a,Z_i) - dS^*(t;a,Z_i)$. Applying integration by parts to the integral term we have
\ba{
    &\int_0^{\tau} J_i(t;a,\wh{S},\wh{S}_c) \{d\wh{S}(t;a,Z_i) - dS^*(t;a,Z_i)\} \\
    =& \left[ J_i(t;a,\wh{S},\wh{S}_c) \{ \wh{S}(t;a,Z_i) - S^*(t;a,Z_i)\} \right]\bigg|_0^\tau    - \int_0^\tau \frac{ \{ \wh{S}(t;a,Z_i) - S^*(t;a,Z_i)\} dM_{ci}(t;a,\wh{S}_c)}{\wh{S}(t;a,Z_i)\wh{S}_c(t;a,Z_i)},\label{AIPW:l2.int.by.parts}
}
So 
\ba{
Q_{212} =& \frac{1}{n} \sum_{i=1}^n \sum_{a=0,1} \frac{aA_i^a (1-A_i)^{1-a}}{\pi^*(Z_i)^a\{1-\pi^*(Z_i)\}^{1-a}} \left[ J_i(t;a,\wh{S},\wh{S}_c) \{ \wh{S}(t;a,Z_i) - S^*(t;a,Z_i)\} \right]\bigg|_0^\tau \\
&- \frac{1}{n} \sum_{i=1}^n \sum_{a=0,1} \frac{aA_i^a (1-A_i)^{1-a}}{\pi^*(Z_i)^a\{1-\pi^*(Z_i)\}^{1-a}} \int_0^\tau \frac{ \{ \wh{S}(t;a,Z_i) - S^*(t;a,Z_i)\} dM_{ci}(t;a,\wh{S}_c)}{\wh{S}(t;a,Z_i)\wh{S}_c(t;a,Z_i)}.
}
Note that 
$dM_{ci}(t;a,\wh{S}_c) = dN_{ci}(t) - Y_i(t) d\wh{\Lambda}_c(t;a,Z_i)$. Since both $N_{ci}(t)$ and $\wh{\Lambda}_c(t;a,Z_i)$ are monotone functions, we may again apply \eqref{AIPW:sup ineq} on the second term above. The nuisance functions are bounded away from zero a.s., so we have
\ba{
|Q_{212}| \lesssim& \frac{1}{n} \sum_{i=1}^n \sup_{t \in [0,\tau], a \in \{0,1\}}\left|\{\wh{S}(t;a,Z_i) - S^*(t;a,Z_i)\} \right|,
}
so $E(|Q_{212}|) = o(1)$ from Assumption~\ref{assump6} and $Q_{212} = o_p(1)$ by Markov's inequality.

By applying \eqref{AIPW:sup ineq} twice on each of the double integrals in $Q_{213}$, we can show $Q_{213} = o_p(1)$ in exactly the same way. 

Same approach used for $Q_{21}$ also gives $Q_{22} = o_p(1)$ and $Q_{23} = o_p(1)$. We hence have $Q_2 = o_p(1)$.

$Q_3 = o_p(1)$ can again be shown using the same techniques we use for $Q_2$, so we omit the details.

Lastly, we show that $U(\beta, \pi^*, S^*, S_c^*) = \mu(\beta, \pi^*, S^*, S_c^*) + o_p(1)$ for $\beta \in \mathcal{B}$.

From the definition of the AIPW estimating functions $D_{1i}(t; \beta^*, \Lambda^*, \pi, S, S_c)$ and $D_{2i}(\beta^*, \Lambda^*, \pi, S, S_c)$, we have
\ban{
    &d\mathcal{N}_{i}^{(0)}(t;\pi, S, S_c) = D_{1i}(t; \beta^*, \Lambda^*, \pi, S, S_c) + \Gamma_i^{(0)}(t;\beta^*,\pi,S, S_c) d\Lambda^*(t), \\
    &\int_0^\tau d\mathcal{N}_{i}^{(1)}(t;\pi, S, S_c) = D_{2i}(\beta^*, \Lambda^*, \pi, S, S_c) + \int_0^\tau \Gamma_i^{(1)}(t;\beta^*,\pi,S, S_c) d\Lambda^*(t).
}
For $\beta \in \mathcal{B}$, we apply this to $U(\beta,\pi^*, S^*, S_c^*)$ and have 
\ba{
    &U(\beta,\pi^*, S^*, S_c^*)  \\
    =& \frac{1}{n} \sum_{i=1}^n \int_0^\tau d\mathcal{N}_{i}^{(1)}(t;\pi^*, S^*, S_c^*) - \bar{A}(t;\beta,\pi^*, S^*, S_c^*)d\mathcal{N}_{i}^{(0)}(t;\pi^*, S^*, S_c^*)  \\
    =& \frac{1}{n} \sum_{i=1}^n \bigg[ D_{2i}(\beta^*, \Lambda^*, \pi^*, S^*, S_c^*) + \int_0^\tau  \Gamma_i^{(1)}(t;\beta^*,\pi^*,S^*, S_c^*) d\Lambda^*(t)  \\
    &- \int_0^\tau \bar{A}(t;\beta,\pi^*, S^*, S_c^*)\{\Gamma_i^{(0)}(t;\beta^*,\pi^*,S^*, S_c^*) d\Lambda^*(t) 
 + D_{1i}(t;\beta^*, \Lambda^*, \pi^*, S^*, S_c^*) \}\bigg] \\
    =& \int_0^\tau \frac{1}{n} \sum_{i=1}^n \{\Gamma_i^{(1)}(t;\beta^*,\pi^*,S^*, S_c^*) - \bar{A}(t;\beta,\pi^*, S^*, S_c^*) \Gamma_i^{(0)}(t;\beta^*,\pi^*,S^*, S_c^*)\}d\Lambda^*(t)  \\
    &+ \frac{1}{n} \sum_{i=1}^n D_{2i}(\beta^*, \Lambda^*, \pi^*, S^*, S_c^*) \\
    &- \frac{1}{n} \sum_{i=1}^n \int_0^\tau \bar{A}(t;\beta,\pi^*,S^*, S_c^*) D_{1i}(t;\beta^*, \Lambda^*, \pi^*, S^*, S_c^*)  \\
    =& \int_0^\tau \{\mathcal{S}^{(1)}(t;\beta^*,\pi^*, S^*, S_c^*) - \bar{A}(t;\beta,\pi^*, S^*, S_c^*)\mathcal{S}^{(0)}(t;\beta^*,\pi^*, S^*, S_c^*)\}d\Lambda^*(t)  \\
    &+ \frac{1}{n} \sum_{i=1}^n D_{2i}(\beta^*, \Lambda^*, \pi^*, S^*, S_c^*) \\
    &- \frac{1}{n} \sum_{i=1}^n \int_0^\tau \bar{A}(t;\beta,\pi^*,S^*, S_c^*) D_{1i}(t;\beta^*, \Lambda^*, \pi^*, S^*, S_c^*) \\
    =& \int_0^\tau \left\{\bar{A}(t;\beta^*,\pi^*, S^*, S_c^*) - \bar{A}(t;\beta,\pi^*, S^*, S_c^*)\right\}\mathcal{S}^{(0)}(t;\beta^*,\pi^*, S^*, S_c^*)d\Lambda^*(t)  \\
    &+ \frac{1}{n} \sum_{i=1}^n D_{2i}(\beta^*, \Lambda^*, \pi^*, S^*, S_c^*) \\
    &- \frac{1}{n} \sum_{i=1}^n \int_0^\tau \bar{A}(t;\beta,\pi^*,S^*, S_c^*) D_{1i}(t;\beta^*, \Lambda^*, \pi^*, S^*, S_c^*).
 }
Next, for each of $\bar{A}$ and $\mathcal{S}^{(0)}$, we add and subtract its limits and have
\ba{
   & U(\beta,\pi^*, S^*, S_c^*)  \\
    = & \quad \mu(\beta, \pi^*, S^*, S_c^*)  \\
    &+ \int_0^\tau \{\bar{A}(t;\beta^*,\pi^*, S^*, S_c^*) - \bar{A}(t;\beta,\pi^*, S^*, S_c^*) - \bar{\alpha}(t;\beta^*,\pi^*, S^*, S_c^*) + \bar{\alpha}(t;\beta,\pi^*, S^*, S_c^*)\}  \\
    &\quad\times \mathcal{S}^{(0)}(t;\beta^*,\pi^*, S^*, S_c^*)d\Lambda^*(t)  \label{AIPW:l2.5}   \\
    &+ \int_0^\tau \{\bar{\alpha}(t;\beta^*,\pi^*, S^*, S_c^*) - \bar{\alpha}(t;\beta,\pi^*, S^*, S_c^*) \} \\
    &\quad \times \{\mathcal{S}^{(0)}(t;\beta^*,\pi^*, S^*, S_c^*)  - \mathpzc{s}^{(0)}(t;\beta^*,\pi^*, S^*, S_c^*)\}   d\Lambda^*(t) \label{AIPW:l2.6} \\
    &+ \frac{1}{n} \sum_{i=1}^n D_{2i}(\beta^*, \Lambda^*, \pi^*, S^*, S_c^*) \label{AIPW:l2.7}\\
    &- \frac{1}{n} \sum_{i=1}^n \int_0^\tau \bar{\alpha}(t;\beta,\pi^*,S^*, S_c^*) D_{1i}(t;\beta^*, \Lambda^*, \pi^*, S^*, S_c^*) \label{AIPW:l2.8} \\
    &+ \frac{1}{n} \sum_{i=1}^n \int_0^\tau \{\bar{\alpha}(t;\beta,\pi^*,S^*, S_c^*) - \bar{A}(t;\beta,\pi^*,S^*, S_c^*)\} D_{1i}(t;\beta^*, \Lambda^*, \pi^*, S^*, S_c^*). \label{AIPW:l2.9}
}
For \eqref{AIPW:l2.5}, since $\Lambda^*(t)$ is an increasing function and $\mathcal{S}^{(0)}(t;\beta^*,\pi^*, S^*, S_c^*)$ is bounded a.s., we can apply  \eqref{AIPW:sup ineq} to it and have
\ba{
    &\bigg| \int_0^\tau \{\bar{A}(t;\beta^*,\pi^*, S^*, S_c^*) - \bar{A}(t;\beta,\pi^*, S^*, S_c^*) - \bar{\alpha}(t;\beta^*,\pi^*, S^*, S_c^*) + \bar{\alpha}(t;\beta,\pi^*, S^*, S_c^*)\}  \\
    &\quad\times \mathcal{S}^{(0)}(t;\beta^*,\pi^*, S^*, S_c^*)d\Lambda^*(t) \bigg| \\
    \lesssim& \sup_{t \in [0,\tau]}\left|\bar{A}(t;\beta^*,\pi^*, S^*, S_c^*) - \bar{A}(t;\beta,\pi^*, S^*, S_c^*) - \bar{\alpha}(t;\beta^*,\pi^*, S^*, S_c^*) + \bar{\alpha}(t;\beta,\pi^*, S^*, S_c^*) \right|,
}
which is $o_p(1)$ from Assumption~\ref{assumpA1}. Similarly, \eqref{AIPW:l2.6} is $o_p(1)$.

Next, we note that the increments 
in $D_{1i}(t;\beta^*, \Lambda^*, \pi^*, S^*, S_c^*)$ are $dN_i(t)$, $dS^*(t;A_i,Z_i)$, $dS^*(t;a,Z_i)$ and $d\Lambda(t)$, all of which are increments of monotone functions. So similar to \eqref{AIPW:l2.5} and \eqref{AIPW:l2.6}, we can apply \eqref{AIPW:sup ineq}, the strict positivity Assumptions~\ref{assump4} and Assumption~\ref{assumpA1} to show that \eqref{AIPW:l2.9} is $o_p(1)$.

Since we have $S^* = S^o$ or $(\pi^*, S_c^*) = (\pi^o, S_c^o)$, Theorem~\ref{thm:dr:avebeta} gives that both \eqref{AIPW:l2.7} and \eqref{AIPW:l2.8} are sums of i.i.d. mean zero terms. The strict positivity Assumption~\ref{assump4} ensures that these i.i.d. mean zero terms are also bounded, hence having bounded variance. So $\eqref{AIPW:l2.7} = o_p(1)$ and $\eqref{AIPW:l2.8} = o_p(1)$ by the weak law of large numbers.

The second part of the Lemma, 
\ba{
    \frac{\partial}{\partial \beta}U(\beta, \wh{\pi}, \wh{S}, \wh{S}_c) \overset{p}{\to} -\nu(\beta,\pi^*, S^*, S_c^*),
}
can be shown using exactly the same arguments as how we proved $U(\beta, \wh{\pi}, \wh{S}, \wh{S}_c) \overset{p}{\to} \mu(\beta, \pi^*, S^*, S_c^*)$ above, which completes the proof.

\qed

\noindent {\bf Proof of Lemma~\ref{AIPW:lem:U_root_n}}
First, write
\ba{
    \sqrt{n}U(\beta^*, \wh{\pi}, \wh{S}, \wh{S}_c) =& \sqrt{n}U(\beta^*, \pi^o,S^o, S_c^o) + Q_4 +Q_5 + Q_6,
}
where
\ba{
    Q_4 =& \sqrt{n} \{U(\beta^*, \wh{\pi},\wh{S},\wh{S}_c) - U(\beta^*, \pi^o,\wh{S}, S_c^o) \} - \sqrt{n}\{U(\beta^*, \wh{\pi},S^o,\wh{S}_c) - U(\beta^*, \pi^o S^o, S_c^o)\},\\
    Q_5 =& \sqrt{n}\{U(\beta^*, \wh{\pi}, S^o, \wh{S}_c) - U(\beta^*, \pi^o, S^o,S_c^o)\}, \\
    Q_6 =& \sqrt{n}\{U(\beta^*, \pi^o,  \wh{S}, S_c^o) - U(\beta^*, \pi^o,  S^o,S_c^o)\}.
}

The structure of the proof is as follows: we first show that using the rate condition Assumption~\ref{assump7} among other assumptions that $Q_4$, which is a difference in differences, is $o_p(1)$. Next, we show that $Q_5$ and $Q_6$ are $o_p(1)$, which uses, among other assumptions, the independence between in-fold and out-of-fold data induced by cross-fitting. Finally, we show that $\sqrt{n}U(\beta^*, \pi^o,S^o, S_c^o)$ is asymptotically equivalent to a sum of i.i.d. terms.

We first show that $Q_4 = o_p(1)$. For any fixed nuisance function $S$, we have 
\ban{
   & \sqrt{n} \{U_1(\beta^*, \wh{\pi}, S,\wh{S}_c) - U_1(\beta^*, \pi^o, S, S_c^o) \} \\
   =& \frac{1}{\sqrt{n}} \sum_{i=1}^n \int_0^\tau d\mathcal{N}_i^{(1)}(t; \wh{\pi}, S, \wh{S}_c) - d\mathcal{N}_i^{(1)}(t; \pi^o, S, S_c^o) \\
   &- \frac{1}{\sqrt{n}} \sum_{i=1}^n \int_0^\tau \bar{A}(t; \beta^*, \pi^o, S, S_c^o)\{d\mathcal{N}_i^{(0)}(t; \wh{\pi}, S, \wh{S}_c) - d\mathcal{N}_i^{(0)}(t; \pi^o, S, S_c^o)\}\\
   &- \frac{1}{\sqrt{n}}  \sum_{i=1}^n \int_0^\tau \{\bar{A}(t; \beta^*, \wh{\pi}, S, \wh{S}_c) - \bar{A}(t; \beta^*, \pi^o, S, S_c^o) \}d\mathcal{N}_i^{(0)}(t; \wh{\pi}, S, \wh{S}_c).
}
So we can write
\ba{
    Q_4 = Q_{41}- Q_{42} - Q_{43} - Q_{44},
}
where
\ba{
    Q_{41} =& \frac{1}{\sqrt{n}}  \sum_{i=1}^n \int_0^\tau d\mathcal{N}_i^{(1)}(t; \wh{\pi}, \wh{S}, \wh{S}_c) - d\mathcal{N}_i^{(1)}(t; \pi^o, \wh{S}, S_c^o) - d\mathcal{N}_i^{(1)}(t; \wh{\pi}, S^o, \wh{S}_c) + d\mathcal{N}_i^{(1)}(t; \pi^o, S^o, S_c^o) \\
    &- \frac{1}{\sqrt{n}}  \sum_{i=1}^n \int_0^\tau \bar{A}(t; \beta^*, \pi^o, S^o, S_c^o) \\
    &\quad \times \{ d\mathcal{N}_i^{(0)}(t; \wh{\pi}, \wh{S}, \wh{S}_c) - d\mathcal{N}_i^{(0)}(t; \pi^o, \wh{S}, S_c^o) - d\mathcal{N}_i^{(0)}(t; \wh{\pi}, S^o, \wh{S}_c) + d\mathcal{N}_i^{(0)}(t; \pi^o, S^o, S_c^o) \} \\
    Q_{42} =& \frac{1}{\sqrt{n}} \sum_{i=1}^n \int_0^\tau \{\bar{A}(t; \beta^*, \wh{\pi}, S^o, \wh{S}_c) - \bar{A}(t; \beta^*, \pi^o, S^o, S_c^o) \}\{d\mathcal{N}_i^{(0)}(t; \wh{\pi}, \wh{S}, \wh{S}_c) - d\mathcal{N}_i^{(0)}(t; \wh{\pi}, S^o, \wh{S}_c) \}\\
    Q_{43} =& \frac{1}{\sqrt{n}}  \sum_{i=1}^n \int_0^\tau  \{\bar{A}(t; \beta^*, \pi^o, \wh{S}, S_c^o) - \bar{A}(t; \beta^*, \pi^o, S^o, S_c^o)\}\{d\mathcal{N}_i^{(0)}(t; \wh{\pi}, \wh{S}, \wh{S}_c) - d\mathcal{N}_i^{(0)}(t; \pi^o, \wh{S}, S_c^o)\} \\
    Q_{44} =&  \frac{1}{\sqrt{n}}  \sum_{i=1}^n \int_0^\tau d\mathcal{N}_i^{(0)}(t; \wh{\pi}, \wh{S}, \wh{S}_c) \\
    &\times  \{ \bar{A}(t; \beta^*, \wh{\pi}, \wh{S}, \wh{S}_c) - \bar{A}(t; \beta^*, \pi^o, \wh{S}, S_c^o) - \bar{A}(t; \beta^*, \wh{\pi}, S^o, \wh{S}_c) + \bar{A}(t; \beta^*, \pi^o, S^o, S_c^o)   \} \label{AIPW:l4.Q15}.
}

Consider $Q_{41}$, which can be written as $Q_{41} = - Q_{411} + Q_{412} - Q_{413} - Q_{414}$, where
\ba{
    Q_{411} =& \frac{1}{\sqrt{n}}  \sum_{i=1}^n\sum_{a=0,1} \frac{A_i^a(1-A_i)^{1-a}}{\pi^o(Z_i)^a\{1-\pi^o(Z_i)\}^{1-a}} \int_0^\tau \Bigg[ \{ a -  \bar{A}(t; \beta^*, \pi^o, S^o, S_c^o) \} \\
    &\times \int_0^t \left \{ \frac{d\wh{S}(t;a,Z_i)}{\wh{S}(u;a,Z_i)} - \frac{dS^o(t;a,Z_i)}{S^o(u;a,Z_i)}  \right \} \left \{  \frac{dM_{ci}(u;a,\wh{S}_c)}{\wh{S}_c(u;a,Z_i)} - \frac{dM_{ci}(u;a,S_c^o)}{S_c^o(u;a,Z_i)} \right \} \Bigg] \label{AIPW:l4.4} \\
    Q_{412} =& \frac{1}{\sqrt{n}} \sum_{i=1}^n \{A_i -  \bar{A}(\tau; \beta^*, \pi^o, S^o, S_c^o)\} \left \{ \frac{1}{\wh{\pi}(Z_i)^{A_i}\{1-\wh{\pi}(Z_i)\}^{1-A_i}} - \frac{1}{\pi^o(Z_i)^{A_i}\{1-\pi^o(Z_i)\}^{1-A_i}}     \right \}  \\ 
    & \times \left \{ \wh{S}(\tau;A_i,Z_i) - S^o(\tau;A_i,Z_i)    \right \} \\
    Q_{413} =&  \frac{1}{\sqrt{n}}  \sum_{i=1}^n\sum_{a=0,1}  A_i^a(1-A_i)^{1-a}  \left \{ \frac{1}{\wh{\pi}(Z_i)^a\{1-\wh{\pi}(Z_i)\}^{1-a}} - \frac{1}{\pi^o(Z_i)^a\{1-\pi^o(Z_i)\}^{1-a}}     \right \} \\
    &\times \int_0^\tau  dS^o(t;a,Z_i) \{a -  \bar{A}(t; \beta^*, \pi^o, S^o, S_c^o)\}  \int_0^t \left \{ \frac{1}{\wh{S}(u;a,Z_i)} - \frac{1}{S^o(u;a,Z_i)}  \right \} \frac{dM_{ci}(u;a,\wh{S}_c)}{\wh{S}_c(u;a,Z_i)} \\
    Q_{414} =& \frac{1}{\sqrt{n}}  \sum_{i=1}^n\sum_{a=0,1}  A_i^a(1-A_i)^{1-a}  \left \{ \frac{1}{\wh{\pi}(Z_i)^a\{1-\wh{\pi}(Z_i)\}^{1-a}} - \frac{1}{\pi^o(Z_i)^a\{1-\pi^o(Z_i)\}^{1-a}}     \right \} \\
    &\times \int_0^\tau \{d\wh{S}(t;a,Z_i) - dS^o(t;a,Z_i)  \} \{a -  \bar{A}(t; \beta^*, \pi^o, S^o, S_c^o)\}  J_i(t;a,\wh{S}, \wh{S}_c).
}

For $Q_{411}$, we first notice that by the strict positivity Assumption~\ref{assump4}, 
$A_i^a(1-A_i)^{1-a}/\{\pi^o(Z_i)^a\{1-\pi^o(Z_i)\}^{1-a}\}$ is bounded a.s.. The expectation of the absolute value of the double integral in $Q_{411}$ can be bounded directly using $\mathcal{D}^\dag_1$ defined in Assumption~\ref{assump7}, which leads to
\ba{
E(|Q_{411}|) \lesssim& \sqrt{n} \mathcal{D}^\dag_1 = o(1),
}
where the last equality follows from rate condition Assumption~\ref{assump7}.

As discussed in the Asymptotic Properties Section, integral remainders $\mathcal{D}^\dag_1$ is specific to our case because both nuisance functions $S(t;a,z)$ and $S_c(t;a,z)$ are time-dependent, which can lead to a product between the differences $\wh{S}_c - S_c$ and differences of increments $d\wh{S} - dS^o$, like in \eqref{AIPW:l4.4}. To the best of our knowledge, remainder terms like this can not be sufficiently controlled using existing tools, which requires us to make additional assumptions, such as $\mathcal{D}_1^\dag(\wh{S}, \wh{S}_c;S^o,S_c^o) = o(n^{-1/2})$ in the rate condition Assumption~\ref{assump7}. 


For $Q_{412}$, recall that $A_i -  \bar{A}(\tau; \beta^*, \pi^o, S^o, S_c^o)$ is bounded a.s. and $\wh{\pi}(Z_i)$, $\pi^o(Z_i)$, $1 - \wh{\pi}(Z_i)$ and $1 - \pi^o(Z_i)$ are bounded away from zero a.s., so we have
\ba{
|Q_{412}| \le& \frac{1}{\sqrt{n}} \sum_{i=1}^n |A_i -  \bar{A}(\tau; \beta^*, \pi^o, S^o, S_c^o)| \cdot \frac{|\wh{\pi}(Z_i) - \pi^o(Z_i)| \cdot  \left | \wh{S}(\tau;A_i,Z_i) - S^o(\tau;A_i,Z_i)    \right | }{ | \wh{\pi}(Z_i)^{A_i}\{1-\wh{\pi}(Z_i)\}^{1-A_i} \pi^o(Z_i)^{A_i}\{1-\pi^o(Z_i)\}^{1-A_i} |} \\
\lesssim& \frac{1}{\sqrt{n}}  \sum_{i=1}^n |\wh{\pi}(Z_i) - \pi^o(Z_i)| \cdot  \sup_{t \in [0,\tau], a \in \{0,1\}}\left|\wh{S}(t;a,Z_i) - S^o(t;a,Z_i) \right|.
}
Therefore
\ba{
    E(|Q_{412}|) \lesssim& \sqrt{n} E^\dag\Bigg \{ E \Bigg [ |\wh{\pi}(Z) - \pi^o(Z) | \cdot \sup_{t \in [0,\tau], a \in \{0,1\}} \left| \wh{S}(t;a,Z) - S^o(t;a,Z) \right| \Bigg ] \Bigg \}  \\
   \le &   \sqrt{n} \left\|\wh{\pi} - \pi^o\right\|_\dag \cdot \left\|\wh{S} - S^o\right\|_\dag \label{AIPW:l4.25} \\
   =&  o(1), \label{AIPW:l4.26}
}
where \eqref{AIPW:l4.25} follows from the Cauchy-Schwartz inequality $|E(AB)|^2 \leq E(A^2)E(B^2)$, while \eqref{AIPW:l4.26} uses the rate condition Assumption~\ref{assump7}.

$Q_{413}$ can be bounded similarly with the help of \eqref{AIPW:sup ineq}. First we note that $S^o(t;a,Z_i)$ is a monotone function by assumption. Recall that $dM_{ci}(u;a,\wh{S}_c) = dN_{ci}(u) - Y_i(u) d\wh{\Lambda}_c(u;a,Z_i)$ is also a sum of two terms, each being the product of a term bounded a.s. and an increment of a monotone function. We therefore apply \eqref{AIPW:sup ineq} twice to each of the double integral in $Q_{413}$ and have
\ba{
|Q_{413}| \lesssim&  \frac{1}{\sqrt{n}} \sum_{i=1}^n \sum_{a = 0, 1}  \frac{|\wh{\pi}(Z_i) - \pi^o(Z_i)| }{ | \wh{\pi}(Z_i)^a\{1-\wh{\pi}(Z_i)\}^{1-a} \pi^o(Z_i)^a\{1-\pi^o(Z_i)\}^{1-a} |} \\
&\times \sup_{t \in [0,\tau]} \left| \{a -  \bar{A}(t; \beta^*, \pi^o, S^o, S_c^o)\}  \int_0^t \left \{ \frac{1}{\wh{S}(u;a,Z_i)} - \frac{1}{S^o(u;a,Z_i)}  \right \} \frac{dM_{ci}(u;a,\wh{S}_c)}{\wh{S}_c(u;a,Z_i)} \right| \\
\lesssim& \frac{1}{\sqrt{n}} \sum_{i=1}^n  |\wh{\pi}(Z_i) - \pi^o(Z_i)| \\
&\times \sup_{t \in [0,\tau], a \in \{0,1 \}} \left\{  \sup_{u \in [0,t]}  \left | \frac{\wh{S}(u;a,Z_i) - S^o(u;a,Z_i)}{\wh{S}(u;a,Z_i)S^o(u;a,Z_i)\wh{S}_c(u;a,Z_i)}  \right|  \right \} \\
\lesssim & \frac{1}{\sqrt{n}}  \sum_{i=1}^n |\wh{\pi}(Z_i) - \pi^o(Z_i)| \cdot  \sup_{t \in [0,\tau], a \in \{0,1\}}\left|\wh{S}(t;a,Z_i) - S^o(t;a,Z_i) \right|.
}
So we again have
\ba{
    E(|Q_{413}|) \lesssim& \sqrt{n} E^\dag\Bigg \{ E \Bigg [ |\wh{\pi}(Z) - \pi^o(Z) | \cdot \sup_{t \in [0,\tau], a \in \{0,1\}} \left| \wh{S}(t;a,Z) - S^o(t;a,Z) \right| \Bigg ] \Bigg \}  \\
   \le &   \sqrt{n} \left\|\wh{\pi} - \pi^o\right\|_\dag \cdot \left\|\wh{S} - S^o\right\|_\dag \\
   =&  o(1)
}
from the Cauchy-Schwartz inequality and the rate condition Assumption~\ref{assump7}.

The integral in $Q_{414}$ involves a difference in increments $d\wh{S}(t;a,Z_i) - dS^o(t;a,Z_i) $, so we apply integration by parts  and have
\ba{
    & \int_0^\tau \{d\wh{S}(t;a,Z_i) - dS^o(t;a,Z_i)  \} \{a -  \bar{A}(t; \beta^*, \pi^o, S^o, S_c^o)\}  J_i(t;a,\wh{S}, \wh{S}_c) \\
    =& \left[\{\wh{S}(t;a,Z_i) - S^o(t;a,Z_i)  \} \{a -  \bar{A}(t; \beta^*, \pi^o, S^o, S_c^o)\} J_i(t;a,\wh{S}, \wh{S}_c) \right] \bigg|_0^\tau \\
    &- \int_0^\tau \{\wh{S}(t;a,Z_i) - S^o(t;a,Z_i)\} \frac{\partial}{\partial t}\left[ \{a -  \bar{A}(t; \beta^*, \pi^o, S^o, S_c^o)\} J_i(t;a,\wh{S}, \wh{S}_c) \right] \\
    =& \{\wh{S}(\tau;a,Z_i) - S^o(\tau;a,Z_i)  \} \{a -  \bar{A}(\tau; \beta^*, \pi^o, S^o, S_c^o)\} J_i(\tau;a,\wh{S}, \wh{S}_c) \\
    &- \int_0^\tau \{\wh{S}(t;a,Z_i) - S^o(t;a,Z_i)\} \{a -  \bar{A}(t; \beta^*, \pi^o, S^o, S_c^o)\} \cdot \frac{dM_{ci}(t;a,\wh{S}_c) }{\wh{S}(t;a,Z_i)\wh{S}_c(t;a,Z_i)} \\
    &+ \int_0^\tau \{\wh{S}(t;a,Z_i) - S^o(t;a,Z_i)\} \frac{\partial}{\partial t}\left[ \frac{\mathcal{S}^{(1)}(t; \beta^*, \pi^o, S^o, S_c^o)}{\mathcal{S}^{(0)}(t; \beta^*, \pi^o, S^o, S_c^o)}  \right]  \cdot  J_i(t;a,\wh{S}, \wh{S}_c)   \\
    =& \{\wh{S}(\tau;a,Z_i) - S^o(\tau;a,Z_i)  \} \{a -  \bar{A}(\tau; \beta^*, \pi^o, S^o, S_c^o)\} J_i(\tau;a,\wh{S}, \wh{S}_c) \\
    &\- \int_0^\tau \{\wh{S}(t;a,Z_i) - S^o(t;a,Z_i)\} \{a -  \bar{A}(t; \beta^*, \pi^o, S^o, S_c^o)\} \cdot \frac{dM_{ci}(t;a,\wh{S}_c) }{\wh{S}(t;a,Z_i)\wh{S}_c(t;a,Z_i)} \\
    &+ \int_0^\tau \{\wh{S}(t;a,Z_i) - S^o(t;a,Z_i)\} J_i(t;a,\wh{S}, \wh{S}_c) \frac{1}{\mathcal{S}^{(0)}(t; \beta^*, \pi^o, S^o, S_c^o)} \cdot \frac{1}{n} \sum_{j=1}^n  d\Gamma_j^{(1)}(t; \beta^*, \pi^o, S^o, S^o_c)  \\
    &- \int_0^\tau \{\wh{S}(t;a,Z_i) - S^o(t;a,Z_i)\} J_i(t;a,\wh{S}, \wh{S}_c) \frac{\mathcal{S}^{(1)}(t; \beta^*, \pi^o, S^o, S_c^o)}{\mathcal{S}^{(0)}(t; \beta^*, \pi^o, S^o, S_c^o)^2} \cdot \frac{1}{n} \sum_{j=1}^n  d\Gamma_j^{(0)}(t; \beta^*, \pi^o, S^o, S^o_c), \label{AIPW:l4.int.by.parts.2}
}
where the last two equalities follow from the product rule.
For $l = 0,1$, we again apply the product rule and have
\ba{
& d\Gamma_j^{(l)}(t; \beta^*, \pi^o, S^o, S^o_c) \label{AIPW:l4.2} \\
=&  \frac{A^l_j e^{\beta^* A_j}}{\pi^o(Z_j)^{A_j}\{1-\pi^o(Z_j)\}^{1-A_j}S^o_c(t;A_j,Z_j)}dY_j(t) \\
&- \frac{A^l_j Y_j(t)e^{\beta^* A_j}}{\pi^o(Z_j)^{A_j}\{1-\pi^o(Z_j)\}^{1-A_j}S^o_c(t;A_j,Z_j)^2} dS^o_c(t;A_j,Z_j) - \frac{A^l_j e^{\beta^* A_j}}{\pi^o(Z_j)^{A_j}\{1-\pi^o(Z_j)\}^{1-A_j}} dS^o(t;A_j,Z_j) \\
&+ \sum_{a=0,1} a^l \left\{1 + \frac{A_j^a (1-A_j)^{1-a}}{\pi^o(Z_j)^a \{1 - \pi^o(Z_j)\}^{1-a}}J_j(t;a,S^o,S^o_c)\right\}e^{\beta^* a} dS^o(t;a,Z_j)\\
&+ \sum_{a=0,1} a^l \frac{A_j^a (1-A_j)^{1-a}}{\pi^o(Z_j)^a \{1 - \pi^o(Z_j)\}^{1-a}}\frac{S^o(t;a,Z_j)e^{\beta^* a}}{S^o(t;a,Z_j)S^o_c(t;a,Z_j)} dM_{cj}(u;a,S^o_c).
}
Since $dM_{cj}(u;a,S^o_c) = dN_{cj}(u) - Y_j(u) d\Lambda_c^o(u;a,Z_j)$, we can now see that $d\Gamma_j^{(l)}(t; \beta^*, \pi^o, S^o, S^o_c)$ is once again a sum of terms, each being a product between a term that is bounded a.s. and an increment of a monotone function. Therefore, applying \eqref{AIPW:sup ineq}, we have 
\ba{
    & \left|\int_0^\tau \{d\wh{S}(t;a,Z_i) - dS^o(t;a,Z_i)  \} \{a -  \bar{A}(t; \beta^*, \pi^o, S^o, S_c^o)\}  J_i(t;a,\wh{S}, \wh{S}_c) \right| \\
    \lesssim& | \wh{S}(\tau;a,Z_i) - S^o(\tau;a,Z_i)  | \\
    &+ \sup_{t \in [0,\tau], a \in \{0,1\}}\left|\wh{S}(t;a,Z_i) - S^o(t;a,Z_i) \right| \\
    &+ \frac{1}{n} \sum_{j=1}^n \sup_{t \in [0,\tau], a \in \{0,1\}}\left|\wh{S}(t;a,Z_i) - S^o(t;a,Z_i) \right|\\
    &+ \frac{1}{n} \sum_{j=1}^n \sup_{t \in [0,\tau], a \in \{0,1\}}\left|\wh{S}(t;a,Z_i) - S^o(t;a,Z_i) \right|\\
    \lesssim& \sup_{t \in [0,\tau], a \in \{0,1\}}\left|\wh{S}(t;a,Z_i) - S^o(t;a,Z_i) \right|.
}
So
\ba{
    |Q_{414}| \lesssim \frac{1}{\sqrt{n}}  \sum_{i=1}^n |\wh{\pi}(Z_i) - \pi^o(Z_i)| \cdot \sup_{t \in [0,\tau], a \in \{0,1\}}\left|\wh{S}(t;a,Z_i) - S^o(t;a,Z_i) \right|,
}
and we again have
\ba{
    E(|Q_{414}|) \lesssim& \sqrt{n} E^\dag\Bigg \{ E \Bigg [ |\wh{\pi}(Z) - \pi^o(Z) | \cdot  \sup_{t \in [0,\tau], a \in \{0,1\}} \left| \wh{S}(t;a,Z) - S^o(t;a,Z) \right| \Bigg ] \Bigg \}  \\
   \le &   \sqrt{n} \left\|\wh{\pi} - \pi^o\right\|_\dag \cdot \left\|\wh{S} - S^o\right\|_\dag \label{AIPW:l4.5} \\
   =&  o(1), \label{AIPW:l4.6}
}
from the Cauchy-Schwartz inequality and the rate condition Assumption~\ref{assump7}.

Therefore, we have 
\ba{
     E(|Q_{41}|) \le  E(|Q_{411}|) + E(|Q_{412}|) + E(|Q_{413}|) + E(|Q_{414}|) \lesssim o(1),
}
so $Q_{41} = o_p(1)$ by Markov's inequality.

Next, we bound $Q_{42}$, which involves the use of $\mathcal{D}^\dag_2$. First, we let $Q_{42} = Q_{421} + Q_{422}$, where 
\ba{
Q_{421} =& \frac{1}{\sqrt{n}} \sum_{i=1}^n \int_0^\tau \{\bar{A}(t; \beta^*, \wh{\pi}, S^o, \wh{S}_c) - \bar{A}(t; \beta^*, \pi^o, S^o, \wh{S}_c) \}\{d\mathcal{N}_i^{(0)}(t; \wh{\pi}, \wh{S}, \wh{S}_c) - d\mathcal{N}_i^{(0)}(t; \wh{\pi}, S^o, \wh{S}_c) \} \\
Q_{422} =& \frac{1}{\sqrt{n}} \sum_{i=1}^n \int_0^\tau \{\bar{A}(t; \beta^*, \pi^o, S^o, \wh{S}_c) - \bar{A}(t; \beta^*, \pi^o, S^o, S_c^o) \}\{d\mathcal{N}_i^{(0)}(t; \wh{\pi}, \wh{S}, \wh{S}_c) - d\mathcal{N}_i^{(0)}(t; \wh{\pi}, S^o, \wh{S}_c) \}.
}
Note that like how we bounded $|Q_{12}|$ earlier, we also have
\ba{
&  \bar{A}(t; \beta^*, \wh{\pi}, S^o, \wh{S}_c) - \bar{A}(t; \beta^*, \pi^o, S^o, \wh{S}_c) \\
=& \frac{\mathcal{S}^{(1)}(t; \beta^*, \wh{\pi}, S^o, \wh{S}_c)\mathcal{S}^{(0)}(t; \beta^*, \pi^o, S^o, \wh{S}_c) - \mathcal{S}^{(1)}(t; \beta^*, \pi^o, S^o, \wh{S}_c)\mathcal{S}^{(0)}(t; \beta^*, \wh{\pi}, S^o, \wh{S}_c)}{\mathcal{S}^{(0)}(t; \beta^*, \wh{\pi}, S^o, \wh{S}_c)\mathcal{S}^{(0)}(t; \beta^*, \pi^o, S^o, \wh{S}_c)} \\
=& \frac{\{\mathcal{S}^{(1)}(t; \beta^*, \wh{\pi}, S^o, \wh{S}_c) - \mathcal{S}^{(1)}(t; \beta^*, \pi^o, S^o, \wh{S}_c)\}\mathcal{S}^{(0)}(t; \beta^*, \pi^o, S^o, \wh{S}_c)}{\mathcal{S}^{(0)}(t; \beta^*, \wh{\pi}, S^o, \wh{S}_c)\mathcal{S}^{(0)}(t; \beta^*, \pi^o, S^o, \wh{S}_c)} \\
& - \frac{\mathcal{S}^{(1)}(t; \beta^*, \pi^o, S^o, \wh{S}_c) \{\mathcal{S}^{(0)}(t; \beta^*, \wh{\pi}, S^o, \wh{S}_c) - \mathcal{S}^{(0)}(t; \beta^*, \pi^o, S^o, \wh{S}_c)\}  }{\mathcal{S}^{(0)}(t; \beta^*, \wh{\pi}, S^o, \wh{S}_c)\mathcal{S}^{(0)}(t; \beta^*, \pi^o, S^o, \wh{S}_c)} \\
=& \frac{\mathcal{S}^{(0)}(t; \beta^*, \pi^o, S^o, \wh{S}_c)}{\mathcal{S}^{(0)}(t; \beta^*, \wh{\pi}, S^o, \wh{S}_c)\mathcal{S}^{(0)}(t; \beta^*, \pi^o, S^o, \wh{S}_c)} \cdot \frac{1}{n}\sum_{j=1}^n \{\Gamma^{(1)}_j(t; \beta^*, \wh{\pi}, S^o, \wh{S}_c) - \Gamma^{(1)}_j(t; \beta^*, \pi^o, S^o, \wh{S}_c)\} \\
& - \frac{\mathcal{S}^{(1)}(t; \beta^*, \pi^o, S^o, \wh{S}_c)  }{\mathcal{S}^{(0)}(t; \beta^*, \wh{\pi}, S^o, \wh{S}_c)\mathcal{S}^{(0)}(t; \beta^*, \pi^o, S^o, \wh{S}_c)} \cdot \frac{1}{n}\sum_{j=1}^n \{\Gamma^{(0)}_j(t; \beta^*, \wh{\pi}, S^o, \wh{S}_c) - \Gamma^{(0)}_j(t; \beta^*, \pi^o, S^o, \wh{S}_c)\}\\
=& \frac{1}{n}\sum_{j=1}^n C_j(t) \{\wh{\pi}(Z_j) - \pi^o(Z_j) \}, \label{AIPW:l4.1}
}
where $C_j(t)$ are some functions bounded a.s.. Similarly, we have
\ba{
    \{\bar{A}(t; \beta^*, \pi^o, S^o, \wh{S}_c) - \bar{A}(t; \beta^*, \pi^o, S^o, S_c^o) \} = \frac{1}{n}\sum_{j=1}^n C_j'(t) \{\wh{S}_c(t;a,Z_i) - S^o_c(t;a,Z_i) \}, \label{AIPW:l4.22}
}
where $C_j'(t)$ are some other functions bounded a.s.. 


Next, let $d\mathcal{N}_i^{(0)}(t; \wh{\pi}, \wh{S}, \wh{S}_c) - d\mathcal{N}_i^{(0)}(t; \wh{\pi}, S^o, \wh{S}_c) = K_{1i} + K_{2i}$, 
where
\ban{
K_{1i} =& \frac{d\wh{S}(t;A_i,Z_i) - dS^o(t;A_i,Z_i)}{ \wh{\pi}(Z_i)^{A_i}\{1- \wh{\pi}(Z_i)\}^{1-A_i}} \\
&- \sum_{a=0,1} \left\{1 + \frac{A_i^a (1-A_i)^{1-a}}{\wh{\pi}(Z_i)^a\{1-\wh{\pi}(Z_i)\}^{1-a}}J_i(t;a,S^o,\wh{S}_c) \right\} \{d\wh{S}(t;a,Z_i) - dS^o(t;a,Z_i)\} \\
K_{2i}=& - \sum_{a=0,1} \frac{A_i^a (1-A_i)^{1-a}}{\wh{\pi}(Z_i)^a\{1-\wh{\pi}(Z_i)\}^{1-a}}\int_0^t \left\{ \frac{1}{\wh{S}(u;a,Z_i)} - \frac{1}{S^o(u;a,Z_i)} \right\} \frac{dM_{ci}(u;a,Z_i)}{\wh{S}_c(u;a,Z_i)} \cdot d\wh{S}(t;a,Z_i).
}
We now have $Q_{421} = Q_{4211} + Q_{4212}$, where
\ba{
Q_{4211} =& \frac{1}{n^{3/2}} \sum_{i=1}^n \sum_{j=1}^n  \{\wh{\pi}(Z_j) - \pi^o(Z_j) \} \int_0^\tau  C_j(t) K_{1i} \\
Q_{4212} =& \frac{1}{n^{3/2}} \sum_{i=1}^n \sum_{j=1}^n  \{\wh{\pi}(Z_j) - \pi^o(Z_j) \} \int_0^\tau  C_j(t) K_{2i}.
}
For $Q_{4212}$, we can apply \eqref{AIPW:sup ineq}, the rate Condition Assumption~\ref{assump7} and the boundedness of appropriate terms to show that 
\ba{
E(|Q_{4212}|) \lesssim \sqrt{n}\|\wh{\pi} - \pi^o\|_\dag \cdot \left\|\wh{S} - S^o\right\|_\dag = o(1).
}

$\int_0^\tau  C_j(t) K_{1i}$ in $Q_{4211}$ involves 
stochastic differences $d\wh{S}(t;a,Z_i) - dS^o(t;a,Z_i)$, so like in \eqref{AIPW:l2.int.by.parts} and \eqref{AIPW:l4.int.by.parts.2} we first apply integration by parts to turn $d\wh{S} - dS^o$ into $\wh{S} - S^o$.
Like \eqref{AIPW:l4.2}, the $dC_j(t)$ term we have as a result of integration by parts can again be shown to be a sum of terms, each being a product between a term that is bounded a.s. and an increment of a monotone function. This allows us to apply \eqref{AIPW:sup ineq}, the rate Condition Assumption~\ref{assump7} and the boundedness of appropriate terms, which leads to $E(|Q_{4211}|) \lesssim \sqrt{n}\|\wh{\pi} - \pi^o\|_\dag \cdot \left\|\wh{S} - S^o\right\|_\dag = o(1)$. We therefore have $E(|Q_{421}|) = o(1)$ and $Q_{421} = o_p(1)$ by Markov's inequality.

For term $Q_{422}$, we first write $Q_{422} = Q_{4221} + Q_{4222}$, where
\ba{
    Q_{4221} =& \frac{1}{\sqrt{n}} \sum_{i=1}^n  \int_0^\tau \{\bar{A}(t; \beta^*, \pi^o, S^o, \wh{S}_c) - \bar{A}(t; \beta^*, \pi^o, S^o, S_c^o) \} K_{1i}, \\
    Q_{4222} =& \frac{1}{\sqrt{n}} \sum_{i=1}^n  \int_0^\tau \{\bar{A}(t; \beta^*, \pi^o, S^o, \wh{S}_c) - \bar{A}(t; \beta^*, \pi^o, S^o, S_c^o) \} K_{2i}.
}
$E(|Q_{4221}|)$ involves a product between $d\wh{S}(t;a,Z_i) - dS^o(t;a,Z_i)$ and $\wh{S}_c(t;a,Z_i) - S_c^o(t;a,Z_i)$, which can not be bounded using any existing tools we have. Therefore, we directly bound it using $\mathcal{D}^\dag_2$ in Assumption~\ref{assump7}, which gives
\ba{
E(|Q_{4221}|) \lesssim \sqrt{n} \mathcal{D}^\dag_2 = o(1).
}

Next, using \eqref{AIPW:l4.22}, we have
\ba{
    Q_{4222} =& \frac{1}{n^{3/2}} \sum_{i=1}^n \sum_{j=1}^n \int_0^\tau  C_j'(t) \{\wh{S}_c(t;a,Z_i) - S^o_c(t;a,Z_i) \} K_{2i}.
}
By again applying \eqref{AIPW:sup ineq}, the rate condition Assumption~\ref{assump7}  and the boundedness of appropriate terms to $Q_{4222}$, we have
\ba{
E(|Q_{4222}|) \lesssim \sqrt{n}\left\|\wh{S}_c - S_c^o\right\|_\dag \cdot \left\|\wh{S} - S^o\right\|_\dag = o(1).
}

Therefore, $E(|Q_{422}|) = o(1)$ from rate condition Assumption~\ref{assump7}.

Combining our results, we have
\ba{
    E(|Q_{42}|) \leq  E(|Q_{421}|) +  E(|Q_{422}|) = o(1).
}

Using the same techniques we used for $Q_{41}$ and $Q_{42}$ above, with the rate condition Assumption~\ref{assump7} and without using $\mathcal{D}^\dag$, we can show that $E(|Q_{43}|) = o(1)$ and $E(|Q_{44}|) = o(1)$.

Hence we conclude that $E(|Q_4|) \le$ $E(|Q_{41}|) + E(|Q_{42}|) + E(|Q_{43}|) + E(|Q_{44}|) = o(1)$. Then by Markov's inequality, $Q_4 = o_p(1)$.

Next, we show that $Q_5 = o_p(1)$.

Using the definition of $D_{1i}(\beta, \Lambda, \pi, S, S_c), D_{2i}(\beta, \Lambda, \pi, S, S_c)$ defined in Supplementary Material~\ref{AIPW:appendix:notations},
it can be verified that
\ba{
    U(\beta^*, \pi, S, S_c) = \frac{1}{n}\sum_{i=1}^n \left[ D_{2i}(\beta^*, \Lambda^*, \pi, S, S_c) - \int_0^\tau \bar{A}(t;\beta^*, \pi, S, S_c) D_{1i}(t; \beta^*, \Lambda^*, \pi, S, S_c) \right] \label{AIPW:l4.7}.
}
So we have $Q_5 = Q_{51} - Q_{52} - Q_{53} - Q_{54}$, where
\ba{
Q_{51} =& \frac{1}{\sqrt{n}} \sum_{i=1}^n \{D_{2i}(\beta^*, \Lambda^*, \wh{\pi}, S^o, \wh{S}_c) - D_{2i}(\beta^*, \Lambda^*, \pi^o, S^o, S_c^o)\} \\
Q_{52} =& \frac{1}{\sqrt{n}} \sum_{i=1}^n  \int_0^\tau \bar{\alpha}(t;\beta^*, \wh{\pi}, S^o, \wh{S}_c) \{D_{1i}(t; \beta^*, \Lambda^*, \wh{\pi}, S^o, \wh{S}_c) - D_{1i}(t; \beta^*, \Lambda^*, \pi^o, S^o, S_c^o) \} \\
Q_{53} =& \frac{1}{\sqrt{n}} \sum_{i=1}^n  \int_0^\tau \{\bar{A}(t;\beta^*, \wh{\pi}, S^o, \wh{S}_c) - \bar{\alpha}(t;\beta^*, \wh{\pi}, S^o, \wh{S}_c)\} \\
&\times \{D_{1i}(t; \beta^*, \Lambda^*, \wh{\pi}, S^o, \wh{S}_c) - D_{1i}(t; \beta^*, \Lambda^*, \pi^o, S^o, S_c^o) \} \\
Q_{54} =& \int_0^\tau \{\bar{A}(t;\beta^*, \wh{\pi}, S^o, \wh{S}_c) - \bar{A}(t;\beta^*, \pi^o, S^o, S_c^o) \} \cdot \frac{1}{\sqrt{n}} \sum_{i=1}^n D_{1i}(t; \beta^*, \Lambda^*, \pi^o, S^o, S_c^o).
}
First, consider $Q_{51}$.
By the law of total variance, we have 
\ba{
\mbox{Var}(Q_{51}) = \mbox{Var}\{E(Q_{51}|O^\dag)\} + E\{\mbox{Var}(Q_{51}|O^\dag)\}.
}
We note from Theorem~\ref{thm:dr:avebeta} that $E\{D_{2i}(\beta^*, \Lambda^*, \wh{\pi}, S^o, \wh{S}_c) - D_{2i}(\beta^*, \Lambda^*, \pi^o, S^o, S_c^o)|O^\dag\} = 0$ for each $i$, 
where $O^\dag$ is the sample independent from $O$ that is used for estimating the nuisance functions, so $E(Q_{51}|O^\dag) = 0$. Moreover, when conditional on $O^\dag$, $Q_{51}$ is a sample average of mean-zero i.i.d terms, so we have
\ba{
\mbox{Var}(Q_{51}|O^\dag) =& \frac{n}{n} E\left[ \{D_{2}(\beta^*, \Lambda^*, \wh{\pi}, S^o, \wh{S}_c) - D_{2}(\beta^*, \Lambda^*, \pi^o, S^o, S_c^o)\}^2 | O^\dag \right].
}
Expand $D_{2}(\beta^*, \Lambda^*, \wh{\pi}, S^o, \wh{S}_c) - D_{2}(\beta^*, \Lambda^*, \pi^o, S^o, S_c^o)$, we have
\ba{
&D_{2}(\beta^*, \Lambda^*, \wh{\pi}, S^o, \wh{S}_c) - D_{2}(\beta^*, \Lambda^*, \pi^o, S^o, S_c^o) \label{AIPW:l4.37} \\
=& - \int_0^\tau \frac{A\{\wh{\pi}(Z) - \pi^o(Z)\}}{\wh{\pi}(Z)\pi^o(Z)} \cdot \{ dS^o(t;A,Z) + S^o(t;A,Z)e^{\beta^* A}d\Lambda^*(t) \}\\
&+ \int_0^\tau \frac{A S_c^o(t;A,Z) \{\pi^o(Z) - \wh{\pi}(Z)\}}{\wh{\pi}(Z)S_c^o(t;A,Z)\wh{S}_c(t;A,Z)\pi^o(Z)^A\{1-\pi^o(Z)\}^{1-A}} \cdot \{dN(t) - Y(t)e^{\beta^*}d\Lambda^*(t)\}\\
&+ \int_0^\tau \frac{A \wh{\pi}(Z) \{S_c^o(t;A,Z) - \wh{S}_c(t;A,Z)\}}{\wh{\pi}(Z)S_c^o(t;A,Z)\wh{S}_c(t;A,Z)\pi^o(Z)^A\{1-\pi^o(Z)\}^{1-A}} \cdot \{dN(t) - Y(t)e^{\beta^*}d\Lambda^*(t)\}\\
&- \int_0^\tau \frac{AJ(t;1,S^o, S_c^o)\{\wh{\pi}(Z) - \pi^o(Z)\}\}}{\wh{\pi}(Z)\pi^o(Z)} \cdot \{dS^o(t;1,Z) + S^o(t;1,Z)e^{\beta^*}d\Lambda^*(t) \}\\
&+ \int_0^\tau \frac{A \{ J(t;1,S^o, \wh{S}_c) -  J(t;1,S^o, S_c^o) \}}{\wh{\pi}(Z)} \cdot \{dS^o(t;1,Z) + S^o(t;1,Z)e^{\beta^*}d\Lambda^*(t) \}.
}
We now see that $D_{2}(\beta^*, \Lambda^*, \wh{\pi}, S^o, \wh{S}_c) - D_{2}(\beta^*, \Lambda^*, \pi^o, S^o, S_c^o)$ consists of several terms, where each term is an integral of a difference in nuisance functions with respect to a monotone function. This allows us to apply \eqref{AIPW:sup ineq} to each of the terms and have
\ba{
|D_{2}(\beta^*, \Lambda^*, \wh{\pi}, S^o, \wh{S}_c) - D_{2}(\beta^*, \Lambda^*, \pi^o, S^o, S_c^o)| \lesssim |\wh{\pi}(Z) - \pi^o(Z)| + \sup_{t \in [0,\tau], a \in \{0,1\}}|S_c^o(t;a,Z) - \wh{S}_c(t;a,Z)|.
}
From the inequality $(a+b)^2 \le 2a^2 + 2b^2$, we have 
\ba{
\mbox{Var}(Q_{51}|O^\dag) \lesssim& E\left[ \left(|\wh{\pi}(Z) - \pi^o(Z)| + \sup_{t \in [0,\tau], a \in \{0,1\}}|S_c^o(t;a,Z) - \wh{S}_c(t;a,Z)|\right)^2   \bigg|O^\dag \right] \\
\le& 2 E [ \{\wh{\pi}(Z) - \pi^o(Z)\}^2 |O^\dag ] + 2 E \left[ \left\{\sup_{t \in [0,\tau], a \in \{0,1\}}|S_c^o(t;a,Z) - \wh{S}_c(t;a,Z)| \right\}^2  \bigg |O^\dag \right].
}
So 
\ba{
&\mbox{Var}(Q_{51})\\
=& \mbox{Var}^\dag\{E(Q_{51}|O^\dag)\} + E^\dag\{\mbox{Var}(Q_{51}|O^\dag)\} \\
\lesssim& 0 + E^\dag( E [  \{ \wh{\pi}(Z) - \pi^o(Z)\}^2 |O^\dag ]) + E^\dag \left(  E \left[ \left\{\sup_{t \in [0,\tau], a \in \{0,1\}}|S_c^o(t;a,Z) - \wh{S}_c(t;a,Z)| \right\}^2  \bigg |O^\dag \right] \right) \\
=& \|\wh{\pi} - \pi^o\|_\dag^2 + \|\wh{S}_c - S_c^o\|_\dag^2 \\
=& o(1).
}
Therefore, $Q_{51} = o_p(1)$ by Chebyshev's inequality.

Conditional on $O^\dag$, we also have from Theorem~\ref{thm:dr:avebeta} that $E\{D_{1i}(t;\beta^*, \Lambda^*, \wh{\pi}, S^o, \wh{S}_c) - D_{1i}(t;\beta^*, \Lambda^*, \pi^o, S^o, S_c^o)|O^\dag\} = 0$ for each $t$ and $i$, so $Q_{52}$ is again a sample average of i.i.d. mean-zero terms when conditional on $O^\dag$, and we can show $Q_{52}=o_p(1)$ in the same way as for $Q_{51}$ above.

Consider $Q_{53}$. Just like the expansion of $D_{2}(\beta^*, \Lambda^*, \wh{\pi}, S^o, \wh{S}_c) - D_{2}(\beta^*, \Lambda^*, \pi^o, S^o, S_c^o)$ in \eqref{AIPW:l4.37} above, we also have $D_{1i}(t;\beta^*, \Lambda^*, \wh{\pi}, S^o, \wh{S}_c) - D_{1i}(t;\beta^*, \Lambda^*, \pi^o, S^o, S_c^o)$ as a sum of terms, where each term is a product between a difference in nuisance functions and an increment of a monotone function. So same as in $Q_{51}$, we apply \eqref{AIPW:sup ineq} to each of the terms and have
\ba{
    |Q_{53}| \lesssim&  \sqrt{n} \sup_{t \in [0,\tau]} \left| \bar{A}(t;\beta^*, \wh{\pi}, S^o, \wh{S}_c) - \bar{\alpha}(t;\beta^*, \wh{\pi}, S^o, \wh{S}_c) \right| \\
    &\cdot \left\{ \frac{1}{n} \sum_{i=1}^n |\wh{\pi}(Z_i) - \pi^o(Z_i)| + \frac{1}{n} \sum_{i=1}^n \sup_{t \in [0,\tau], a \in \{0,1\}}|S_c^o(t;a,Z_i) - \wh{S}_c(t;a,Z_i)| \right\}.
}
From the uniform convergence Assumption~\ref{assump6} and the Markov's inequality, we have
\ba{
\frac{1}{n} \sum_{i=1}^n  \left|\wh{\pi}(Z_i) - \pi^o(Z_i) \right| + \frac{1}{n} \sum_{i=1}^n  \sup_{t \in [0,\tau], a\in\{0,1 \} }  \left|S_c^o(t;a,Z_i) -\wh{S}_c(t;a,Z_i) \right| = o_p(1).
}
From \eqref{C3second} of Assumption~\ref{assumpA4}, we have
\ba{
\sqrt{n} \sup_{t \in [0,\tau]} \left| \bar{A}(t;\beta^*, \wh{\pi}, S^o, \wh{S}_c) - \bar{\alpha}(t;\beta^*, \wh{\pi}, S^o, \wh{S}_c) \right| = O_p(1).
}
We therefore have $Q_{53} = o_p(1)$.

For $Q_{54}$, we have $Q_{54} = Q_{541} - Q_{542} + Q_{543}$, where
\ban{
    Q_{541} =& \int_0^\tau \{\bar{A}(t;\beta^*, \wh{\pi}, S^o, \wh{S}_c) - \bar{\alpha}(t;\beta^*, \wh{\pi}, S^o, \wh{S}_c)\} \cdot \frac{1}{\sqrt{n}} \sum_{i=1}^n D_{1i}(t; \beta^*, \Lambda^*, \pi^o, S^o, S_c^o), \\
    Q_{542} =& \int_0^\tau \{\bar{A}(t;\beta^*, \pi^o, S^o, S_c^o) - \bar{\alpha}(t;\beta^*, \pi^o, S^o, S_c^o) \} \cdot \frac{1}{\sqrt{n}} \sum_{i=1}^n D_{1i}(t; \beta^*, \Lambda^*, \pi^o, S^o, S_c^o),\\
    Q_{543} =& \frac{1}{\sqrt{n}} \sum_{i=1}^n  \int_0^\tau \{\bar{\alpha}(t;\beta^*, \wh{\pi}, S^o, \wh{S}_c) - \bar{\alpha}(t;\beta^*, \pi^o, S^o, S_c^o)\} D_{1i}(t; \beta^*, \Lambda^*, \pi^o, S^o, S_c^o)
}
By \eqref{C3third} of Assumption~\ref{assumpA4}, we have $Q_{541} = o_p(1)$ and $Q_{542} = o_p(1)$. 
$Q_{543}$ is again a sample average of i.i.d. terms when conditional on $O^\dag$, and each of the increments in  $D_{1i}(t; \beta^*, \Lambda^*, \pi^o, S^o, S_c^o)$ is an increment of a monotone function. So like $Q_{51}$, we apply \eqref{AIPW:sup ineq}, followed by the law of total variance and have
\ba{
 \mbox{Var}(Q_{543}) \lesssim 0 + \frac{n}{n} E^\dag \left( E \left[ \left\{\sup_{t\in[0,\tau]} \left|\bar{\alpha}(t;\beta^*, \wh{\pi}, S^o, \wh{S}_c) - \bar{\alpha}(t;\beta^*, \pi^o, S^o, S_c^o) \right| \right\}^2 \right] \right)= o(1),
}
where $o(1)$ follows from \eqref{C3first} of Assumption~\ref{assumpA4}. Therefore, $Q_{543} = o_p(1)$ by Chebyshev's inequality and  $Q_{54} = o_p(1)$.

Combining our results on $Q_{51}$ to $Q_{54}$ ,we have $Q_5 = o_p(1)$.

Same as how we dealt with $Q_5$, we can decompose $Q_6$ in a similar way and show that each of the terms is $o_p(1)$, so we omit the details here.

Lastly, we consider $\sqrt{n}U(\beta^*, \pi^o,S^o, S_c^o)$. Using \eqref{AIPW:l4.7},
we have
\ba{
    &\sqrt{n}U(\beta^*, \pi^o, S^o, S_c^o) \\
    =& \frac{1}{\sqrt{n}}\sum_{i=1}^n \left[ D_{2i}(\beta^*, \Lambda^*, \pi^o, S^o, S_c^o) - \int_0^\tau \bar{A}(t;\beta^*, \pi^o, S^o, S_c^o) D_{1i}(t;\beta^*, \Lambda^*, \pi^o, S^o, S_c^o) \right] \\
    =& \frac{1}{\sqrt{n}}\sum_{i=1}^n \psi_i(\beta^*,\Lambda^*,\pi^o, S^o, S_c^o)  \\
    &+ \int_0^\tau \{ \bar{\alpha}(t;\beta^*, \pi^o, S^o, S_c^o) -  \bar{A}(t;\beta^*, \pi^o, S^o, S_c^o)\}\cdot \frac{1}{\sqrt{n}}\sum_{i=1}^n  D_{1i}(t;\beta^*, \Lambda^*, \pi^o, S^o, S_c^o). \\ \label{AIPW:l4.3}
}
From \eqref{C3third} of Assumption~\ref{assumpA4}, we have $\eqref{AIPW:l4.3} = o_p(1)$, so we have
\ba{
    \sqrt{n}U(\beta^*, \pi^o,S^o, S_c^o)  = \frac{1}{\sqrt{n}}\sum_{i=1}^n \psi_i(\beta^*,\Lambda^*,\pi^o, S^o, S_c^o) + o_p(1).
}
\qed

\subsection{\texorpdfstring{Consistency of {$\hat{\Lambda}(t)$}}{Consistency of Lambda(t)}}
Given the consistency of $\whb$, Assumptions 1-6 and C1-C3 from Cox MSM, when either $S = S^o$, or $(\pi, S_c) = (\pi^o, S_c^o)$, we show that for each $t$,
\ba{
\wh{\Lambda}(t; \whb, \wh{\pi}, \wh{S}, \wh{S}_c) \overset{p}{\to} \Lambda^*(t),
}
where
\ba{
    \wh{\Lambda}(t;\beta, \pi, S, S_c) = \frac{1}{n}\sum_{i =1}^n \int_0^t \frac{ d\mathcal{N}^{(0)}_{i}(u;\pi,S,S_c)}{\mathcal{S}^{(0)}(u; \beta,\pi, S,S_c)}.
}
and 
\ba{
\Lambda^*(t) = \int_0^t \frac{\sum_{a=0,1} dF_a(t)}{ \sum_{a=0,1} S_a(t)e^{\beta^* a}} = - \int_0^t \frac{\sum_{a=0,1} dS_a(t)}{ \sum_{a=0,1} S_a(t)e^{\beta^* a}}
}
{\it Proof.}

Let $\wh{\Lambda}(t; \whb, \wh{\pi}, \wh{S}, \wh{S}_c) - \Lambda^*(t) = L_1 + L_2 + L_3$, where
\ba{
&L_1 = \wh{\Lambda}(t; \whb, \wh{\pi}, \wh{S}, \wh{S}_c) -  \wh{\Lambda}(t; \beta^*, \wh{\pi}, \wh{S}, \wh{S}_c)\\
&L_2 = \wh{\Lambda}(t; \beta^*, \wh{\pi}, \wh{S}, \wh{S}_c) - \wh{\Lambda}(t; \beta^*, \pi^*, S^*, S_c^*) \\
&L_3 = \wh{\Lambda}(t; \beta^*, \pi^*, S^*, S_c^*) - \Lambda^*(t).
}

Consider $L_1$, which can be written as
\ba{
L_1 =& \frac{1}{n}\sum_{i =1}^n \int_0^t \frac{ d\mathcal{N}^{(0)}_i(u;\wh{\pi}, \wh{S}, \wh{S}_c)}{\mathcal{S}^{(0)}(u; \whb,\wh{\pi}, \wh{S}, \wh{S}_c)\mathcal{S}^{(0)}(u; \beta^*,\wh{\pi}, \wh{S}, \wh{S}_c)} \cdot \{ \mathcal{S}^{(0)}(u; \beta^*,\wh{\pi}, \wh{S}, \wh{S}_c) - \mathcal{S}^{(0)}(u; \whb,\wh{\pi}, \wh{S}, \wh{S}_c) \} \\
=&\frac{1}{n}\sum_{i =1}^n \int_0^t \frac{ d\mathcal{N}^{(0)}_i(u;\wh{\pi}, \wh{S}, \wh{S}_c)}{\mathcal{S}^{(0)}(u; \whb,\wh{\pi}, \wh{S}, \wh{S}_c)\mathcal{S}^{(0)}(u; \beta^*,\wh{\pi}, \wh{S}, \wh{S}_c)}\\
& \times \frac{1}{n} \sum_{j =1}^n \Bigg[ \frac{A^l_j Y_j(t)\{e^{\beta^* A_j} -  e^{\whb A_j} \}}{\pi(Z_j)^{A_j}\{1-\pi(Z_j)\}^{1-A_j}S_c(t;A_j,Z_j)} - \frac{A^l_j S(t;A_j,Z_j)\{e^{\beta^* A_j} - e^{\whb A_j} \}}{\pi(Z_j)^{A_j}\{1-\pi(Z_j)\}^{1-A_j}} \\
    &\quad + \sum_{a=0,1} a^l \left\{1 + \frac{A_j^a (1-A_j)^{1-a}}{\pi(Z_j)^a \{1 - \pi(Z_j)\}^{1-a}}J_j(t;a,S,S_c)\right\}S(t;a,Z_j)\{e^{\beta^* a} - e^{\whb a}\} \Bigg]
}
Since $e^{\beta^* a} - e^{\whb a} = o_p(1)$ and $e^{\beta^* A_j} - e^{\whb A_j} = o_p(1)$, while the other terms are bounded a.s., we can apply the inequality
\ba{
   \left |\int_a^b f(t) dG(t) \right| \le \sup_{t \in [a,b]} |f(t)| \cdot |G(b) - G(a)|
}
and have $L_1 = o_p(1)$.

Consider $L_2$, which can be written as
\ba{
L_2 =& \frac{1}{n}\sum_{i =1}^n \int_0^t \frac{ d\mathcal{N}^{(0)}_i(u;\wh{\pi}, \wh{S}, \wh{S}_c) - d\mathcal{N}^{(0)}_i(u;\pi^*, S^*, S_c^*)}{\mathcal{S}^{(0)}(u; \beta^*,\wh{\pi}, \wh{S}, \wh{S}_c)} \\
+& \frac{1}{n}\sum_{i =1}^n \int_0^t \frac{ d\mathcal{N}^{(0)}_i(u;\pi^*, S^*, S_c^*)}{\mathcal{S}^{(0)}(u; \beta^*,\wh{\pi}, \wh{S}, \wh{S}_c)\mathcal{S}^{(0)}(u; \beta^*,\pi^*, S^*, S_c^*)} \cdot \{ \mathcal{S}^{(0)}(u; \beta^*,\wh{\pi}, \wh{S}, \wh{S}_c) - \mathcal{S}^{(0)}(u; \beta^*,\pi^*, S^*, S_c^*)\}.
}
As we can see, we are dealing with differences between $\wh{\pi}, \wh{S}, \wh{S}_c$ and $\pi^*, S^*, S_c^*$, which is the same as in Proof of Lemma \ref{AIPW:lem:U_to_mu and dU_to_nu}, 
so we omit the details for showing $L_2 = o_p(1)$ here.

Lastly, for $L_3$, we first note that from the definition of $D_1$, we have
\ba{
L_3 =& \frac{1}{n}\sum_{i =1}^n \int_0^t \frac{ 
d\mathcal{N}^{(0)}_{i}(u;\pi^*, S^*, S_c^*)}{\mathcal{S}^{(0)}(u; \beta^*,\pi^*, S^*, S_c^*)} - \Lambda^*(t) \\
=& \frac{1}{n}\sum_{i =1}^n \int_0^t \frac{
d\mathcal{N}^{(0)}_{i}(u;\pi^*, S^*, S_c^*)
}{\mathcal{S}^{(0)}(u; \beta^*,\pi^*, S^*, S_c^*)} 
-\frac{1}{n}\sum_{i =1}^n  \int_0^t \frac{\Gamma_i^{(0)}(u;\beta^*,\pi^*, S^*, S_c^*)}{\mathcal{S}^{(0)}(u; \beta^*,\pi^*, S^*, S_c^*)} d\Lambda^*(t) 
\\
=& \frac{1}{n}\sum_{i =1}^n \int_0^t \frac{D_{1i}(u; \beta^*, \Lambda^*, \pi^*, S^*, S_c^*)}{\mathcal{S}^{(0)}(u; \beta^*,\pi^*, S^*, S_c^*)} \\
=& \frac{1}{n}\sum_{i =1}^n \int_0^t \frac{D_{1i}(u; \beta^*, \Lambda^*, \pi^*, S^*, S_c^*)}{\mathpzc{s}^{(0)}(u; \beta^*,\pi^*, S^*, S_c^*)}  \label{TATE:consistency.1} \\
+& \frac{1}{n}\sum_{i =1}^n \int_0^t \frac{D_{1i}(u; \beta^*, \Lambda^*, \pi^*, S^*, S_c^*)}{\mathcal{S}^{(0)}(u; \beta^*,\pi^*, S^*, S_c^*) \mathpzc{s}^{(0)}(u; \beta^*,\pi^*, S^*, S_c^*)} \cdot \{ \mathpzc{s}^{(0)}(u; \beta^*,\pi^*, S^*, S_c^*) -  \mathcal{S}^{(0)}(u; \beta^*,\pi^*, S^*, S_c^*)\}. \label{TATE:consistency.2}
}
From the Theorem of DR, $D_{1i}(u; \beta^*, \Lambda^*, \pi^*, S^*, S_c^*)$ is mean-zero, while $\mathpzc{s}^{(0)}(u; \beta^*,\pi^*, S^*, S_c^*)$ is a fixed function, so the entire integral in \eqref{TATE:consistency.1}  is mean-zero, and by the weak law of large numbers \eqref{TATE:consistency.1}  is $o_p(1)$.
\eqref{TATE:consistency.2} is $o_p(1)$ because $\sup_{t \in [0,\tau]}\left| \mathpzc{s}^{(0)}(u; \beta^*,\pi^*, S^*, S_c^*) -  \mathcal{S}^{(0)}(u; \beta^*,\pi^*, S^*, S_c^*) \right| = o_p(1)$. So $L_3 = o_p(1)$.

\qed

\end{document}